 \documentclass[a4paper,onecolumn,notitlepage,nofootinbib,superscriptaddress,prd]{revtex4-1}
 
\usepackage{amsmath}
\usepackage{graphicx}
\usepackage{color}
\usepackage{ulem}
\usepackage{bm}

\usepackage{natbib,hyperref,ifthen}
\newcommand{\cdoth}{\!\cdot\!}
\renewcommand{\mathbf}{\vec}

\newcommand{\bk}{\mathbf{k}}
\newcommand{\bkappa}{\mathbf{\kappa}}
\newcommand{\bq}{\mathbf{q}}
\newcommand{\bp}{\mathbf{p}}
\newcommand{\bx}{\mathbf{x}}
\newcommand{\by}{\mathbf{y}}
\newcommand{\ra}{\mathrm{a}}
\newcommand{\rb}{\mathrm{b}}

\newcommand{\rn}{\mathrm{n}}
\newcommand{\tf}{\tilde{F}}
\newcommand{\be}{\begin{equation}}
\newcommand{\ee}{\end{equation}}
\newcommand{\hMpc}{h^{-1}\ \text{Mpc}}
\newcommand{\hMpcsq}{h^{-2}\ \text{Mpc}^2}
\newcommand{\ihMpc}{h\ \text{Mpc}^{-1}}

\renewcommand{\vec}{\bm}
\newcommand{\derd}{\text{d}}
\newcommand{\deltadir}{\delta^\text{(D)}}
\newcommand{\deltakron}{\delta^\text{(K)}}
\newcommand{\vhk}{\vec{\hat k}}

\newcommand{\sigmadsq}{\sigma_\text{d}^2}

\newcommand{\cssq}{c_\text{s}^2}

\newcommand{\lcdm}{$\Lambda$CDM}

\allowdisplaybreaks

\begin{document}
\title{Precise Calibration of the One-Loop Bispectrum in the Effective Field Theory of Large Scale Structure}
\author{Theodore Steele}
\email{ts715@cam.ac.uk}
\affiliation{Department of Applied Mathematics and Theoretical Physics, University of Cambridge}
\author{Tobias Baldauf}
\affiliation{Department of Applied Mathematics and Theoretical Physics, University of Cambridge}

\date{\today}
\begin{abstract}
The bispectrum is the leading non-Gaussian statistic in Large-Scale Structure (LSS) clustering and encodes the interactions in the underlying field.  It is thus an important diagnostic for  primordial non-Gaussianity and higher order galaxy biasing.
In this paper we present a detailed test and calibration of the matter bispectrum counterterms in the Effective Field Theory of LSS against a suite of $N$-body simulations. We are going beyond previous studies in employing realisation based perturbation theory that allows for a significant reduction in cosmic variance error bars. This enables the measurement of the low-energy constants on large scales before two-loop corrections become relevant, around $k<0.09 \ihMpc$ at $z=0$.  We also go beyond previous work in using bispectrum propagator terms, i.e. correlators with linear and second order fields, to quantify the two new counterterms in isolation and to establish consistency with the power spectrum counterterm.
By investigating the fully non-linear bispectrum, $B_{\mathrm{nnn}}$, as well as the terms $B_{\mathrm{n}11}$ and $B_{\mathrm{n}21}$, we find evidence for the new counterterms deviating from the shape suggested by the UV-limit of the relevant bispectrum contributions.
We also show that the commonly used Einstein-de Sitter approximation for the time dependence of the tree-level bispectrum is insufficient for precise studies of the one-loop bispectrum and that it is necessary to use \lcdm~growth factors in order to obtain meaningful one-loop counterterm constraints. Finally, we also find evidence for small deviations in the growth factors that arise from time integration inaccuracies in the $N$-body simulations.
\end{abstract}
\maketitle

\tableofcontents

\section{Introduction}
Cosmological probes are observables that can be used to constrain the parameter space of fundamental theories; examples include the Cosmic Microwave Background (CMB) \cite{Aghanim:2018eyx,Akrami:2018vks} and the distribution of matter and galaxies across the universe, known as cosmological Large-Scale Structure (LSS).  Since its discovery, the CMB has proven to be a rich source of cosmological information.
While it has provided many valuable insights and tight parameter constraints on the \lcdm\ model, its statistical power has been mostly exhausted. Yet many open questions remain, such as the mass of neutrinos or the dynamics and field content of inflation. LSS offers to provide significantly more information than the CMB on account of its three dimensional nature.  However, modelling LSS is more difficult than modelling the CMB due to non-linearities in the late-time Universe.

Until the beginning of the last decade, there was a strong ambition to describe LSS in a parameter free, deterministic approach referred to as Standard Perturbation Theory (SPT) \cite{rev}.  This model is based on a fluid approximation to the Vlaslov equation describing an ensemble of collisionless particles and solves the equations perturbatively using a power series in the density and velocity divergence fields.  However, the results of SPT beyond tree-level are in tension with simulations as SPT does not take into account the small-scale effects of gravitational collapse. These cannot be described perturbatively and, through their resultant effects on the gravitational field, lead to non-perturbative effects on cosmic structure at larger scales.  

To solve this problem, a modification to SPT, called the Effective Field Theory of Large Scale Structure (EFTofLSS) \cite{Baumann:2010tm,Carrasco:2012cv,Carrasco:2013mua,Pajer:2013jj}, has been developed.  The EFTofLSS introduces an arbitrarily chosen coarse graining or regulation scale $\Lambda$ that allows for a convergent series expansion for wavenumbers below the non-linear wavenumber $k_\text{NL}$, beyond which physics is non-perturbative.  The coarse grained equations of motion are now $\Lambda$-dependent and contain an effective stress term that encapsulates microscopic velocity dispersion and products of small scale modes.  As these effects are non-perturbative, they need to be parametrised in a broad manner that allows for all operators that are compatible with the symmetries of the model.  The new terms this introduces into the perturbative expansion are called counterterms and account for small scale deviations from the pressureless perfect fluid assumed by SPT, absorbing divergences and the dependence on $\Lambda$, making the overall model cutoff independent.

Recently, the potential of the EFTofLSS to deliver unbiased cosmological constraints has been proven in a blind analysis of a simulation suite \cite{Nishimichi:2020tvu}.  The same approach has then been used to perform an analysis of galaxy correlators in the BOSS survey \cite{Ivanov:2019pdj,DAmico:2019fhj}, which can be connected to the overall matter and dark matter correlators by taking into account the appropriate biasing terms. The latter account for the fact that the galaxy distribution correlates with but does not perfectly mimic the underlying dark matter distribution \cite{Desjacques:2016bnm}.

A lot of attention has been devoted to the EFT at the level of the power spectrum \cite{Carrasco:2012cv,Carrasco:2013mua,Baldauf:2015aha}, the two-point correlator and therefore the simplest statistic in LSS, while less attention has been devoted to the bispectrum \cite{Angulo:2014tfa,Baldauf:2014qfa} and trispectrum \cite{Bertolini:2015fya,Bertolini:2016bmt}. The galaxy bispectrum at one-loop has recently been studied in \cite{Eggemeier:2018qae}. The higher order statistics offer additional information as well as the opportunity to perform consistency checks with the EFT parameters calculated from the power spectrum.  Further tests of the EFT bispectrum have also been performed in \cite{Lazanu:2015rta} and their calculations have been pushed to two-loop order in \cite{Lazanu:2018yae}.  We improve upon these results by implementing the full \lcdm\ time dependence rather than the commonly employed Einstein-de Sitter approximation for cosmological perturbation theory and cancelling cosmic variance through realisation perturbation theory. We analyse the effects of these corrections for the power spectrum, auto bispectrum, and bispectrum propagators. 

This paper will use a simulation suite previously studied in \cite{Baldauf:2014qfa,Baldauf:2015aha}. It is based on a $\Omega_\text{m,0}=0.272$, $\sigma_8=0.81$, $n_\text{s}=0.967$ cosmology with $h=0.724$. The $N_\text{p}=1024^3$ particles in a cubic box of dimension $L=1500\hMpc$ are set up at initial redshift $z_\text{i}=99$ using 2LPT \cite{Crocce:2006ve} and then evolved to $z=0$ using Gadget 2 \cite{Springel:2005mi}. We are considering 14 realisations of this simulation volume. The realization perturbation theory approach has previously been used to evaluate and test SPT \cite{Roth:2011test,Taruya:2018jtk,Taruya:2020qoy}, but not to our knowlegde to constrain the EFT counterterms.

This paper is structured as follows:
We first review the EFT and the corresponding counterterms in Sec.~\ref{EFT}, before reviewing the power spectrum and bispectrum as well as their \lcdm\ time dependence and our realisation perturbation theory implementation in Sec.~\ref{grid}.  
We then explain the various fitting procedures we used to estimate the counterterm amplitudes before presenting parameter constraints in Sec.~\ref{results}.  
We finally conclude and offer a discussion of our various results and their implications in Sec.~\ref{sec:conc}.
\section{Perturbative Modelling of the Bispectrum}
\label{EFT}
\subsection{The Effective Field Theory}
Cosmological perturbation theory is based on an ensemble of $N$ collisionless point particles \cite{rev} whose overall phase space density is given by $f(\bx,\bp)=\sum_{i=1}^{N}f_{i}(\bx,\bp)$, where $\vec p$ is the canonical momentum and  $f_{i}(\mathbf{x},\mathbf{p})=\deltadir(\mathbf{x}-\mathbf{x}_{i})\deltadir(\mathbf{p}-ma\mathbf{v}_{i})$.  Such an ensemble obeys the collisionless Boltzmann equation
\begin{equation}
\frac{\text{D}f}{\text{D}t}=\frac{\partial f}{\partial t}+\frac{\mathbf{p}}{ma^{2}}\cdot \frac{\partial f}{\partial \mathbf{x}}- m\frac{\partial \phi}{\partial \mathbf{x}}\cdot \frac{\partial f}{\partial \mathbf{p}}=0~\, ,
\label{EFTBoltz}
\end{equation}
where $\phi$ is the gravitational potential.
We coarse grain the phase space by introducing a smoothing function, $W_{\Lambda}(k)$. The presence of this function differentiates the EFT from SPT, and it is defined by 
\begin{equation}
W_{\Lambda}(k)=e^{-\frac{1}{2}\frac{k^{2}}{\Lambda^{2}}}~,
\end{equation}
for some cutoff $\Lambda<k_\text{NL}$. Here $k_\text{NL}$ is the non-linear scale beyond which the physics is non-perturbative. Throughout this paper we choose $\Lambda=0.3\ihMpc$ unless explicitly stated otherwise.  Note that the theory is independent of the cutoff and that the need for an explicit cutoff is of a numerical nature as discussed in detail below in Sec.~\ref{eq:realpt}.  

Taking the first two moments of Eq.~\eqref{EFTBoltz} gives us definitions for the matter and momentum densities,
\begin{equation}
\rho(\bx,\tau)=\frac{m}{a^{3}}\int \derd^{3}y \int d^{3}q~ W_{\Lambda}(\bx-\by)~f(\by,\bp)
\end{equation}
and
\begin{equation}
\boldsymbol{\pi}(\bx,\tau)=\frac{1}{a^{4}}\int \derd^{3}y \int d^{3} \bp~ W_{\Lambda}(\bx-\by)~ \bp~ f(\by,\bp)~.
\end{equation}
Applying these to the Boltzmann equation allows us to derive the equations of motion
\begin{equation}
\dot{\rho}+3H\rho+\frac{1}{a}\partial_{i}(\rho v^{i})=0
\label{rho}
\end{equation}
and
\begin{equation}
\dot{v}^{i}+Hv^{i}+\frac{1}{a}v^{j}\partial_{j}v^{i}+\frac{1}{a}\partial_{i}\phi=-\frac{1}{a\rho}\partial_{j}[\tau^{ij}]_{\Lambda},
\label{tau}
\end{equation}
where $\dot{x}=\partial_{t}x$, $H=\dot{a}/a$ is the Hubble rate, $\mathbf{v}=\vec{\pi}/\rho+\mathrm{counterterms}$ is the fluid velocity \cite{Mercolli:2013bsa}, and the stress-energy tensor $\tau^{ij}$ describes the non-perturbative effects of small scale physics. The stress-energy tensor is set to zero in SPT.  Taking the derivative of Eq.~\eqref{tau} and defining the velocity divergence field $\theta=\partial_{i}v^{i}$, we obtain
\begin{equation}
\partial_{\eta}\theta+\mathcal{H}\theta+v^{j}\partial_{j}\theta+\partial_{i}v^{j}\partial_{j}v^{i}+\triangle\phi=\tau_{\theta}~,
\label{eq2}
\end{equation}
where $\eta$ is the conformal time defined through $a \derd \eta=\derd t$ and
\begin{equation}
\tau_{\theta}=-\partial_{i}\left[\frac{1}{\rho}\partial_{j}\tau^{ij}\right]\, .
\end{equation}

Defining the density contrast $\delta=\rho/\bar{\rho}-1$, Eq.~\eqref{rho} becomes
\begin{equation}
\partial_{\eta}\delta+\nabla \cdot \left[\left(1+\delta\right)v^{i}\right]=0~.
\label{delta}
\end{equation}
Rewriting Eqs.~\eqref{tau} and \eqref{delta} in Fourier space gives us
\begin{equation}
\partial_{\eta}\delta(\mathbf{k},\eta)+\theta(\mathbf{k},\eta)=\mathcal{S}_{\alpha}(\mathbf{k},\eta),
\label{eom1}
\end{equation}
and
\begin{equation}
\partial_{\eta}\theta(\mathbf{k},\eta)+\mathcal{H}\theta(\mathbf{k},\eta)+\frac{3}{2}\Omega_{\mathrm{m}}\mathcal{H}^{2}\delta(\mathbf{k},\eta)=\mathcal{S}_{\beta}(\mathbf{k},\eta)~.
\label{eom2}
\end{equation}
The two non-linear source terms $S_{\alpha}$ and $S_{\beta}$ may be defined as
\begin{equation}
S_{\alpha}(\mathbf{k},\eta)=-\int \frac{\mathrm{d}^{3}\mathbf{p}}{(2\pi)^{3}}\alpha(\mathbf{q},\mathbf{k}-\mathbf{q})\theta(\mathbf{q},\eta)\delta(\mathbf{k}-\mathbf{q},\eta)~,
\end{equation}
\begin{equation}
S_{\beta}(\mathbf{k},\eta)=-\int \frac{\mathrm{d}^{3}\mathbf{p}}{(2\pi)^{3}}\beta(\mathbf{q},\mathbf{k}-\mathbf{q})\theta(\mathbf{q},\eta)\theta(\mathbf{k}-\mathbf{q},\eta)+\tau_{\theta}(\mathbf{k},\eta)~.
\label{eq:Sbeta}
\end{equation}
The kernels $\alpha$ and $\beta$ encapsulating the coupling between modes are defined as
\begin{equation}
\alpha(\mathbf{k}_{1},\mathbf{k}_{2})=\frac{\mathbf{k}_{1}\cdot (\mathbf{k}_{1}+\mathbf{k}_{2})}{k_{1}^{2}}
\end{equation}
and
\begin{equation}
\beta(\mathbf{k}_{1},\mathbf{k}_{2})=\frac{1}{2}(\mathbf{k}_{1}+\mathbf{k}_{2})^{2}\frac{\mathbf{k}_{1}\cdot\mathbf{k}_{2}}{k_{1}^{2}k_{2}^{2}}\, .
\end{equation}

We can solve the equations of motion using a power series ansatz $\delta_{n}(\vec k,a)=D_{n}\delta_{n}(\vec k)$ for growth factor $D_{n}$, where in a matter dominated universe we simply have $D_{n}=a^{n}$.  
The $n$-th order density field is given by
\begin{equation}
\delta_{\mathrm{n}}(\bk)=\int_{\bq_{1}}^{\Lambda}\cdots\int_{\bq_{\mathrm{n}}}^{\Lambda}\deltadir \Biggl(\bk-\sum_{i=1}^{n}\bq_{i}\Biggr)
F_{\mathrm{n}}(\bq_{1},...,\bq_{\mathrm{n}})\prod_{i=1}^{n}\delta_{1}(\bq_{i})~,
\label{deltan}
\end{equation}
where $\int_{\bq}=\int \derd^{3}\bq / (2\pi)^{3}$ and the functions of momenta $F_{\mathrm{n}}$ are the gravitational coupling kernels of the density field.  The corresponding coupling kernels for the velocity divergence are referred to as $G_n$.  The $F_n$ and $G_n$ kernels are given by the recursion relations
\begin{align}
F_{\mathrm{n}}(\bk_{1},...,\bk_{\mathrm{n}})=&\sum^{n-1}_{m=1}\frac{G_{m}(\bk_{1},...,\bk_{m})}{(2n+3)(n-1)}[(2n+1)\alpha(\bkappa_{1}^m,\bkappa_{m+1}^n)F_{\mathrm{n}-m}(\bk_{m+1},...,\bk_{\mathrm{n}})\nonumber\\&+2\beta(\bkappa_{1}^m,\bkappa_{m+1}^n)G_{\mathrm{n}-m}(\bk_{m+1},...,\bk_{\mathrm{n}})]~,\label{eq:recursf}\\
G_{\mathrm{n}}(\bk_{1},...,\bk_{\mathrm{n}})=&\sum^{n-1}_{m=1}\frac{G_{m}(\bk_{1},...,\bk_{m})}{(2n+3)(n-1)}[3\alpha(\bkappa_{1}^m,\bkappa_{m+1}^n)F_{\mathrm{n}-m}(\bk_{m+1},...,\bk_{\mathrm{n}})\nonumber\\&+2n\beta(\bkappa_{1}^m,\bkappa_{m+1}^n)G_{\mathrm{n}-m}(\bk_{m+1},...,\bk_{\mathrm{n}})]~\label{eq:recursg},
\end{align}
where $\kappa_{a}^b=\sum_{i=a}^b k_i$.  In clustering statistics these kernels are used in their symmetrised form referred to as $F_n^\text{(s)}$ and $G_n^\text{(s)}$.


\subsection{The Effective Stress-Energy Tensor}
To satisfy the symmetries of the cosmological model, the components of the effective stress-energy tensor $\tau_{ij}$ must be compatible with homogeneity and isotropy as well as Galilean invariance. The only gravitational terms which are compatible with these symmetries are the second derivatives of the gravitational potential, $\nabla_{i}\nabla_{j}\phi$, and of the velocity potential, $\nabla_{i}\nabla_{j}u$.

The various terms in the stress-energy tensor can be grouped into two categories: viscosity terms, which provide the correlated corrections to the perturbative terms, and noise terms, which account for the self-coupling of non-perturbative (strongly coupled) small scale modes analogous to the one-halo term in the halo model.

At cubic order, when the stress-tensor regularises only the cubic couplings at one-loop, there are two terms in the source term for the velocity divergence shown in Eq.~\eqref{eq:Sbeta}:
\begin{equation}
    \tau_\theta\vert_{1}=-d_{\phi}^{2}\partial^{i}\Delta\Phi+\frac{d_{u}^{2}}{\mathcal{H}f}\partial^{i}\Delta u~,
\end{equation}
for free parameters $d_{\phi}^{2}$ and $d_{u}^{2}$ which we combine in the new variable $d^{2}\equiv d_{\phi}^{2}+d_{u}^{2}$.  

At quartic order, where we have to regularise $\delta_{4}$,  this expands to become
\begin{equation}
\tau^{ij}_\theta\vert_{2}=-d^{2}\delta^{ij}\delta_{2}+c_{1}\delta^{ij}(\Delta\Phi)^{2}+c_{2}\partial^{i}\partial^{j}\Phi\Delta\Phi+c_{3}\partial^{i}\partial_{k}\Phi\partial^{j}\partial^{k}\Phi~.
\end{equation}
We will find it useful to integrate these over time to define the counterterm density field at scale factor $a$ as\footnote{By combining Equations~\eqref{eom1} and \eqref{eom2} into a second order equation we can obtain the Green's function \cite{Baldauf:2016sjb}
\begin{equation}
        G_{\delta}(a,a')=\Theta(a-a')\frac{2}{5}\frac{1}{\mathcal{H}_{0}^{2}\Omega_\text{m}^{0}}\frac{D_{1}(a')}{a'}\left[\frac{D_{1-}(a)}{D_{1-}(a')}-\frac{D_{1}(a)}{D_{1}(a')}\right]~,
\end{equation}
for first order growing mode $D_{1}$ and decaying mode $D_{1-}(a)=H(a)$.}
\begin{equation}
    \tilde{\delta}_{n}=\int \derd a' G_{\delta}(a,a')\tau_{\theta}\vert_{n}~.
    \label{intdelta}
\end{equation}

Defining the tidal tensor
\begin{equation}
    s_{ij}\equiv \partial^{i}\partial^{j}\Phi - \frac{1}{3}\delta^{ij}\Delta\Phi~,
\end{equation}
one can obtain 
\begin{align}
    \tau_{\theta}\vert_{1}&=-d^{2}\Delta\delta_{1}~,\\
    \label{eq:ctrtermsreal}
    \tau_{\theta}\vert_{2}&=-d^{2}\Delta\delta_{2}-e_{1}\Delta\delta^{2}_{1}-e_{2}\Delta s^{2}-e_{3}\partial_{i}\left[s^{ij}\partial_{j}\delta_{1}\right]~,
\end{align}
where $e_{i}$ are functions of $c_{i}$ and $d_{u}^{2}$, leaving us with four free parameters: $d^{2}$, $e_{1}$, $e_{2}$, and $e_{3}$.

We can define the counterterm density field analogously to Eq.~\eqref{deltan} by defining a set of counterkernels $\tilde{F}$, such that
\begin{equation}
        \tilde{\delta}_{\mathrm{n}}(\bk)\equiv \int_{\bq_{1}}...\int _{\bq_{\mathrm{n}}} \deltadir\left(\bk-\sum_{i=1}^{n}\bq_{i}\right)\tilde{F}_{\mathrm{n}}(\bq_{1},...,\bq_{\mathrm{n}})\prod_{i=1}^{n}\delta_{1}(\bq_{i})~.
\end{equation}

\subsection{The Counterterms}
We wish to define the $n$th-order counterkernels in terms of the free parameters of the $n$th order stress-energy tensor.  The parameters $d^{2}$ and $e_{1,2,3}$ are time dependent; we can approximate them as scaling as $D_{1}^{m}(a)$ for $m$-dependent power series growth factors $D^{m}$.
The value for $m$ can be inferred from approximating the linear power spectrum by a power law $k^{n_\text{NL}}$ around the non-linear scale and assuming self similarity, leading to $m=(1-n_\text{NL})/(n_\text{NL}+3)$ \cite{Pajer:2013jj,Baldauf:2014qfa}. For the small-scale terms in the EFT, we estimate $n_\text{NL}\approx -3/2$, yielding $m=5/3$. The part of the counterterm that is capturing the cutoff dependence of the SPT loop integrals needs to have $m=1$ instead.
We can therefore separate the time dependent and time independent components by defining the new parameters $\bar{d}^{2}$ and $\bar{e}_{i}$ as
\begin{align}
d^{2}&\equiv D^{m}_{1}(a)\mathcal{H}_{0}^{2}\Omega_\text{m}^{0}\overline{d}~,\\
e_{i}^{2}&\equiv D^{m}_{1}(a)\mathcal{H}_{0}^{2}\Omega_\text{m}^{0}\overline{e}_{i}~.
\end{align}
In order to obtain the cubic order counterterm, we insert this definition of $d^{2}$ into the cubic order stress-energy tensor and insert the result into Eq.~\eqref{intdelta}.  The time dependency arising from the Green's function in the integral is \cite{Baldauf:2014qfa}
\begin{equation}
g_{1}(a,m)=-\frac{2}{(m+1)(2m+7)}D_{1}^{m+1}(a)~,
\end{equation}
such that
\begin{align}
    \tilde{\delta}_{1}(\bk,a)=g_{1}(a,m)\bar{d}^{2}k^{2}\delta_{1}(\bk,a)~,
\end{align}
which gives us
\begin{equation}
    \tilde{F}_{1}(\bk,a)= g_{1}(a,m)\bar{d}^{2}  k^{2}~.
    \label{gamma1}
\end{equation}

In order to define a counterkernel from the quartic order stress-energy tensor, we will split it into three components \cite{Baldauf:2014qfa}:
\begin{equation}
    \tilde{F}_{2}(\bk_{1}\bk_{2})\equiv \tilde{F}_{2}^{\tau}(\bk_{1},\bk_{2})+\tilde{F}_{2}^{\alpha\beta}(\bk_{1},\bk_{2})+\tilde{F}_{2}^{\delta}(\bk_{1},\bk_{2})~,
\label{F2t1}
\end{equation}
where $\tilde{F}_{2}^{\tau}$ stems from the three terms containing $e_{i}$, $\tilde{F}_{2}^{\alpha\beta}$ comes from inserting $\tilde{\delta}_{1}$ and $\tilde{\theta}_{1}$ into the source terms, and $\tf_{2}^{\delta}$ comes from replacing $\delta_{1}$ with $\delta_{2}$ in the $d^{2}$ term in Eq.~\eqref{eq:ctrtermsreal}.
Solving the equations of motion with $k^2 \delta_1$ as one of the source terms in the coupling kernel and considering $k^2 \delta_2$ in the source terms and integrating these sources with the Green's function leads to

\begin{equation}
\tilde{F}_{2}(\bk_{1},\bk_{2})=\left[\sum_{i=1}^{3}g_{2}^{e}\bar{e}_{i}E_{i}(\bk_{1},\bk_{2})+g_{1}\bar{d}^{2}\Gamma(\bk_{1},\bk_{2})\right]~,
\end{equation}
where
\begin{equation}
g_{2}^{\mathrm{e}}(a,m)=-\frac{2}{(m+2)(2m+9)}D_{1}^{m+1}(a)~,
\label{g2e}
\end{equation}
and the $E_i$ are shape functions which, combined with the free parameters $e_{i}$, regulate the shape dependence of the counterterms and arise from the symmetry inspired terms quadratic terms in the gravitational potential:
\begin{align}
E_1(\vec k_1,\vec k_2)=&(k_1+k_2)^2~,\\
E_2(\vec k_1,\vec k_2)=&(k_1+k_2)^2\left(\frac{(\vec k_1\cdot \vec k_2)^2}{k_1^2k_2^2}-\frac13 \right)~,\\
E_3(\vec k_1,\vec k_2)=&-\frac{1}{6}(k_{1}+k_{2})^{2}+\frac{1}{2}\bk_{2}\cdot\bk_{3}\left(\frac{(\bk_{2}+\bk_{3})\cdot\bk_{3}}{k_{3}^{2}}+\frac{(\bk_{2}+\bk_{3})\cdot\bk_{2}}{k_{2}^{2}}\right)~.
\end{align}
Once integrated over time, the $\Gamma$ function is given by
\begin{equation}
    \Gamma(\vec k_1,\vec k_2)=\left(\frac{g^{e}_{2}(a,m)}{g_{1}(a,m)}(\bk_{1}+\bk_{2})^{2}F_{2}(\bk_{1},\bk_{2})+\tilde{F}_{2}^{\alpha\beta}(\bk_{1},\bk_{2})\right)~,
\end{equation}
with
\begin{equation}
\begin{split}
    \tilde{F}_{2}^{\alpha\beta}(\bk_{1},\bk_{2})=&k_{2}^{2}\left[\frac{g_{2}^{\ra}(a,m)}{g_{1}(a,m)}\alpha(\bk_{2},\bk_{1})+\frac{\tilde{g}_{2}^{\ra}(a,m)}{g_{1}(a,m)}\alpha(\bk_{1},\bk_{2})+\frac{g_{2}^{\rb}(a,m)}{g_{1}(a,m)}\beta(\bk_{1},\bk_{2})\right]\\
    &+k_{1}^{2}\left[\frac{g_{2}^{\ra}(a,m)}{g_{1}(a,m)}\alpha(\bk_{1},\bk_{2})+\frac{\tilde{g}_{2}^{\ra}(a,m)}{g_{1}(a,m)}\alpha(\bk_{2},\bk_{1})+\frac{g_{2}^{\rb}(a,m)}{g_{1}(a,m)}\beta(\bk_{1},\bk_{2})\right]~\, .
\end{split}
\end{equation}
where
\begin{align}
g_{2}^{\ra}(a,m)&=-\frac{1}{(m+1)(m+2)(2m+9)}D_{1}^{m+1}(a)~,\\
\tilde{g}_{2}^{\ra}(a,m)&=(m+2) g_{2}^{\ra}(a,m)~,\\
g_{2}^{\rb}(a,m)&=-\frac{4}{(m+1)(2m+7)(2m+9)}D_{1}^{m+1}(a)~.
\end{align}
For simplicity, we can then define a new set of parameters
\begin{align}
    \gamma_{1}&=-g_{1}(a,m)\bar{d}^{2}~,\\
    \gamma_{2}&=-g_{1}(a,m)\bar{d}^{2}~,\\
    \epsilon_{i}&=-g_{2}^{e}(a,m)\bar{e}_{i}~,
\end{align}
where the two $\gamma$ terms are defined identically in terms of the model's symmetries but are treated separately when calculating their values and where we have simply incorporated the time dependent terms into our definition of the parameters, giving us
\begin{equation}
    \tilde{F}_{1}=-\gamma_{1}k^{2}~,
    \label{F1t}
\end{equation}
\begin{equation}
\tilde{F}_{2}(\bk_{1},\bk_{2})=-\left[\sum_{i=1}^{3}\epsilon_{i}E_{i}(\bk_{1},\bk_{2})+\gamma_{2}\Gamma(\bk_{1},\bk_{2})\right]\, .
\label{eq:symmf2tilde}
\end{equation}
These definitions will form the basis of our implementation of the EFT to regularise the one-loop bispectrum.  According to convention, $\gamma_{1}$ is referred to as the speed of sound, $\cssq$, when calculated from the power spectrum.

\subsection{The Power Spectrum}
\label{bispectraPTsec}
The power spectrum is the correlator of two fields, $(2\pi)^3\delta^\text{(D)}(\vec k_1+\vec k_2)P_{AB}(k_1)=\langle \delta_A(\vec k_1)\delta_B(\vec k_2) \rangle$, and constitutes the simplest and most easily measurable large-scale structure statistic.  There are two non-linear power spectra of interest to this analysis, the auto power spectrum $P_{\mathrm{nn}}$, and the propagator $P_{\mathrm{n}1}$, given by
\begin{align}
    P_{\mathrm{nn}}=&P_{11}+P_{22}+2P_{31}+2P_{\tilde{1}1}~,\\
    P_{\mathrm{n}1}=&P_{11}+P_{31}+P_{\tilde{1}1}~,
\end{align}
where all power spectra are functions of the magnitude $|\vec k|$ of the wavevector only. Here $P_{11}$ is the linear power spectrum, i.e. the correlator of two Gaussian fields $\delta_1$ and fully describes the initial conditions in the absence of primordial non-Gaussianity. The one-loop contributions are given in perturbation theory by
\begin{equation}
P_{31}(k)=3P_{11}(k)\int_{\mathbf{q}}~F_{3}^\text{(s)}(\mathbf{k},\bq,-\bq)P_{11}(q)~,
\end{equation}
\begin{equation}
P_{22}(k)=2\int_{\mathbf{q}}~F_{2}^\text{(s)}(\mathbf{k-q},\mathbf{q})^{2}P_{11}(|\mathbf{k}-\mathbf{q}|)P_{11}(q)~.
\end{equation}

\begin{figure}[h!]
\centering
\includegraphics[width=16cm]{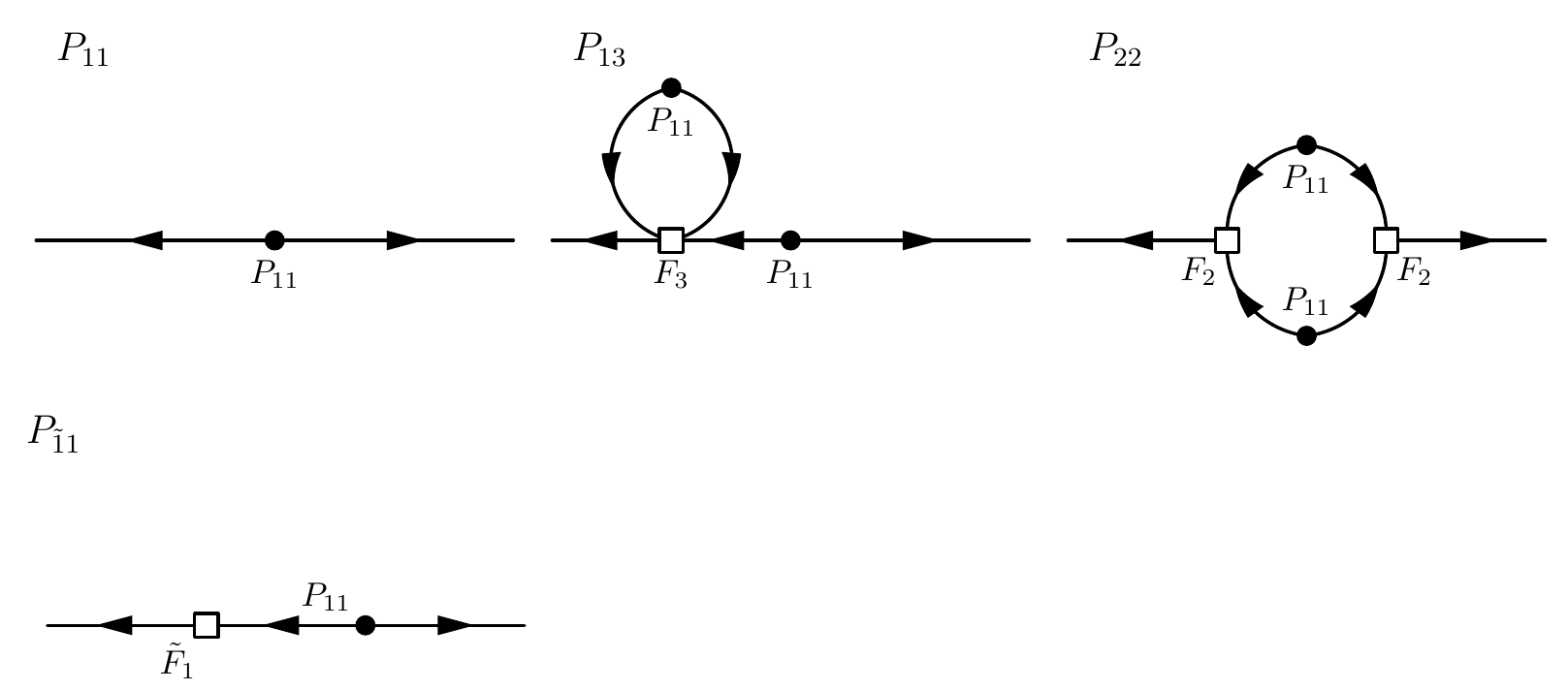}
\caption{Feynman diagram representations of the linear and one-loop contributions to the power spectra of large-scale structure together with the  one-loop counterterm $P_{\tilde{1}1}$.}
\label{FPS}
\end{figure}

The term $P_{31}$ is regularised by the counterterm $P_{\tilde{1}1}=\tilde{F}_{1}P_{11}$, the diagram for which is shown alongside those for $P_{11}$ and the one-loop terms in Fig.~\ref{FPS}. Following from Eq.~\eqref{F1t}, $\tilde{F}_{1}=-\cssq k^{2}$, leaving one free parameter, the speed of sound.

\subsubsection{UV-sensitivity}
The $P_{31}$ term in the one-loop matter power spectrum at small external wavenumbers can be expanded as
\be
\lim_{k\to 0}P_{31}(k)=-\frac{61}{630} k^2 P_{11}(k)\int_{\vec q} \frac{P_{11}(q)}{q^2}=-\frac{61}{210}k^2 P_{11}(k)\sigma_\text{d}^2\, ,
\label{eq:p13uv}
\ee
where the displacement dispersion $\sigmadsq$ is given by
\be
\sigmadsq(\Lambda)=\frac{1}{6\pi^2}\int_0^\Lambda \derd q P_{11}(q)\, .
\label{eq:sigmadsq}
\ee
The regularisation of the propagator term $P_{31}\to P_{31}+P_{\tilde 11}$ suggests the replacement \cite{Baldauf:2015aha}
\be
\sigmadsq\to\sigmadsq+\frac{210}{61}\cssq\, .
\label{eq:sigmadrep}
\ee
For the cosmology considered in this study we have the displacement dispersion of modes up to our cutoff $\sigmadsq(\Lambda=0.3\ihMpc)=32.9\hMpcsq$. The displacement dispersion of all modes would be
$\sigmadsq(\Lambda\to \infty)=36.56\hMpcsq$
The difference between the two corresponds to the running of the EFT counterterm amplitude
\be
\Delta \cssq=\frac{61}{210} \left[\sigmadsq(\Lambda\to \infty)-\sigmadsq(\Lambda =0.3\ihMpc)\right]=1.054 \hMpcsq\, .
\label{eq:csrunhere}
\ee

\subsection{The Bispectrum}

\begin{figure}[h!]
    \centering  
    \includegraphics[width=0.8\textwidth]{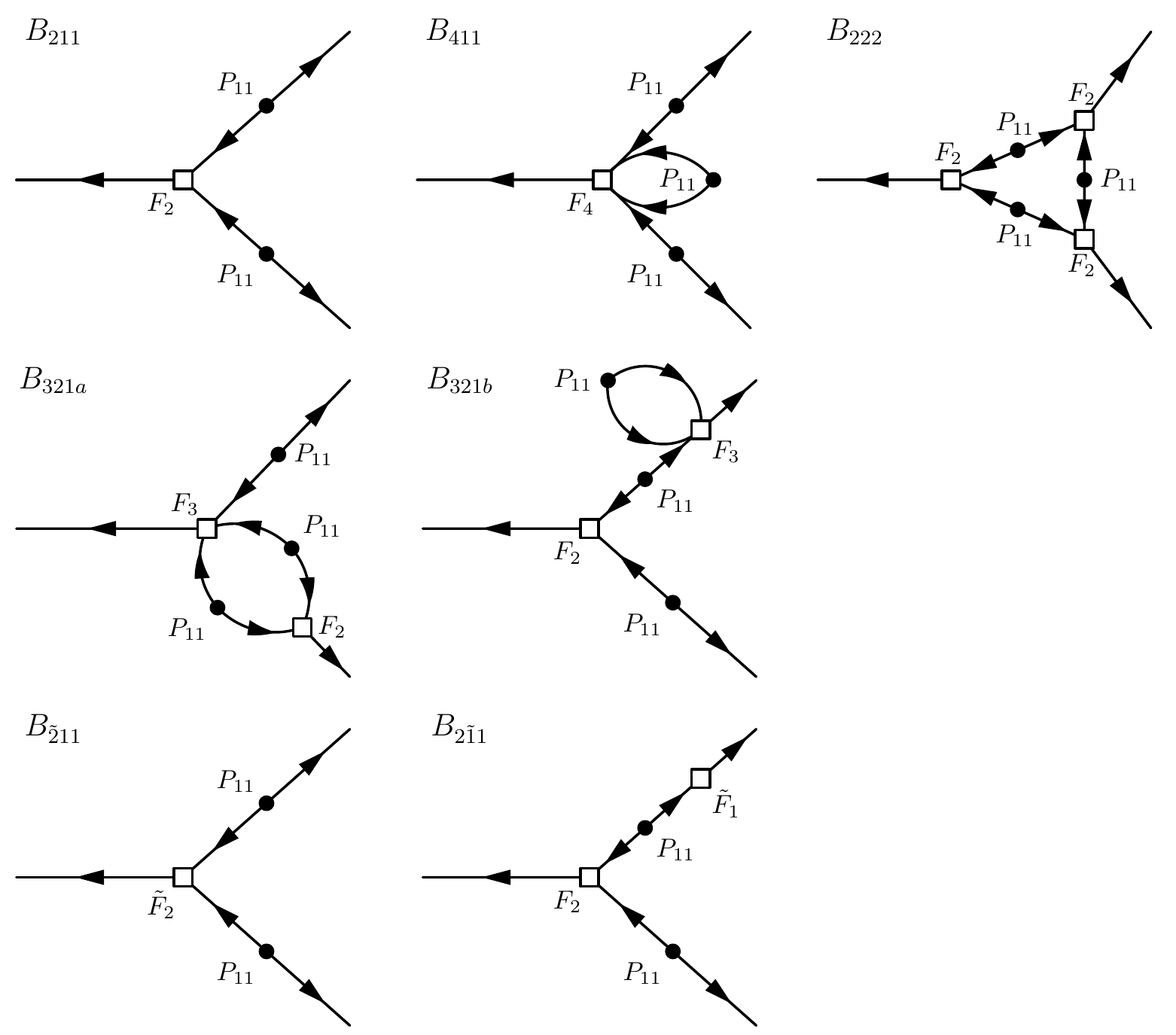}
    \caption{Feynman diagram representations of the contributions to the tree-level and one-loop bispectrum together with the one-loop counterterms.}
    \label{bisfeyn}
\end{figure}

The bispectrum is the correlator of three fields $(2\pi)^3\delta^\text{(D)}(\vec k_1+\vec k_2+\vec k_3)B_{ABC}(k_1,k_2,k_3)=\langle \delta_A(\vec k_1)\delta_B(\vec k_2)\delta_C(\vec k_3)\rangle$ and vanishes in a Gaussian universe, encapsulating information about cosmic non-Gaussianities.  We will be studying three non-linear bispectra: i) the auto bispectrum $B_{\mathrm{nnn}}$, the correlator of three non-linear fields; ii) the propagator $B_{\mathrm{n}11}$, the correlator of one non-linear and two linear fields; and iii) a term which we informally also refer to as a propagator, $B_{\mathrm{n}21}$, the correlator of a non-linear, a second-order, and a linear field. These can be described perturbatively as: 
\begin{align}
    B_{\mathrm{nnn}}=&B_{211}^\text{s}+B_{411}^\text{s}+B_{321}^\text{s}+B_{222}+B_{\tilde{2}11}^\text{s}+B_{\tilde{1}21}^\text{s}\\
    B_{\mathrm{n}11}=&B_{211}+B_{411}+B_{\tilde{2}11}\\
    B_{\mathrm{n}21}=&B_{321}+B_{\tilde{1}21}
\end{align}
where all bispectra are functions of $(k_1,k_2,k_3)$ and 
\begin{align}
&B_{211}(k_{1},k_{2},k_{3})=2F^\text{(s)}_{2}(k_{2},k_{3})P_{11}(k_{2})P_{11}(k_{3})~,\\
&B_{411}(k_{1},k_{2},k_{3})=12\int_{\vec q}F^\text{(s)}_{4}(\bq,-\bq,-\bk_{2},-\bk_{3})P_{11}(q)P_{11}(k_{2})P_{11}(k_{3})~,\\
&B_{321a}(k_{1},k_{2},k_{3})=6\int_{\vec q}F^\text{(s)}_{3}(-\bq,\bq-\bk_{2},-\bk_{3})F^\text{(s)}_{2}(\bq,\bk_{2}-\bq)P_{11}(q)P_{11}(|\bk_{2}+\bq|)P_{11}(k_{3})~,\\
&B_{321b}(k_{1},k_{2},k_{3})=6\int_{\vec q}F^\text{(s)}_{3}(\bq,-\bq,\bk_{3})F^\text{(s)}_{2}(\bk_{2},\bk_{3})P_{11}(q)P_{11}(k_{2})P_{11}(k_{3})~,\label{eq:b321bdef}\\
&B_{222}(k_{1},k_{2},k_{3})=8\int_{\vec q}F^\text{(s)}_{2}(-\bq,\bk_{3}+\bq) F^\text{(s)}_{2}(\bk_{2}-\bq,\bk_{3}+\bq))F^\text{(s)}_{2}(\bk_{2}-\bq)\nonumber\\ &\hspace{3cm}\times P_{11}(q)P_{11}(|\bk_{2}-\bq|)P_{11}(|\bk_{3}-\bq|)~,
\end{align}
with $B_{321}=B_{321a}+B_{321b}$.  These perturbative contributions to the bispectra are represented diagrammatically in Fig.~\ref{bisfeyn}.

For the auto bispectrum, $B_{\mathrm{nnn}}$, we symmetrise over the permutations\footnote{For the sake of definiteness the cyclic permutations are given by
$[\{k_1,k_2,k_3\},\{k_2,k_3,k_1\},\{k_3,k_1,k_2\}]$ and all permutations by $[\{k_1,k_2,k_3\},\{k_2,k_3,k_1\},\{k_3,k_1,k_2\},\{k_2,k_1,k_3\},\{k_1,k_3,k_2\},\{k_3,k_2,k_1\}]$.}
\begin{align}
B_{211}^\text{s}(k_1,k_2,k_3)&=B_{211}(k_1,k_2,k_3)+2\mathrm{~permutations}~,\\
B_{411}^\text{s}(k_1,k_2,k_3)&=B_{411}(k_1,k_2,k_3)+2\mathrm{~permutations}~,\\
B_{321}^\text{s}(k_1,k_2,k_3)&=B_{321}(k_1,k_2,k_3)+5\mathrm{~permutations}~,\\
B_{\tilde{2}11}^\text{s}(k_1,k_2,k_3)&=B_{\tilde{2}11}(k_1,k_2,k_3)+2\mathrm{~permutations}~,\\
B_{\tilde{1}21}^\text{s}(k_1,k_2,k_3)&=B_{\tilde{1}21}(k_1,k_2,k_3)+5\mathrm{~permutations}~.
\end{align}
To retain the configuration dependence we do not perform a symmetrisation for $B_{\mathrm{n}11}$ and $B_{\mathrm{n}21}$.

Up to one-loop, two connected three-field correlators can be constructed which include the viscosity terms from the stress-energy tensor:
\begin{equation}
    \langle \tilde{\delta}_{1}(\bk_{1})\delta_{2}(\bk_{2})\delta_{1}(\bk_{1})\rangle, ~\mathrm{and} ~\langle \tilde{\delta}_{2}(\bk_{1})\delta_{1}(\bk_{2})\delta_{1}(\bk_{3})\rangle~,
\end{equation}
which correspond to the bispectra
\begin{align}
    B_{\tilde{1}21}(\bk_{1},\bk_{2},\bk_{3};\gamma_1)&=2F_{2}(\bk_{2},\bk_{3})P_{\tilde 11}(\bk_{2})P_{11}(\bk_{3})~,\nonumber\\
    &=2\tilde{F}_{1}(\bk_{2};\gamma_1) F_{2}(\bk_{2},\bk_{3})P_{11}(\bk_{2})P_{11}(\bk_{3})~,\\
    B_{\tilde{2}11}(\bk_{1},\bk_{2},\bk_{3};\gamma_2,\epsilon_i)&=2\tilde{F}_{2}(\bk_{2},\bk_{3};\gamma_2,\epsilon_i)P_{11}(\bk_{2})P_{11}(\bk_{3})~,
    \end{align}
where $\tilde{F}_{1}$ and $\tilde{F}_{2}$ are defined in Eqs.~\eqref{F1t} and \eqref{eq:symmf2tilde}, respectively.

\subsubsection{$\Lambda$CDM time dependence}
\label{sec:lcdmtimedep}
For convenience and simplicity, the gravitational coupling kernels in SPT are often derived under the assumption of a matter-only Einstein-de Sitter (EdS) Universe, with scale factors replaced by linear growth factors \cite{rev}.  This approximation is usually considered to be fairly accurate, with the use of more accurate $\Lambda$CDM growth factors having been shown to lead to sub-percent level corrections in the one-loop power spectrum \cite{Exact,Fasiello:2016qpn}.  

While in the power spectrum the leading order term is unaffected by changes in growth factor due to its containing only the linear $\delta_{1}$, for the bispectrum the tree-level result is affected by the \lcdm\ corrections to $\delta_{2}$.  These corrections to the tree-level bispectrum can be encoded by the second order gravitational coupling kernel 
\be
F_{2,\Lambda\text{CDM}}(\vec k_1,\vec k_2)=\frac{5}{7} \frac{D_{2A}(a)}{D_1^2(a)} \alpha(\vec k_1, \vec k_2) + \frac{2}{7} \frac{D_{2B}(a)}{D_1^2(a)} \beta(\vec k_1, \vec k_2)
\label{eq:F2lcdm}
\ee
Here $D_{2,A}$ and $D_{2,B}$ are the exact \lcdm\ growth factors, which would both reduce to $D_{2,A,B}=D_{1}^2$ in EdS (for an explicit expression of the \lcdm~growth factors see App.~\ref{app:exgrowth}).  

\begin{figure}[t]
\includegraphics[width=0.49\textwidth]{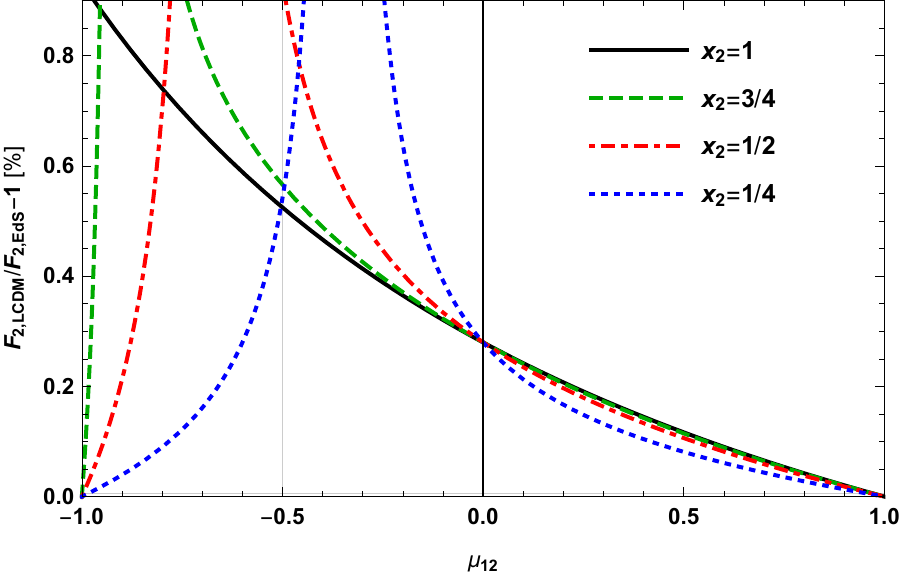}
\includegraphics[width=0.49\textwidth]{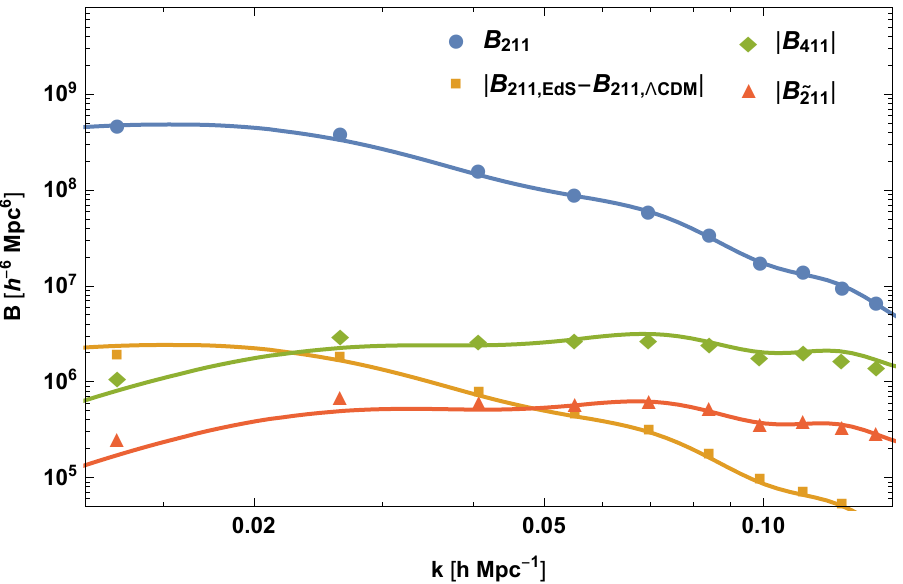}
\caption{
\emph{Left panel:} Fractional deviations between the \lcdm\ and EdS-like quadratic coupling kernel as a function of $x_2=k_2/k_1$ and $\mu_{12}=\vec{\hat k}_1\cdot \vec{\hat k}_2$.  The kernels agree in the aligned configuration $\vec k_1 \parallel \vec k_2$ and show deviations of the order of $0.5 \%$ otherwise.
\emph{Right panel:} Equilateral bispectrum contributions to the propagator $B_{\mathrm{n}11}$.  The difference between the EdS and \lcdm\ tree-level bispectra (orange) exceeds the amplitude of the one-loop bispectrum contribution from $B_{411}$ for $k<0.02\ihMpc$ and the typical size of the EFT counterterms from $B_{\tilde{2}11}$ for $k<0.08 \ihMpc$.  The orange data points show that our grid implementation of the \lcdm~second order density field agrees with the analytical calculation.
}
\label{fig:fracdevf2}
\end{figure}

\begin{figure}[h]
\centering
\includegraphics[width=0.49\textwidth]{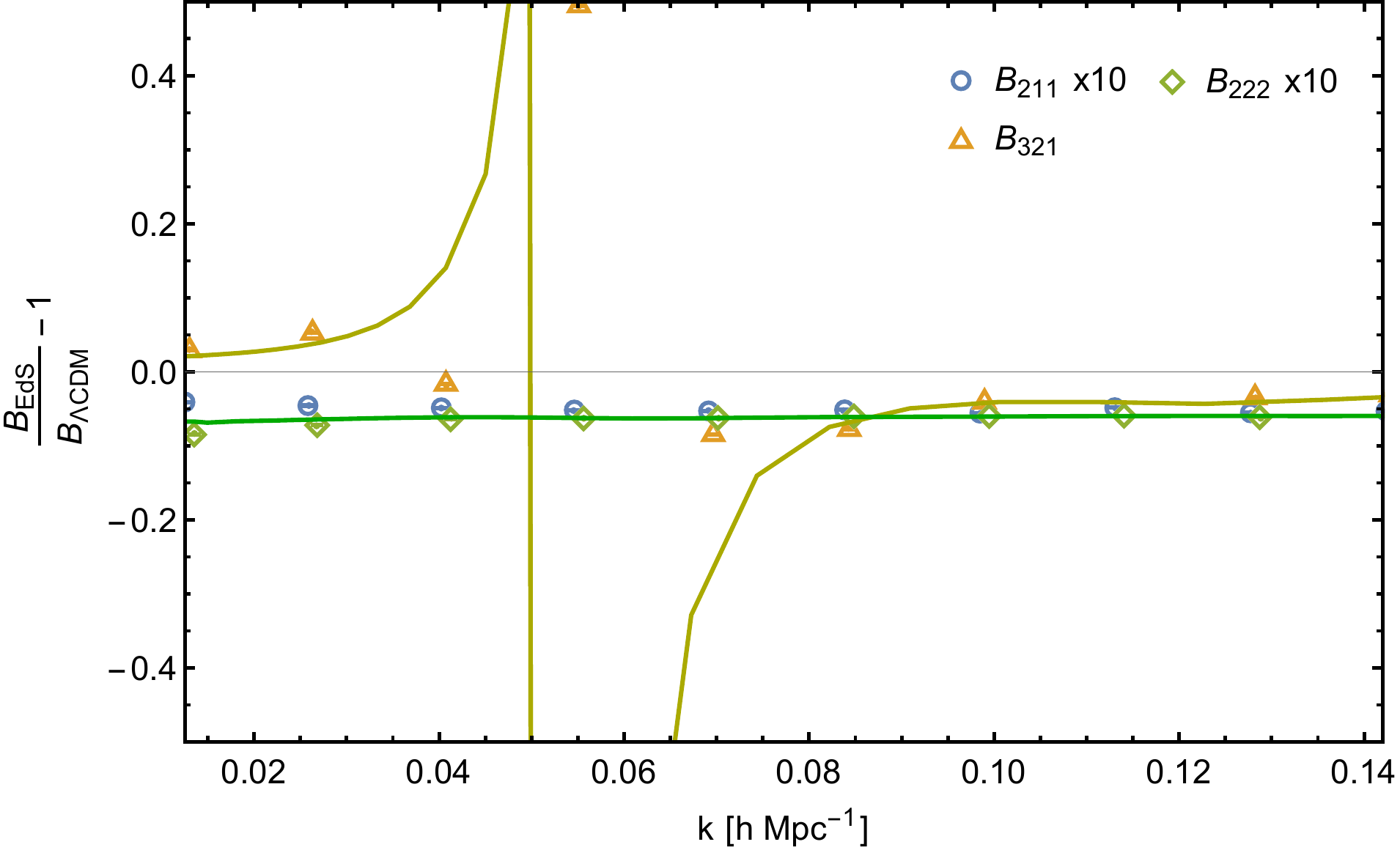}
\includegraphics[width=0.49\textwidth]{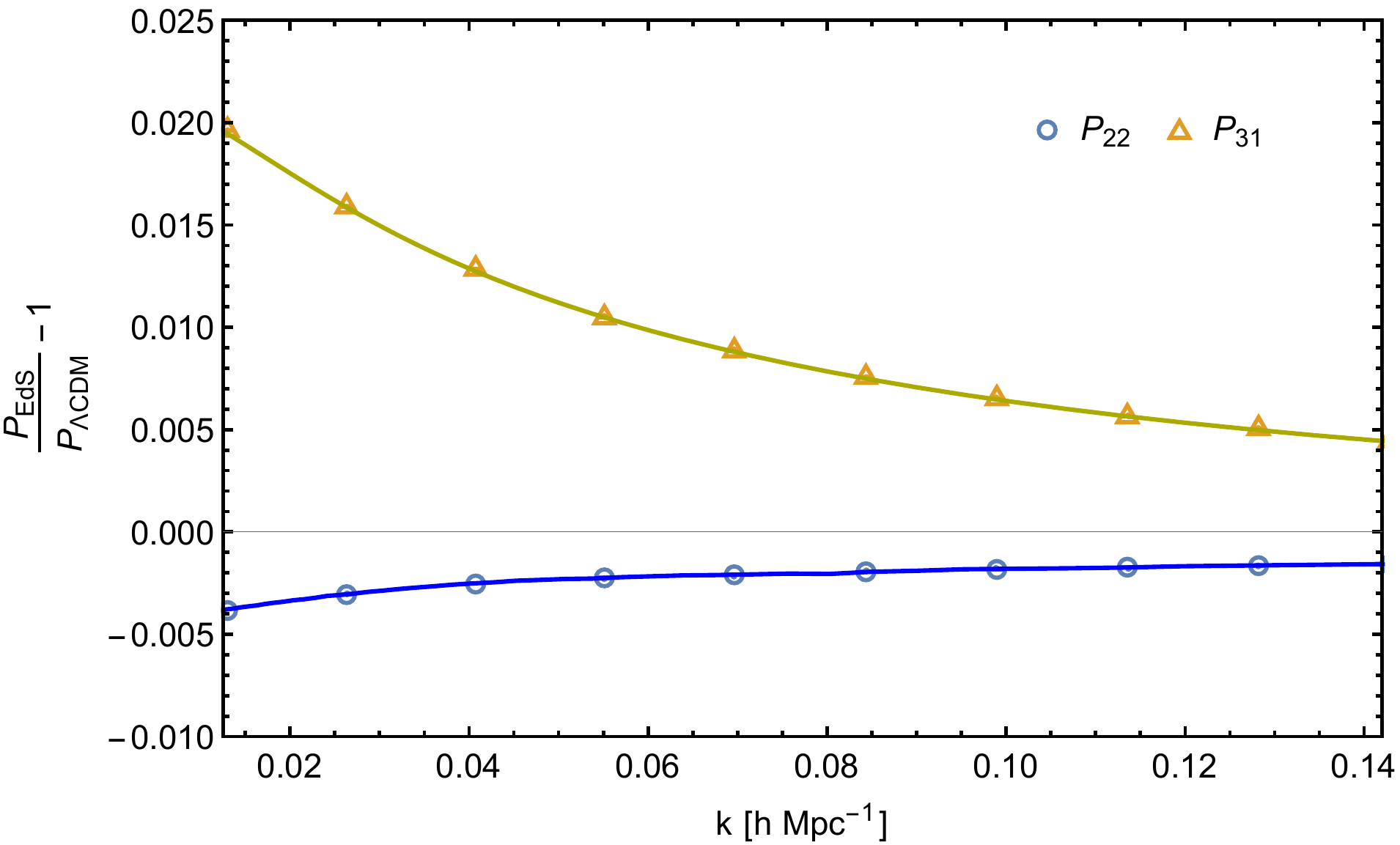}
\caption{\emph{Left panel: }Ratio of the bispectrum contributions up to one-loop in the equilateral configuration with EdS and $\Lambda$CDM growth factors, where the relative deviation of $B_{211}$ and $B_{222}$ has been multiplied by ten to improve visibility.  The pole is due to the sign-change of the sum of the two contributions, $B_{321a}$ and $B_{321b}$.  \emph{Right panel: }The ratio of the power spectrum contributions at one-loop with EdS and $\Lambda$CDM growth factors.  Note that the change in $P_{31}$ leads to a shift in the inferred $\cssq$ as shown in Fig.~\ref{cs}.}
\label{Bgratio}
\end{figure}

For the cosmology under consideration we have at $z=0$ that ${D_{2A}(a)}/{D_1^2(a)}-1\approx 2.7\times 10^{-3}$ and ${D_{2B}(a)}/{D_1^2(a)}-1\approx -7\times 10^{-3}$.
In the left panel of Fig.~\ref{fig:fracdevf2} we show the corrections to the tree-level bispectrum as a function of the cosine of the opening angle $\mu_{12}=\vhk_1\cdot \vhk_2$ and $x_2=k_2/k_1$.  In the equilateral bispectrum ($x_2=1,\  \mu_{12}=-1/2$) this leads to a $0.5\%$ fractional deviation between the \lcdm\ and EdS kernels, which is of the same order of magnitude as the EFT corrections we are trying to constrain here. This can be seen in the right panel of Fig.~\ref{fig:fracdevf2}:  the difference between the EdS and \lcdm\ tree-level bispectra exceeds the amplitude of the one-loop bispectrum contribution from $B_{411}$ for $k<0.02\ihMpc$ and the typical size of the EFT counterterms from $B_{\tilde{2}11}$ for $k<0.08 \ihMpc$.  The data points show that our grid implementation of the \lcdm~second order density field agrees with the analytical calculation.
We thus implemented the exact \lcdm~versions of the $F_2$ and $F_3$ kernels on the grid and will be using them throughout this study.  The details of this implementation are discussed below in Eqs.~\eqref{eq:exdelta2} and \eqref{eq:exdelta3}.  We have validated that the grid implementations of the kernels with the correct growth factors do indeed reproduce the expected deviations from the EdS approximation in the power and bispectrum, as shown in Fig.~\ref{Bgratio}. In the left panel we can see that while the corrections for the $B_{211}$ and $B_{222}$ terms are of order $0.5\%$, the corrections around the zero crossing of the combined $B_{321}$ term are significant. In the right panel we reproduce the $P_{31}$ and $P_{22}$ corrections from \cite{Exact} and compare them to the grid implementation, finding perfect agreement. As well will discuss below in Sec.~\ref{sec:csconsps}, when estimating the amplitude of the power spectrum counterterm, the $2\%$ corrections to $P_{31}$ lead to a $\Delta \cssq=0.2\hMpcsq$ change in the inferred value of of the one-loop power spectrum counterterm.

Based on results for the one-loop power spectrum \cite{Exact,Lewandowski:2017kes} in \lcdm, and the above results, it seems plausible to expect $<1\%$ corrections for the one-loop bispectrum as well.  This makes these corrections typically a factor of 10 smaller than the expected counterterms, such that we can ignore the \lcdm~corrections for the one-loop contributions.  When pushing the accuracy to the next loop order, i.e. the two-loop bispectrum, these corrections might indeed matter.

\begin{figure}[h]
\centering
\includegraphics[width=0.49\textwidth]{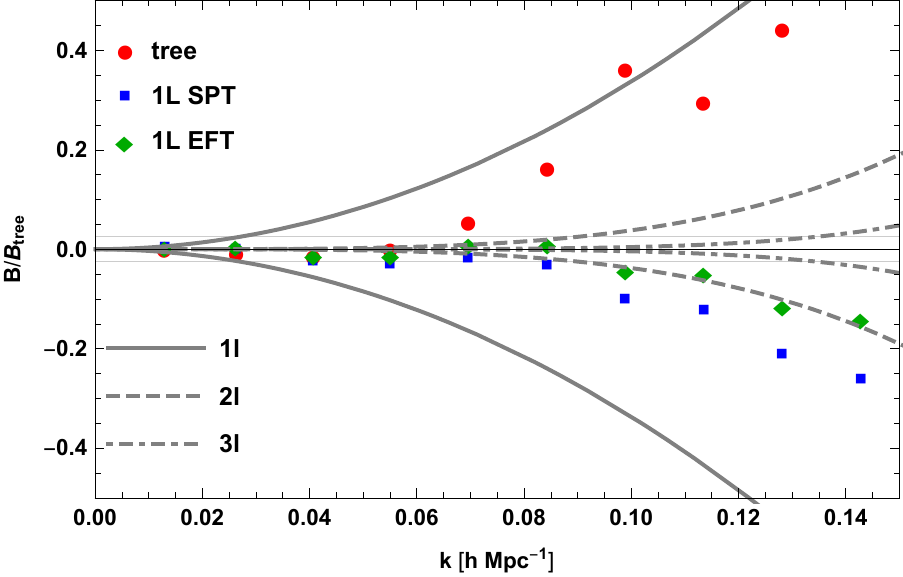}
\includegraphics[width=0.49\textwidth]{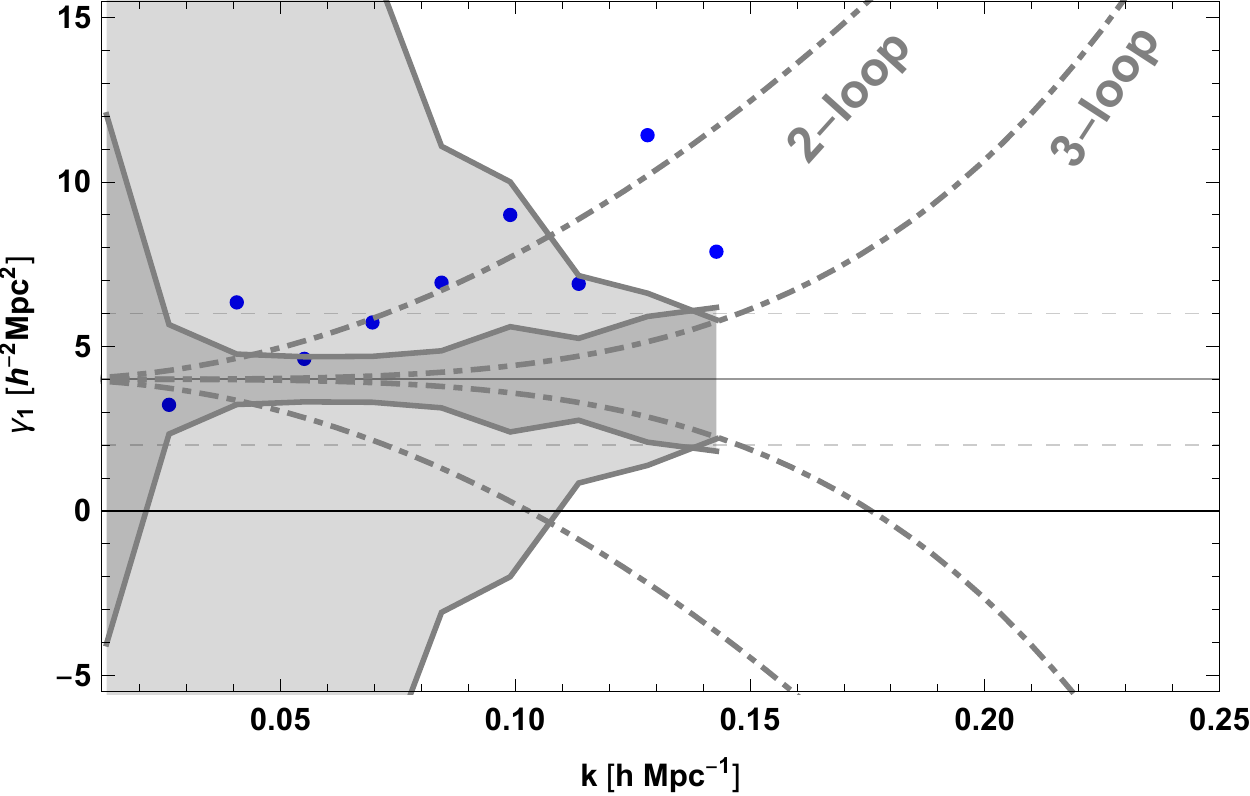}
\caption{\emph{Left panel: }Ratio of the residual bispectra and the tree-level bispectrum in the equilateral configuration.  We show the difference between non-linear bispectrum and tree-level (red), one-loop SPT (blue), and one-loop EFT (green).  We overplot the theoretical errors defined in Eq.~\eqref{eq:theoerrbisp}.  We see that the one-loop theoretical error (solid grey) does indeed form an appropriate envelope for the residual corrections once the tree-level prediction has been subtracted from the data.  \emph{Right panel: } Scale dependence of the bispectrum counterterm amplitude $\gamma_{2}$ in the equilateral configuration.  We show the statistical errors without CVC (light shaded) and with CVC (dark shaded).  Only with CVC can we get close to the true large scale limit before the theoretical error bars induced by two-loop terms take over.}
\label{fig:theoerr}
\end{figure}

\subsubsection{UV-sensitivity}
The bispectrum term $B_{321,b}$, defined in Eq.~\eqref{eq:b321bdef}, can be written in terms of $P_{31}$ as
\be
B_{321,b}(\vec k_1,\vec k_2)=2F_{2}^\text{(s)}(\bk_1,\bk_2)P_{31}(k_1)P_{11}(k_2)\, .
\ee
In analogy to Eq.~\eqref{eq:p13uv}, this leads to the UV-limit
\be
B_{321,\text{UV}}(\vec k_1,\vec k_2)=\lim_{k_1,k_2\to 0}B_{321,b}(\bk_1,\bk_2)=-2\frac{61}{210}k_1^2 \sigmadsq F_2^\text{(s)}(\bk_1,\bk_2)P_{11}(k_1)P_{11}(k_2)=-\frac{61}{210}k_1^2 \sigmadsq B_{211}(\bk_1,\bk_2)\, ,
\label{B321UV}
\ee
based on which we can suggest the UV ansatz for the counterterm
\be
B_{\tilde 121,\mathrm{UV}}(\vec k_1,\vec k_2;\gamma_{1})=- 2 \gamma_{1} k_1^2 F_2(\bk_1,\bk_2)P_{11}(k_1)P_{11}(k_2)=-\gamma_{1} k_1^2B_{211}(\bk_1,\bk_2)\, ,
\label{B1tUV}
\ee
where the fact that we are simply taking the UV-limit of $F_{3}$ as we did for $P_{31}$ means that we expect this $\gamma_{1}$ to be identical to $\cssq$ from the power spectrum.  Thus the UV inspired ansatz agrees with the full symmetry inspired counterterm, ${F}_{3,\mathrm{UV}}=\tilde{F}_{1}$.  The contribution from $B_{411}$ has a strong UV-sensitivity that is given by the limit of the $F_4$ kernel,
\be
\int \frac{\derd \Omega_{\bq}}{4\pi}\lim_{k_1,k_2\to 0}F_4(\bk_1,\bk_2,\bq,-\bq)=\frac{1}{18q^2}F_{4,\mathrm{UV}}(\bk_1,\bk_2)~,
\ee
giving us
\be
\begin{split}
B_{411,\text{UV}}(\bk_1,\bk_2)=\lim_{k_1,k_2\to 0}B_{411}(\bk_1,\bk_2)=&2F_{4,\mathrm{UV}}(\bk_1,\bk_2) P(k_1)P(k_2)\frac{1}{3}\int_{\bq} \frac{P(q)}{q^2}\\
=&2\sigma_\text{d}^2 F_{4,\mathrm{UV}}(\bk_1,\bk_2) P(k_1)P(k_2)~,
\label{B411UV}
\end{split}
\ee
with \cite{Baldauf:2014qfa} 
\be
\begin{split}
F_{4,\mathrm{UV}}(\vec k_1,\vec k_2)=
-\frac{9552 }{18865}(k_1^2+k_2^2)\mu_{12}^2-\frac{61}{420}\mu_{12}\left(\frac{k_2^3}{k_1}+\frac{k_1^3}{k_2}\right)\\-\frac{12409 \mu_{12}^3 k_1
   k_2}{56595}-\frac{115739 \mu_{12} k_1 k_2}{113190}-\frac{4901}{18865}(k_1^2+k_2^2)\, .
\end{split}
\ee
Together with Eq.~\eqref{eq:sigmadrep}, this suggests the UV-ansatz
\be
B_{\tilde 211,\text{UV}}(\vec k_1,\vec k_2;\gamma_{2})=2\frac{210}{61}\gamma_{2} F_{4,\mathrm{UV}}(\bk_1,\bk_2)P(k_1)P(k_2)~,
\label{B2tUV}
\ee
which is less general than the symmetry inspired $\tilde{F}_2$.

If we were to focus entirely upon the stress-energy tensor inspired by symmetry and ignore these UV limits, we would treat $\gamma_{2}$, $\epsilon_{1}$, $\epsilon_{2}$, and $\epsilon_{3}$ in Eq.~\eqref{eq:symmf2tilde} as four independent parameters.  Alternatively, we could fix the ratio of the parameters to match the above UV-limit of $F_{4}$ and approximate $\tilde{F}_{2}\approx 210/61\ F_{4,\text{UV}}$  by setting \cite{Baldauf:2014qfa}
\begin{align}
   \epsilon_{1}=\frac{3466}{14091}\gamma_{2}, \ \ \epsilon_{2}=\frac{7285}{32879}\gamma_{2},\ \ \epsilon_{3}=\frac{41982}{32879}\gamma_{2}~.
   \label{F2tUV}
\end{align}
We will consider both the UV inspired and the more general symmetry motivated counterterm parametrisations in our constraints to follow.

\subsection{Estimation of Clustering Statistics}
\label{grid}
We estimate the non-linear density field by assigning particles to a regular grid using a cloud-in-cell (CIC) mass assignment scheme. The density field is subsequently transformed to Fourier space and divided by the Fourier transform of the CIC window function. The power spectrum is then estimated by averaging products of two density fields over spherical shells.

The bispectra of various combinations of fully non-linear and perturbative matter density fields are estimated using the algorithm previously employed in \cite{Baldauf:2014qfa}.
We are estimating the bispectrum in linearly binned shells in $k$-space using \cite{Scoccimarro:1997st}
\begin{align}
\hat B_{ABC}(k_i,k_j,k_l)=\frac{V_\text{f}}{V_{ijl}}\int_{[\bq_1]_i}\int_{[\bq_2]_j}\int_{[\bq_3]_l}\delta_A(\bq_1)\delta_B(\bq_2)\delta_C(\bq_3)(2\pi)^3\deltadir(\bq_1+\bq_2+\bq_3)\, ,
\label{eq:biest}
\end{align}
where $V_\text{f}$ is the volume of the fundamental cell $V_\text{f}=(2\pi/L)^3$. Here A, B and C stand for various combinations of fully non-linear ($N$-body) and perturbative density fields on the lattice as described in the next section.
The square brackets describe a linear bin ($k$-space interval) around $k_i$ and 
\be
V_{ijl}=\int_{[\bq_1]_i}\int_{[\bq_2]_j}\int_{[\bq_3]_l}(2\pi)^3\deltadir(\bq_1+\bq_2+\bq_3)\approx \frac{8\pi^2}{(2\pi)^6}k_i k_j k_l \Delta k^3
\ee
is the volume of the corresponding Fourier-space shell.

The naive implementation of the above estimator would require a pass through all $N_\text{c}^3$ cells for each of the $N_\text{c}^3$ cells of the grid to ensure the triangle condition, but this quickly becomes unfeasible for small scales.  We thus rewrite the Delta function in Eq.~\eqref{eq:biest} as an integral over plane waves, upon which the expression factorises
\begin{align}
\hat B(k_i,k_j,k_l)=\frac{V_\textit{f}}{V_{ijl}} \int_{\bm x} \prod_{\kappa=i,j,l} \int_{[\bq]_\kappa} \text{exp}[i \bq \cdot \bm x] \delta(\bq)\, .
\end{align}
In a first step, we select density field Fourier modes from a shell in $k$-space, Fourier transform these shells to real space and sum over the product of the Fourier transforms of the three shells.  The error bars of the power spectrum and bispectrum estimators are calculated from the variance over 14 realisations of the simulation box.

\subsection{EFT on the Lattice}
\label{eq:realpt}
\label{methodsec}
\begin{figure}[t]
    \centering  
    \includegraphics[width=0.5\textwidth]{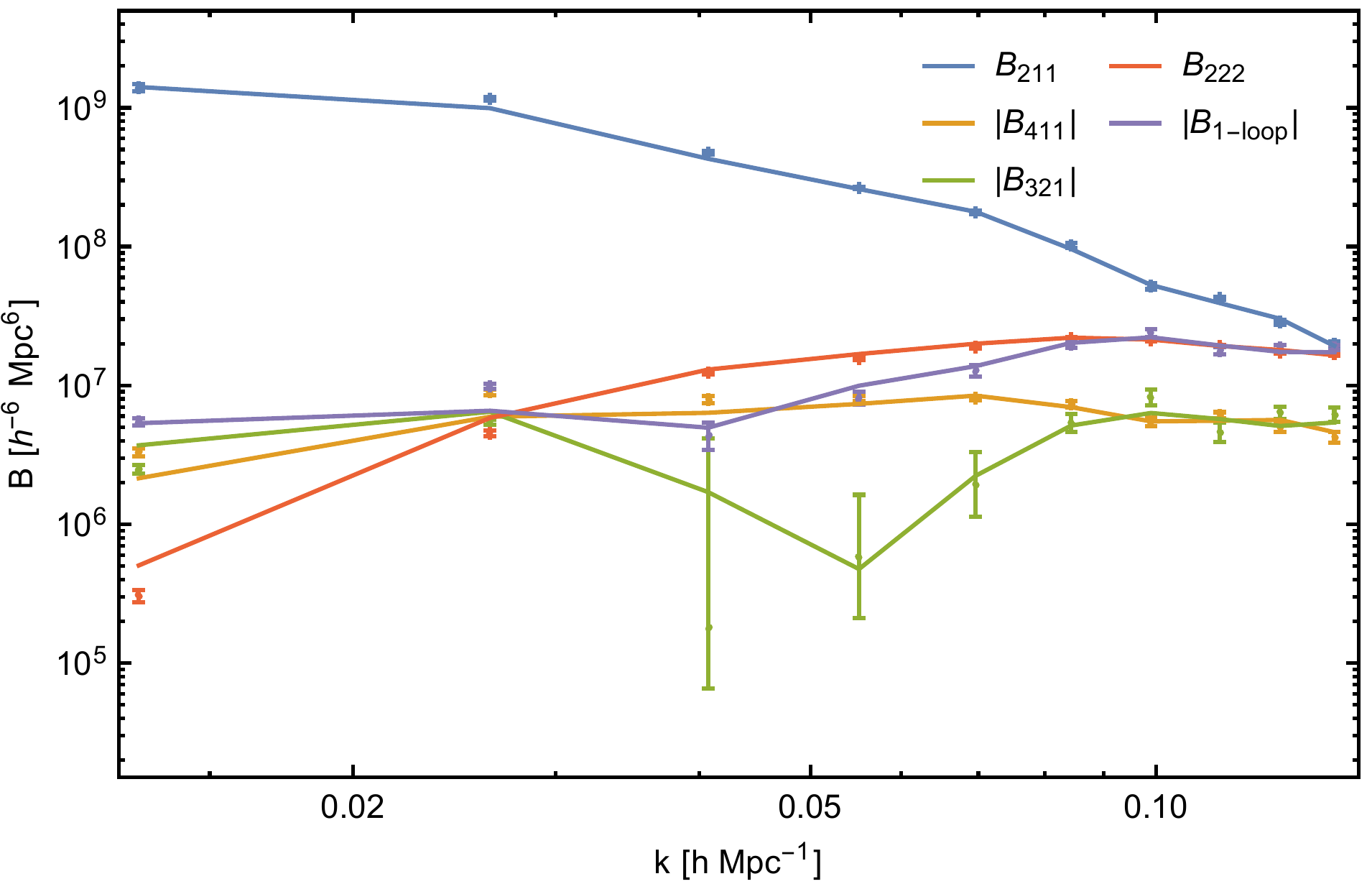}
    \caption{The realisation grid based (points) and analytic (lines) calculations of the one-loop bispectrum contributions together with their sum.  As $B_{411}$ and parts of $B_{321}$ and $B_{\mathrm{1-loop}}$ are negative, the absolute values are shown.  The error bars show the variance of the mean over the fourteen realisations.  The agreement is very good with some small deviations on large scales due to the finite bin width and discrete nature of large-scale modes.}
    \label{BisPT}
\end{figure}

To get to the level of precision required to measure the EFT counterterms on large scales, we need to subtract the leading perturbative orders from the measurements.  If perturbation theory is calculated from the loop integrals, the predictions correspond to the infinite volume limit and would require enormous simulation volumes to beat cosmic variance.  We choose instead to rely on a modest simulation volume and to beat cosmic variance by evaluating the theory for the very modes used to seed the simulation realisation. This approach has previously been used to test SPT \cite{Roth:2011test,Taruya:2018jtk,Taruya:2020qoy}, but not to our knowledge been used to constrain the EFT parameters.
To do so, we take advantage of the fact the the well known recursion relations Eqs.~\eqref{eq:recursf} and \eqref{eq:recursg} for the perturbative density field and velocity dispersion can be rewritten in terms of configuration space fields
\footnote{Note that in what follows, the velocity divergence is rescaled by $-1/\mathcal{H}f$, such that $\theta_1\equiv \delta_1$.}
\begin{equation}
\begin{split}
\delta_{\mathrm{n}}=&\sum^{n-1}_{m=1}\frac{1}{(2n+3)(n-1)}[(2n+1)(\theta_{m}\delta_{\mathrm{n}-m}-\vec \Psi_{\theta_m}\cdot \vec \nabla \delta_{\mathrm{n}-m})+\\&2(-\vec \Psi_{\theta_m}\cdot \vec \nabla \theta_{\mathrm{n}-m}/2 - \vec \Psi_{\theta_{\mathrm{n}-m}}\cdot \vec \nabla \theta_{m}/2 +K_{\theta_m,ij}K_{\theta_{\mathrm{n}-m},ij}+\theta_{m}\theta_{\mathrm{n}-m}/3)]
\label{rr1}
\end{split}
\end{equation}
and
\begin{equation}
\begin{split}
\theta_{\mathrm{n}}=&\sum^{n-1}_{m=1}\frac{1}{(2n+3)(n-1)}[3(\theta_{m}\delta_{\mathrm{n}-m}-\vec \Psi_{\theta_m}\cdot \vec \nabla \delta_{\mathrm{n}-m})+\\&2n(-\vec \Psi_{\theta_m}\cdot \vec \nabla \theta_{\mathrm{n}-m}/2 - \vec \Psi_{\theta_{\mathrm{n}-m}}\cdot \vec \nabla \theta_{m}/2 +K_{\theta_m,ij}K_{\theta_{\mathrm{n}-m},ij}+\theta_{m}\theta_{\mathrm{n}-m}/3)]~.
\label{rr2}
\end{split}
\end{equation}
The displacement fields are given by 
\be
\vec \Psi_{\theta_m}(\vec k)=i\frac{\vec k}{k^2}\theta_m(\vec k)\, 
\ee
and equivalently the tidal tensor is given by
\be
K_{\theta_m,ij}(\vec k)=\left(\frac{k_i k_j}{k^2}-\frac13 \deltakron_{ij}\right)\theta_m(\vec k)\, .
\ee
Starting from the initial Gaussian field $\delta_1=\theta_1$, we can use these relations to generate the higher order density fields one-by-one.  For numerical efficiency, spatial derivatives and inverse Laplacians are calculated in Fourier space and the fields are then transformed to configuration space where the products are evaluated.
In Fig.~\ref{BisPT} we show a comparison of analytic perturbation theory and grid calculations of the contributions to the tree-level and one-loop bispectrum.  We find good agreement on all but the largest scales, where the discreteness of the simulation modes leads to minor deviations from the continuous loop calculations.

The major advantage of working with grid based perturbation theory is that it removes cosmic variance, thus acting as a cosmic variance cancellation (CVC) method. Another issue that the realisation perturbation theory helps with is the dependence of unequal time correlators on the low wavenumber or infrared modes. Equal time correlators do not suffer from this IR-sensitivity \cite{Carrasco:2013sva,Lewandowski:2017kes}. Propagators, such as correlators of the non-linear field with a number of linear fields are IR-sensitive, unequal time correlators. Since both the fully non-linear field and perturbation theory share the very infrared modes this sensitivity is accounted for in realisation perturbation theory. Evaluation of the perturbation theory loop integrals fails to account for the discrete nature of the lowest wavenumbers in the simulation box, even if a cutoff at the fundamental mode of the box is introduced.

As discussed above in Sec.~\ref{sec:lcdmtimedep} the exact \lcdm~time dependence of the SPT density field is relevant to obtain the correct counterterm amplitude on large scales.  To represent the correct evolution from the simulation initial condition realisation and to cancel cosmic variance, we calculate the \lcdm\ second order density field in Eq.~\eqref{eq:F2lcdm} from the Gaussian initial condition $\delta_1$ as
\be
\delta_{2,\Lambda\text{CDM}}=\frac{17}{21}D_{2A}\delta_1\delta_1-D_1^2 \vec \Psi_{\delta_1} \cdoth \vec \nabla \delta_1+\frac{2}{7}D_{2B}K_{\delta_1,ij}K_{\delta_1,ij}\, ,
\label{eq:exdelta2}
\ee
where we used that $5/7D_{2A}+2/7D_{2B}=D_1^2$.  

Taking this to third order, we have (see App.~\ref{app:exgrowth} for the definition of the growth factors)
\be
\begin{split}
\delta_{3,\Lambda\text{CDM}}=&\delta_1 \delta_2 \left[2 \left(\frac{5 D_{3\text{AA},1}}{18}+\frac{D_{3\text{AA},2}}{6}+\frac{2D_{3\text{BA}}}{63}\right)-\frac{3}{2} \left(\frac{D_{3\text{AB},1}}{9}+\frac{2 D_{3\text{AB},2}}{9}+\frac{8
   D_{3\text{BB}}}{189}\right)\right]\\
   &+\delta_1 \theta_2 \left[-\frac{5 D_{3\text{AA},1}}{18}-\frac{D_{3\text{AA},2}}{6}+\frac{5}{2} \left(\frac{D_{3\text{AB},1}}{9}+\frac{2 D_{3\text{AB},2}}{9}+\frac{8
   D_{3\text{BB}}}{189}\right)-\frac{2 D_{3\text{BA}}}{63}\right]\\
   &-\vec \Psi_{\theta_1}\cdoth\vec \nabla\delta_2 \left[2 \left(\frac{5 D_{3\text{AA},1}}{18}+\frac{D_{3\text{BA}}}{21}\right)-\frac{3}{2} \left(\frac{D_{3\text{AB},1}}{9}+\frac{4
   D_{3\text{BB}}}{63}\right)\right]\\
   &-\vec \Psi_{\theta_1} \cdoth \vec \nabla\theta_2 \left[-\frac{5 D_{3\text{AA},1}}{18}+\frac{5}{2} \left(\frac{D_{3\text{AB},1}}{9}+\frac{4 D_{3\text{BB}}}{63}\right)-\frac{D_{3\text{BA}}}{21}\right]\\
   &-\vec \Psi_{\delta_2}\cdoth\vec\nabla  \delta_1 \left[2 \left(\frac{D_{3\text{AA},2}}{6}+\frac{D_{3\text{BA}}}{21}\right)-\frac{3}{2} \left(\frac{2 D_{3\text{AB},2}}{9}+\frac{4 D_{3\text{BB}}}{63}\right)\right]\\
   &-\vec \Psi_{\theta_2}\cdoth \vec \nabla\delta_1
   \left[-\frac{D_{3\text{AA},2}}{6}+\frac{5}{2} \left(\frac{2 D_{3\text{AB},2}}{9}+\frac{4 D_{3\text{BB}}}{63}\right)-\frac{D_{3\text{BA}}}{21}\right]\\
   &+K_{\delta_1,ij} K_{\delta_2,ij} \left(\frac{4 D_{3\text{BA}}}{21}-\frac{4
   D_{3\text{BB}}}{21}\right)+K_{\delta_1,ij} K_{\theta_2,ij} \left(\frac{20 D_{3\text{BB}}}{63}-\frac{2 D_{3\text{BA}}}{21}\right)~,
\end{split}
\label{eq:exdelta3}
\ee
where $\delta_2$ and $\theta_2$ refer to the normal EdS second order terms.
For EdS we would have $D_3\equiv D_1^3$, leading to
\be
\delta_{3,\text{EdS}}=\frac{7 \delta_1 \delta_2}{18}+\frac{25 \delta_1 \theta_2}{54}+\frac{2 K_{\delta_1,ij} K_{\theta_2,ij}}{9}-\frac{7 \vec \Psi_{\delta_1}\cdoth \vec \nabla\delta_2}{18}-\frac{\vec \Psi_{\delta_1}\cdoth\vec\nabla\theta_2}{9}-\frac{\vec \Psi_{\theta_2}\cdoth\vec \nabla\delta_1 }{2}~
\ee
at $z=0$.
{We are also implementing the operators $F_{4,\mathrm{UV}}, \Gamma$, and $E_i$ at the realisation level in analogy to what has been done in \cite{Abidi:2018eyd} for bias parameters.  We have confirmed that their clustering statistics are consistent with the analytical implementation for the counterterms.  The fits to follow are, however, mostly performed with analytical implementations of the counterterms to simplify the error estimation.

\subsection{Theoretical Errors}\label{sec:theoerr}
Regardless of whether we use simulations or perturbative approaches, the precision of our predictions for cosmological clustering statistics is limited.  Often it is assumed that the theory works perfectly up to a certain $k_\text{max}$ and that it cannot be trusted at all beyond that point.  There are a few issues with this approach.  For one, simulations have a limited precision even on the largest scales and perturbative approaches gradually lose precision as we approach the non-linear scale.  It is one of the advantages of the EFTofLSS that it allows us to put an upper bound on subleading corrections, i.e. higher perturbative orders, without explicitly calculating them.
This estimation of higher orders is based on the scaling of loops in scale-free universes.  In a matter-only universe, there is only one scale, the non-linear scale \cite{Baldauf:2016sjb}.  Thus higher loops have to scale as $(k/k_\text{NL})^{3+n_\text{NL}}$, where $n_\text{NL}$ is the slope at the non-linear scale.
\cite{Baldauf:2016sjb} suggested for the bispectrum error
\begin{equation}
    \Delta B_\text{model}(k_1,k_2,k_3)=3 B_{211}(k_1,k_2,k_3)\left(\frac{k_T}{k_\text{NL}}\right)^{(3+n_\text{NL})l}\, ,
    \label{eq:theoerrbisp}
\end{equation}
where $k_\text{T}=(k_1+k_2+k_3)/3$.  The spectral slope of the power spectrum at the non-linear wavenumber $k_\text{NL}=0.3\ihMpc$ is roughly $n_\text{NL}\approx-3/2$.

We show the theoretical errors in the equilateral configuration in Fig.~\ref{fig:theoerr}.  
The residuals between the non-linear bispectrum and the tree-level prediction are within the bounds of the one-loop theoretical error envelope.  The residual with respect to one-loop SPT is outside the two-loop envelope, but once the counterterms are included, the residuals with respect to one-loop EFT are well contained within the envelope.  According to the theoretical error envelope, the two-loop terms lead to corrections of $1\%$ at $k=0.07\ihMpc$, of $2\%$ at $k=0.083\ihMpc$ and of $5\%$ at $k=0.11\ihMpc$.  We will consider a $2\%$ error as acceptable and thus limit our fits to scales larger than $k_\text{max}=0.083\ihMpc$.

In the right panel of Fig.~\ref{fig:theoerr} we show the effect of higher order terms on the constraints on low-energy constants or EFT parameters.  As we can see, the higher order loops start to become important even at fairly low wavenumbers, stressing the need for reducing the statistical error bars on large scales.  The reduction in statistical error bars through realisation perturbation theory is demonstrated through the gray shaded regions.  Only through this approach is it possible to access the regime where the one-loop counterterms asymptote to actual low-energy constants.  In \cite{Baldauf:2015aha}, a one-parameter ansatz for the two-loop power spectrum allowed to capture some of the scale dependence exhibited in the estimation of the power spectrum counterterm amplitude $c_\text{s}^2$.  We have some indications that a similar approach might work here but will leave its detailed discussion for future work.


\section{Constraining the EFT parameters}
\label{results}

\subsection{Power Spectrum}
\label{sec:csconsps}
As a first step we present the inference of the power spectrum low-energy constant $\cssq$ from the auto power spectrum and propagator.  This measurement has been previously presented in \cite{Baldauf:2015aha}, but their constraints were based on the EdS assumption.  
We are measuring the clustering statistics from a numerical simulation, which by definition has finite numerical accuracy. One of the assumptions employed in the discussion so far is that the $N$-body solver correctly reproduces linear growth in the power spectrum on large scales. If the leading order growth in the simulation were off, even by a small amount, this would definitely affect our measurements of the one-loop EFT counterterms. As code comparison studies have shown \cite{Schneider:2015yka}, large-scale linear growth deviations are present due to time stepping and round off errors. To account for the possibility of the leading order linear power spectrum being slightly off, we allow for a correction term $(1+2\Delta D_1)$ in front of the leading-order $P_{11}$ contribution and fit for the free parameter (see App.~\ref{app:lingrowcorr} for details).
The full $\chi^2$ for the auto power spectrum is thus given by
\begin{equation}
    \chi^{2}_{\mathrm{nn}}=\sum_{k=k_{\mathrm{min}}}^{k_{\mathrm{max}}}\frac{\left[P_{\mathrm{nn}}(k)-(1+2\Delta D_1)P_{11}(k)-2P_{31}(k)-2P_{21}(k)-P_{22}(k)+2\cssq k^{2}P_{11}(k)\right]^{2}}{\Delta P^2_{\mathrm{nn}}(k)}~,
\end{equation}
where $\Delta P^2_{\mathrm{nn}}(k)$ is the variance of the residual $P_{\mathrm{nn}}-2P_{31}-2P_{21}-P_{22}-P_{11}$.  All the terms in the above equation are evaluated using realisation perturbation theory and so share their initial conditions with the simulations.  The variance of the estimator is reduced both due to the subtraction of the odd correlator $P_{21}$ that would vanish in the infinite volume limit and due to the variance in the even correlators like $P_{11}$ that matches the associated terms in $P_{\mathrm{nn}}$.  

In the propagator we have fixed one of the linear fields, that we are cross-correlating with, thus the correction factor reduces to $(1+\Delta D_1)$. To constrain the speed-of-sound from the propagator, we minimise
\begin{equation}
    \chi^{2}_{\mathrm{n}1}=\sum_{k=k_{\mathrm{min}}}^{k_{\mathrm{max}}}\frac{\left[P_{\mathrm{n}1}(k)-(1+ \Delta D_1)P_{11}(k)-P_{31}(k)-P_{21}(k)+\cssq k^{2}P_{11}(k)\right]^{2}}{\Delta P^2_{\mathrm{n}1}(k)}~,
\end{equation}
where $\Delta P^2_{\mathrm{n}1}(k)$ is the variance of the residual $P_{\mathrm{n}1}-P_{31}-P_{21}-P_{11}$.  The two calculations of $\cssq$ are shown in Fig.~\ref{cs}. 

As discussed in detail in \cite{Baldauf:2015aha}, the oscillating $k_{\mathrm{max}}$ dependence seen in $\cssq$ as calculated from $P_{\mathrm{nn}}$ is removed when one takes into account two-loop terms which play a larger role at increasingly small physical scales (higher wavenumbers).  The power spectrum and propagator constraints asymptote to the same value on large scales, roughly $k<0.06\ihMpc$, where the one-loop approximation is sufficiently accurate. Note that without CVC, it would have been impossible to constrain the counterterm from our simulations on large scales before the loop corrections become important \cite{Baldauf:2015aha}.
In the same figure, we also show the constraints that would have been obtained if we had made the EdS approximation employed in \cite{Baldauf:2015aha}.  We see that the inferred value of $\cssq$ is reduced by $\Delta \cssq=0.2\hMpcsq$.
In Fig.~\ref{cs}, we show the counterterm amplitude corresponding to our cutoff $\Lambda=0.3\ihMpc$.  According to Eq.~\eqref{eq:csrunhere}, the extrapolation to infinite cutoff requires a subtraction of $\Delta \cssq\approx 1\hMpcsq$.  Thus we have $c_{\text{s},\infty}^2=1.27\hMpcsq$, which is $20\%$ higher than the value reported in \cite{Baldauf:2015aha} due to the \lcdm~corrections which were not considered in that study.

Setting the $\Delta D_1$ terms to zero leads to deviations from the asymptotically flat behaviour of the estimator on large scales. We find that the $\Delta D_1$ constraints themselves asymptote to $-2.5\times  10^{-4}$ on large scales and start to deviate past $k_\text{max}=0.07 \ihMpc$ similar to the $\cssq$ constraints. Note that this constraint is specific for our simulations and dependent on settings of the $N$-body code.

\begin{figure}[t]
    \centering  
    \includegraphics[width=0.5\textwidth]{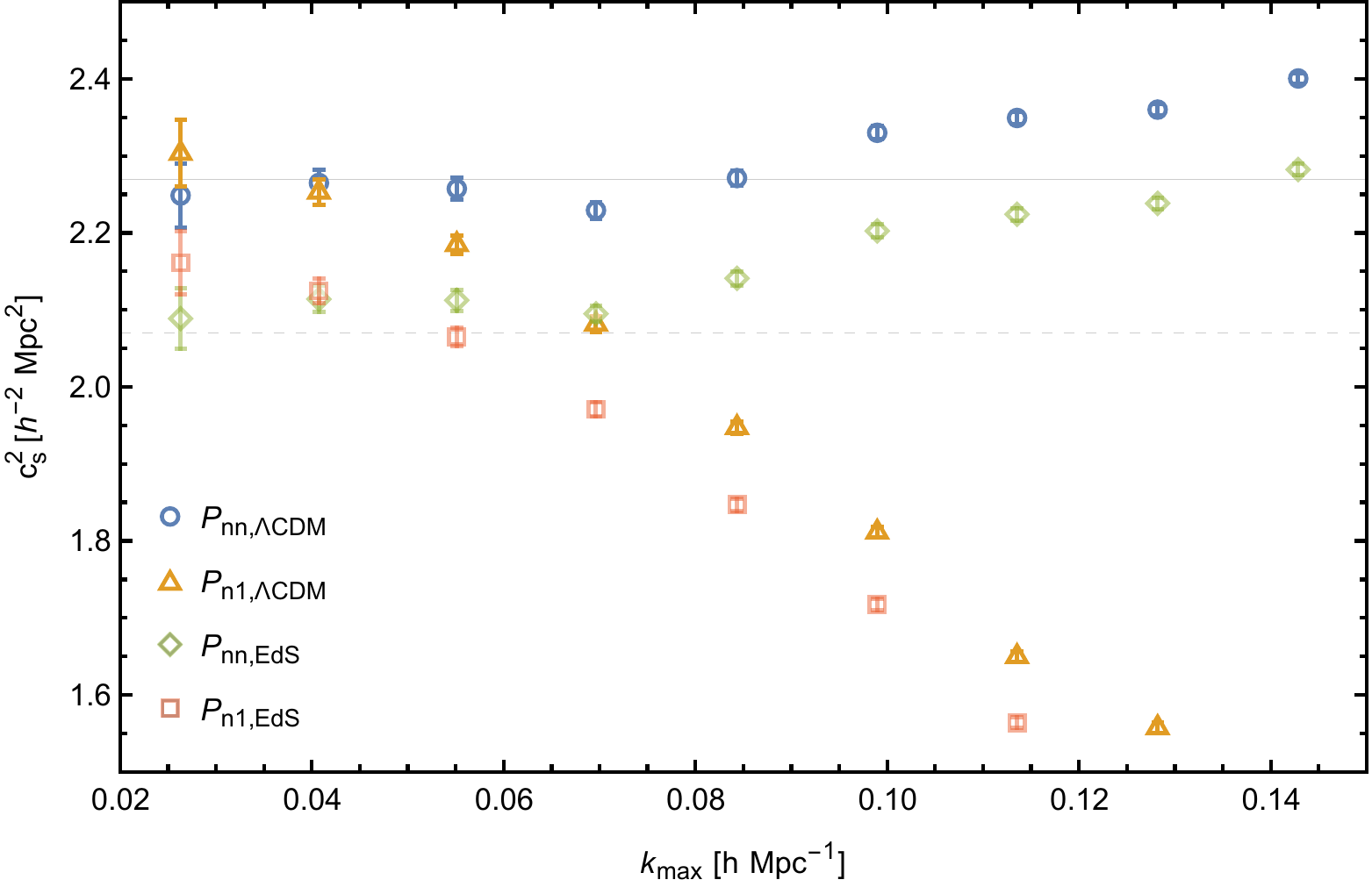}
    \caption{The speed-of-sound parameter $\cssq$ as measured from both the non-linear power spectrum $P_{\mathrm{nn}}$ and the propagator $P_{\mathrm{n}1}$.  We show constraints using the EdS approximation for perturbation theory (diamonds and squares) and using the correct $\Lambda$CDM growth factors (circles and triangles).  The best fit low-energy constant is $\cssq=2.27 \hMpcsq$ (horizontal line).  The difference between $P_{31}$ with exact and EdS growth factors causes a $\Delta \cssq\approx 0.2\hMpcsq$ shift to the value indicated by the horizontal dashed line.  The value shown here is for cutoff $\Lambda=0.3\ihMpc$, and thus needs to be rescaled by Eq.~\eqref{eq:csrunhere} for comparison with values reported for $\Lambda\to \infty$.}
    \label{cs}
\end{figure}

\subsection{Bispectrum}

\subsubsection{Fitting Procedures}
We employ a number of procedures for calculating the parameters that go into the counterkernel calculations.  Firstly, we decide upon a bispectrum from which to do the calculations:
\begin{enumerate}
\item[$B_{\rn\rn\rn}$.] Employing the full non-linear bispectrum, we minimise 
\begin{equation}
\begin{split}
\chi_{\mathrm{nnn}}^{2}(k_{\mathrm{max}})&=\sum_{k_{1,2,3}=k_{\mathrm{min}}}^{k_{\mathrm{max}}}\frac{1}{\Delta B^{2}_{\mathrm{nnn}}(k_{1},k_{2},k_{3})}\Bigl[B_{\mathrm{nnn}}(k_{1},k_{2},k_{3})-B_{\mathrm{SPT}}^\text{s}(k_{1},k_{2},k_{3};\Delta D_{1},\Delta D_{2})\\
&~-B^\text{s}_{\tilde{2}11}(k_{1},k_{2},k_{3};\gamma_{2},\epsilon_i)-B^\text{s}_{\tilde{1}21}(k_{1},k_{2},k_{3};\gamma_{1})\Bigr]^{2}\label{chinnn}~
\end{split}
\end{equation}
with respect to  $\gamma_{1}$, $\gamma_{2}$, $\epsilon_{1}$, $\epsilon_{2}$, $\epsilon_{3}$, and the growth factor corrections $\Delta D_{1}$ and $\Delta D_{2}$, where
\begin{equation}
\begin{split}
B_{\mathrm{SPT}}^\text{s}=(1+3\Delta D_1)B_{111}^\text{s}+(1+2\Delta D_1+\Delta D_2)B_{211}^\text{s}+(1+2\Delta D_1)B_{311}^\text{s}+(1+2\Delta D_1)B_{411}^\text{s}\\+(1+\Delta D_{1}+2\Delta D_2)B_{221}^\text{s}+(1+\Delta D_1+\Delta D_{2})B_{321}^\text{s}+(1+3\Delta D_2)B_{222}^\text{s}~.
\end{split}
\end{equation}
 The terms on the right hand side of Eq.~\eqref{chinnn} are symmetrised with respect to their external momenta.
\item[$B_{\rn 11}$.] Employing the bispectrum propagator,  we minimise
\begin{equation}
\chi_{\mathrm{n}11}^{2}(k_{\mathrm{max}})=\sum_{k_{1,2,3}=k_{\mathrm{min}}}^{k_{\mathrm{max}}}\frac{\left[B_{\mathrm{n}11}(k_{1},k_{2},k_{3})-B_{\mathrm{SPT}}(k_{1},k_{2},k_{3};\Delta D_1,\Delta D_2)-B_{\tilde{2}11}(k_{1},k_{2},k_{3};\gamma_{2},\epsilon_i)\right]^{2}}{\Delta B^{2}_{\mathrm{n}11}(k_{1},k_{2},k_{3})},\label{chin11}
\end{equation}
where
\begin{align}
B_{\mathrm{SPT}}=(1+\Delta D_1)B_{111}+(1+\Delta D_2)B_{211}+B_{311}+B_{411}~,
\end{align}
with respect to  $\gamma_{2}$, $\epsilon_{1}$, $\epsilon_{2}$, $\epsilon_{3}$.
The terms on the right hand side of Eq.~\eqref{chin11} are not symmetrised with respect to their external momenta.  
\item[$B_{\rn 21}$.] Employing our second propagator, we minimise
\begin{equation}
\chi_{\mathrm{n}21}^{2}(k_{\mathrm{max}})=\sum_{k_{1,2,3}=k_{\mathrm{min}}}^{k_{\mathrm{max}}}\frac{\left[B_{\mathrm{n}21}(k_{1},k_{2},k_{3})-B_{\mathrm{SPT}}(k_{1},k_{2},k_{3};\Delta D_1,\Delta D_2)-B_{\tilde{1}21}(k_{1},k_{2},k_{3};\gamma_{1})\right]^{2}}{\Delta B^{2}_{\mathrm{n}21}(k_{1},k_{2},k_{3})}
\label{chin21}
\end{equation}
with respect to $\gamma_{1}$, where
\begin{align}
B_{\mathrm{SPT}}=(1+\Delta D_1)B_{121}+(1+\Delta D_2)B_{221}+B_{321}~.
\end{align}
The terms on the right hand side of Eq.~\eqref{chin21} are not symmetrised with respect to their external momenta.
\end{enumerate}

These functions can be differentiated with respect to the counterparameters $\gamma_{1}$, $\gamma_{2}$, $\epsilon_{1}$, $\epsilon_{2}$, $\epsilon_{3}$ to give linear functions of said parameters.  This makes constraining the parameters by finding the minima of the $\chi^{2}$ functions a simple case of solving a system of linear equations.

Ordinarily, the errors in the denominators of Equations~\eqref{chinnn}, \eqref{chin11}, and \eqref{chin21} would be the variance of the non-linear bispectrum and the SPT would be analytically calculated through perturbation theory.  However, we propose an alternative formulation whereby the SPT components subtracted from the non-linear bispectrum come from perturbation theory on the grid and the denominator is the variance of the ensuing residual. This will reduce the overall variance of the measured parameters as it will remove the variance induced by the individual perturbative contributions to the non-linear bispectrum without affecting the measurements' abilities to constrain the counterparameters.

In addition to choosing which non-linear bispectrum to study, we can define the following fitting procedures to constrain $\gamma_{1}$ and $\gamma_{2}$:
\begin{enumerate}
    \item Fitting for a joint EFT parameter $\gamma$ after setting $\gamma_{1}=\gamma_{2}$.
    \item Fitting $\gamma_{1}=\cssq$ from the power spectrum and using this value in the counterterm $B_{\tilde{1}21}$ while fitting independently for $\gamma_{2}$ in $B_{\tilde{2}11}$.
    \item Fitting for $\gamma_{1}$ and $\gamma_{2}$ independently.
    \item Replacing the analytic counterterms with grid implementations of the counterterms $B_{\tilde{2}11,\text{UV}}$ defined in Eq.~\eqref{B2tUV} and $B_{\tilde{1}21,\mathrm{UV}}$ Eq.~\eqref{B1tUV} and fitting for the contained $\gamma_{1}$ and $\gamma_{2}$.
    \item Replacing the analytic counterterms with grid implementations of the counterterms $B_{\tilde{2}11,\text{UV}}$ defined in Eq.~\eqref{B2tUV} and $B_{\tilde{1}21,\mathrm{UV}}$ Eq.~\eqref{B1tUV} and fitting for a joint $\gamma$ after setting $\gamma_{1}=\gamma_{2}$ .
\end{enumerate}

With fittings $1-3$ for $\gamma_{2}$ we can then we can use the following parametrisations for the $\epsilon$ parameters:

\begin{enumerate}
    \item[U.] Inspired by UV sensitivity such that we assume $F_{4,\mathrm{UV}}\approx\frac{61}{210}\tilde{F}_{2}$, we can use the definitions given in Equations~\eqref{F2tUV} and fit $B_{\tilde{2}11}$ only for $\gamma_{2}$ \cite{Baldauf:2014qfa}.
    \item[S.] Inspired by symmetries, we can fit $B_{\tilde{2}11}$ for all four of its free parameters independently.
\end{enumerate}

This provides us with a wide range of methods with which to constrain the same parameters, allowing us to cross-check the results and to compare the different procedures and comment upon their respective accuracies with reference to their assumptions.
Inclusion of a free $\epsilon_{1}$ parameter in Eqs.~\eqref{chinnn} and \eqref{chin11} results in all parameters being heavily degenerate at most values of $k_{\mathrm{max}}$, as can be quantified from the Fisher matrices of the $\chi^{2}$ functions (see App.~\ref{fm}).  For this reason, for the remainder of the analysis, the UV inspired parametrisations remain unchanged while the symmetry inspired parametrisations only minimise for $\gamma_{1}$, $\gamma_{2}$, $\epsilon_{2}$, and $\epsilon_{3}$, with $\epsilon_{1}$ set to zero.

\subsubsection{Propagator Terms}
\label{propanalysis}
Using the propagator terms, we are able to calculate $\gamma_{2}$ and $\gamma_{1}$ in isolation and compare these results to those we obtain when studying them simultaneously in the auto bispectrum below. In the case of $\gamma_{2}$, we did this with both the UV inspired and symmetry inspired parametrisations.  Naturally, for the propagator terms we are limited to methods 3 and 4 owing to the isolated nature of the $\gamma$ parameters.

\begin{figure}[h]
    \centering  
    \includegraphics[width=0.49\textwidth]{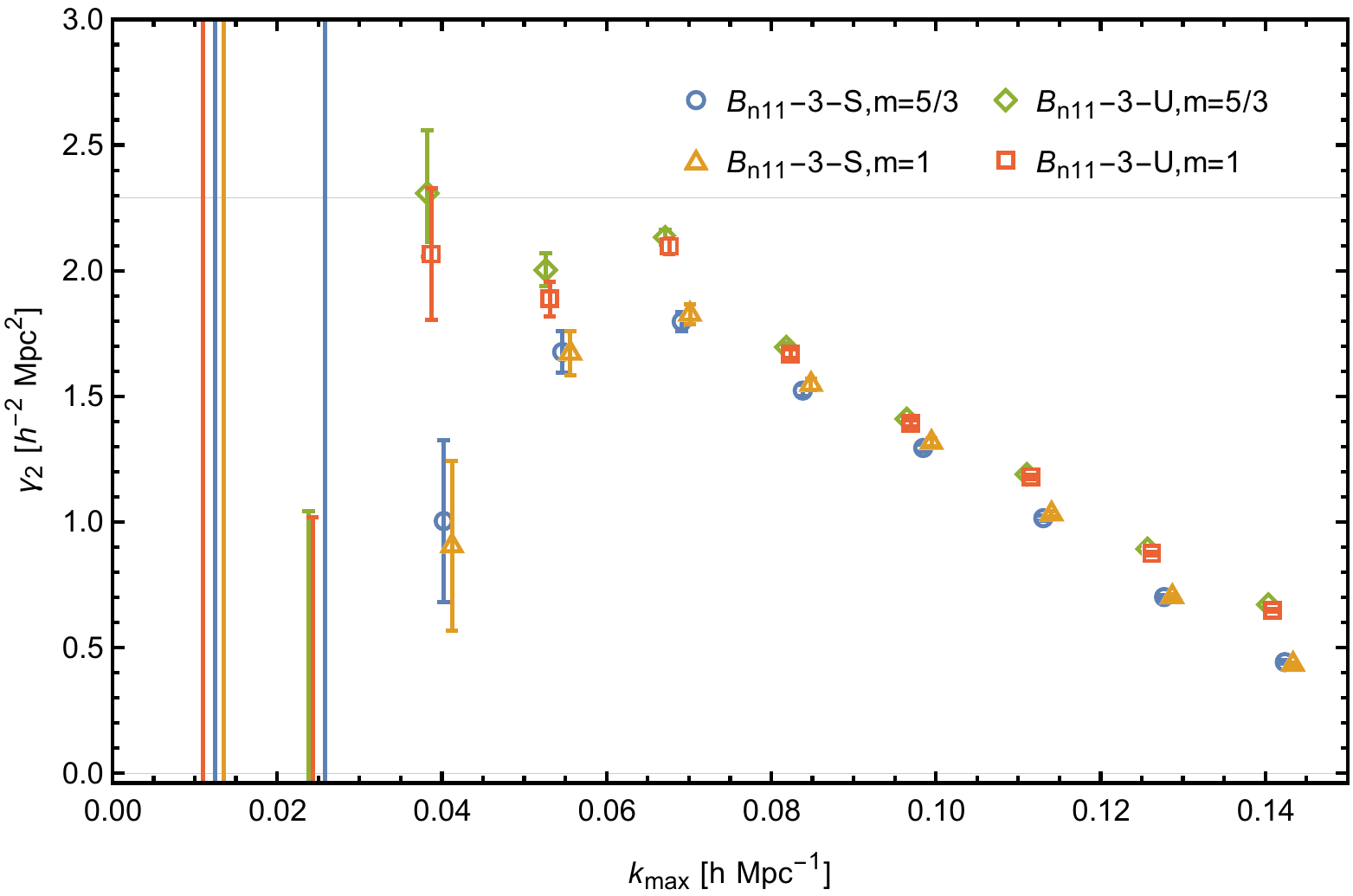}
    \includegraphics[width=0.49\textwidth]{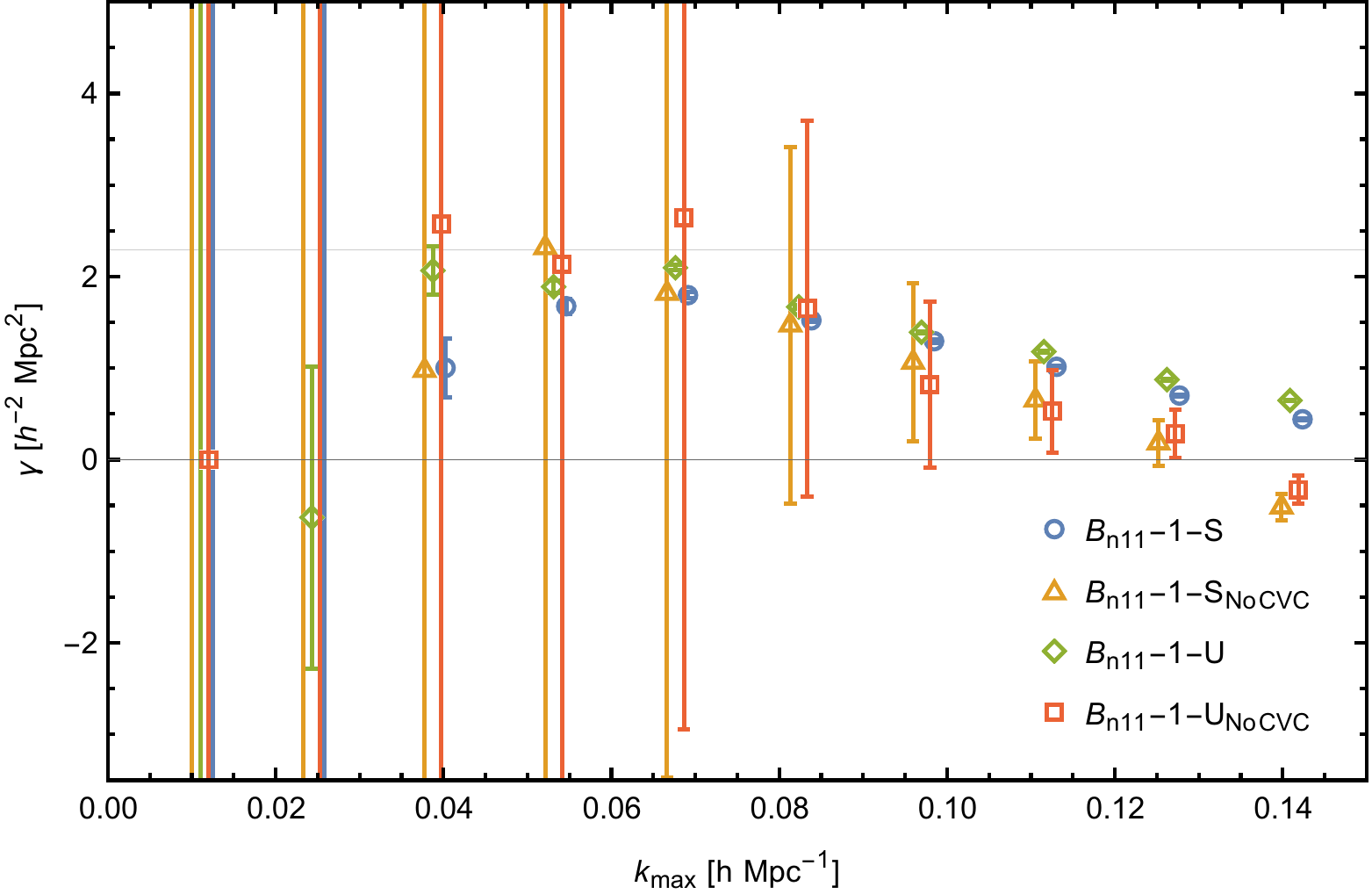}
\caption{\emph{Left panel:} Constraints on $\gamma_{2}$ from $B_{\mathrm{n}11}$  with both symmetry and UV inspired parametrisations for both suggested values of the parameter $m$.  The differences are at the percent level; small, particularly compared to the corrections and errors we would expect from other factors such as the unmodelled contribution of the higher loop terms, but non-zero, such that for the remainder of the study of the propagator we set $m=1$ for the UV inspired parametrisation and $m=5/3$ for the symmetry inspired parametrisation. Note that on large scales the measurements asymptote to the power spectrum $\cssq$ measurement shown by the horizontal line.  \emph{Right panel:} Constraints on $\gamma_{2}$ from $B_{\mathrm{n}11}$  with both symmetry and UV inspired parametrisations with and without cosmic variance cancellation.  In the non-CVC calculations, we continued to use the residuals with subtracted grid perturbation theory and removed CVC only from the variance in the denominator of the $\chi^{2}$ to avoid issues with IR sensitivities of the propagator that are resolved by the subtraction of grid PT from the measured bispectrum propagator.  The results without CVC clearly show significantly larger error bars than those without.}
    \label{propcvc}
\end{figure}

In Fig.~\ref{propcvc} we compare the propagator measurements of $\gamma_{1}$ with both UV and symmetry inspired parametrisations for both suggested values of $m$, as well as with both CVC and without.  These show that there are small but not negligible differences between the values obtained with different values of $m$ and that CVC greatly reduces the error bars of the measurements.  The UV inspired parametrisation replicate the UV limit of the one-loop terms when $m=1$ and the symmetry inspired model represents the symmetries of the EFT irrespective of $m$. Given that there was a measurable difference between the constraints made with the two choices of $m$, albeit a small one, we have elected to perform all future calculations in the propagator with $m=1$ for UV inspired fittings and $m=5/3$ for symmetry inspired fittings, and in all cases we introduce CVC.

For the growth factor corrections with the propagators, we find $\Delta D_1\approx -0.003$ and $\Delta D_2\approx 0.005$ on large scales (see App.~\ref{app:lingrowcorr}). The $\Delta D_1$ constraint seems larger than the one obtained from the power spectrum above, but it has to be noted that it comes primarily from the noise term $B_{111}$.
In App.~\ref{d4test} we validate our fitting procedure on the difference of $B_{411}$ evaluated on the grid for two different cutoffs. By definition this reference field can be fit by the UV-parametrisation of $B_{\tilde{2}11}$, but we also checked its ability to recover the full symmetry inspired parametrisation.

\subsubsection{Auto Bispectrum}
\label{autoanalysis}
Using the full bispectrum $\chi^{2}_{\mathrm{nnn}}$, we are able to calculate $\gamma_{1}$ and $\gamma_{2}$ together, or equate them and calculate a joint $\gamma$.  As we did for the propagator, we will begin by assessing the importance of CVC, the choice of the value of $m$, and of our decision to use $\Lambda$CDM growth factors for $\delta_{2}$ and $\delta_{3}$ instead of the more commonly used EdS approximation.
The effects of incorporating cosmic variance cancellation into the denominator of the $\chi^{2}$ for the full bispectrum are shown in the right hand panel of Fig.~\ref{fullcvc}.  We see that the error bars are significantly reduced in size, as they were for the propagator. As such, we use CVC in all future fittings of the auto bispectrum.  In the left hand panel of the same figure we compare the $B_{\mathrm{nnn}}$-1 fitting for both UV and symmetry inspired parametrisations for both $m=5/3$ and $m=1$.  While the differences between measurements with different values of $m$ are small, they are non-zero, and we choose to set $m=1$ when studying the UV inspired parametrisation and $m=5/3$ when studying the symmetry inspired parametrisations, as we did for the propagator.  

\begin{figure}[t]
    \centering  
    \includegraphics[width=0.49\textwidth]{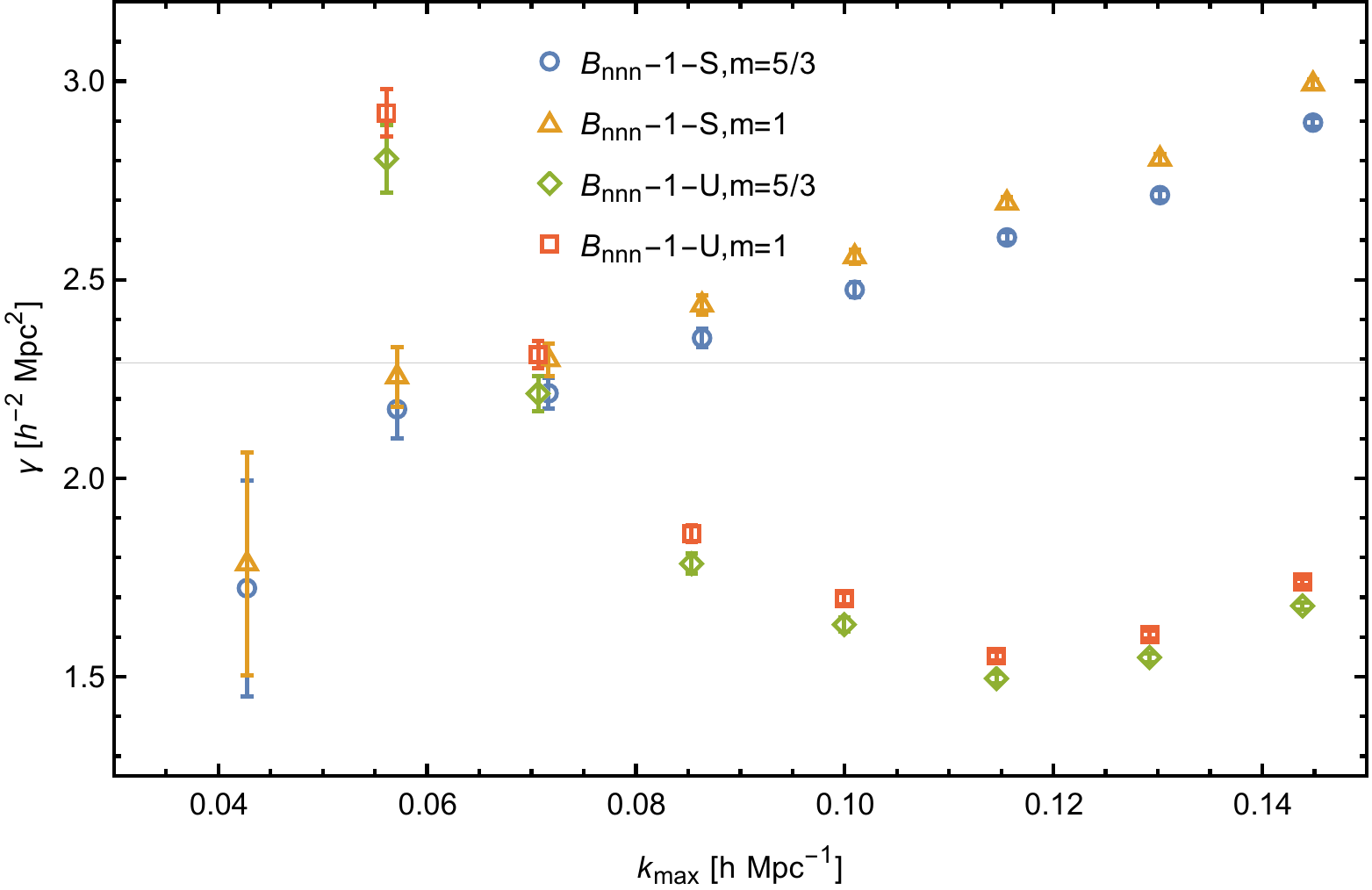}
    \includegraphics[width=0.49\textwidth]{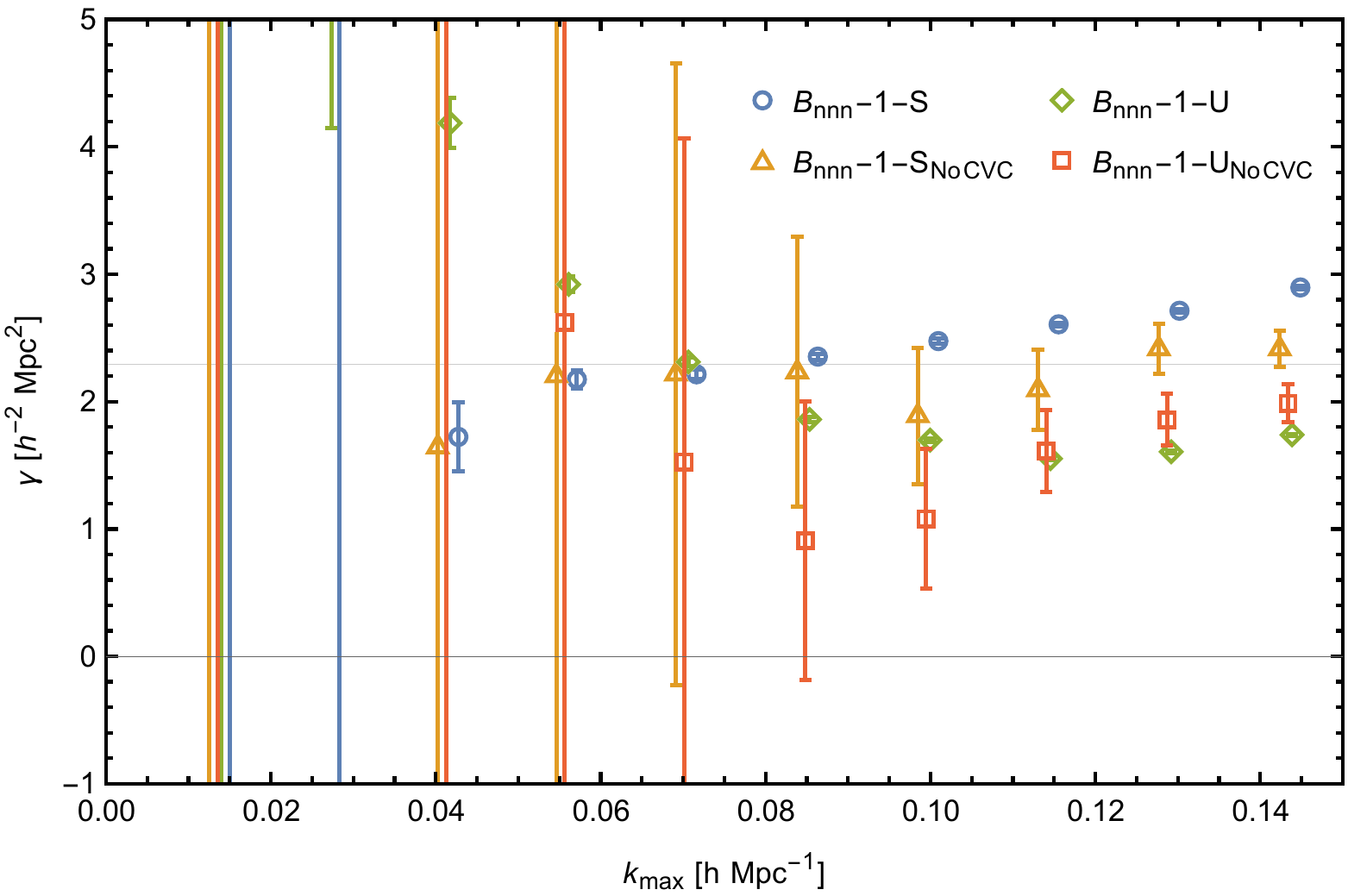}
\caption{\emph{Left panel:} Constraints on a joint $\gamma$ from the auto bispectrum  with both symmetry and UV inspired parametrisations for both suggested values of the parameter $m$.  As with the propagator, the differences are at the percent level.  We set $m=1$ for the UV inspired studies of the auto bispectrum and $m=5/3$ for the symmetry inspired studies.  \emph{Right panel:} Constraints on the joint $\gamma$ from $B_{\mathrm{nnn}}$  with both symmetry and UV inspired parametrisations with and without cosmic variance cancellation.  In the non-CVC case, we continued to use the residual with subtracted grid PT contributions and only changed the variance in the denominator, to avoid issues with IR divergences.  The results without CVC clearly show larger error bars than those without.}
    \label{fullcvc}
\end{figure}

The calculated joint $\gamma$ from $B_{\mathrm{nnn}}$-1-S and $B_{\mathrm{nnn}}$-1-U are plotted in Fig.~\ref{BnnnEdS} with both EdS and $\Lambda$CDM growth factors for $\delta_{2}$ and $\delta_{3}$. As can clearly be seen, the EdS results deviate quite significantly from the more accurate $\Lambda$CDM results.  As such, it is clear that the use of $\Lambda$CDM growth factors is essential for high precision regularisation of the one-loop bispectrum.
\begin{figure}[h]
    \centering  
    \includegraphics[width=0.49\textwidth]{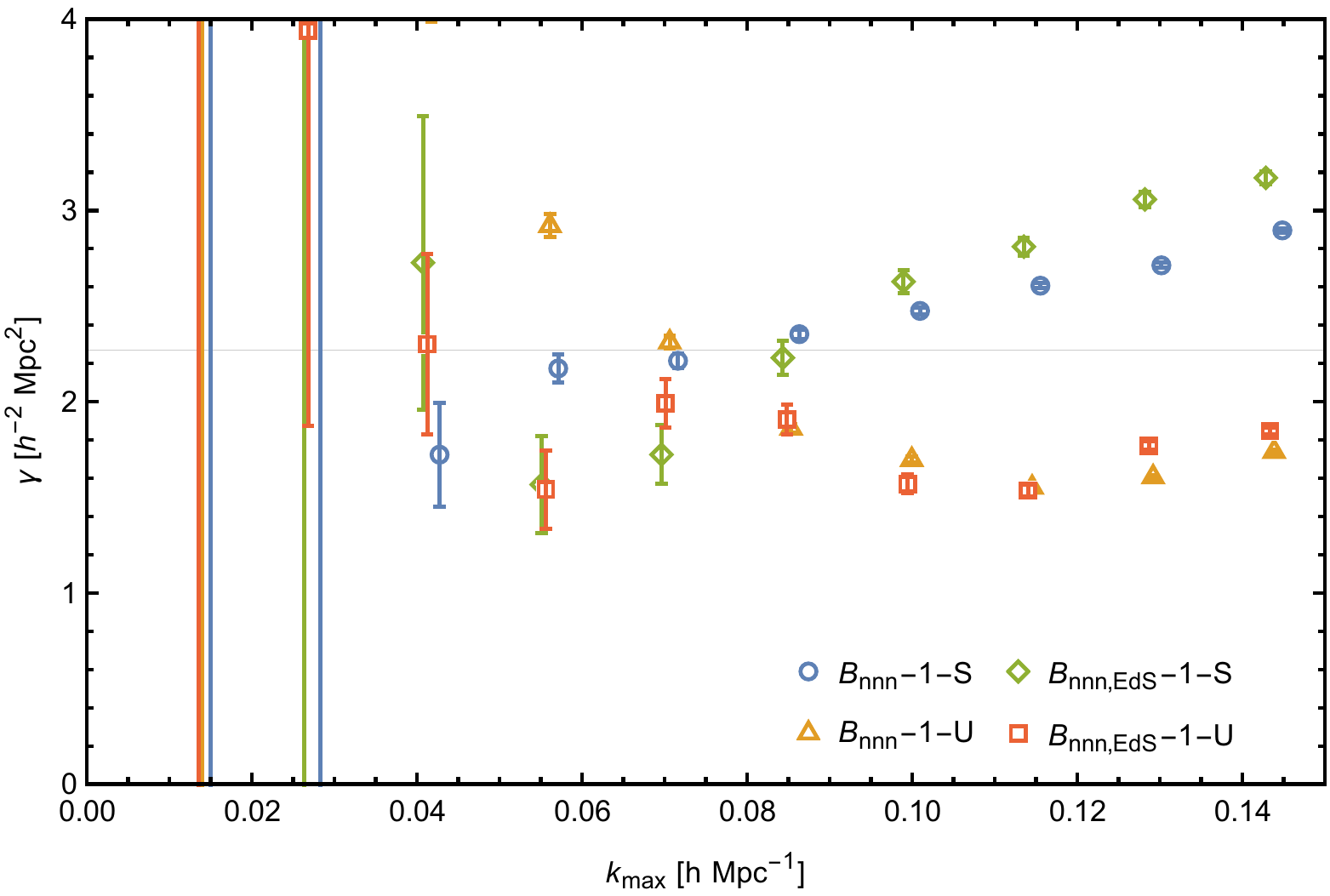}
    \caption{The joint $\gamma$ as calculated from the auto bispectrum using methods $B_{\mathrm{nnn}}$-1-S and $B_{\mathrm{nnn}}$-1-U with $\Lambda$CDM and EdS growth factors.  }
    \label{BnnnEdS}
\end{figure}

For the growth factor corrections with the auto bispectrum, we again find that $\Delta D_1\approx -0.003$ and $\Delta D_2\approx 0.005$ on large scales (see App.~\ref{app:lingrowcorr}). However, the corrections begin to deviate strongly from those of the propagator at $k\sim 0.07\ihMpc$.  

\subsubsection{Counterterm Constraints}
With our growth factors, our use of CVC, and our values of the $m$ parameter chosen for both propagators and the auto bispectrum, we performed a number of different fittings, the calculations of $\gamma$, $\gamma_{1}$, and $\gamma_{2}$ from which are plotted in Fig.~\ref{gammas}.  In the left hand panel we present all of the calculations of the isolated $\gamma_{1}$ and $\gamma_{2}$ from the propagator terms.  The values for the various fitting procedures for $\gamma_{2}$ from $B_{\mathrm{n}11}$ and the $B_{\mathrm{n}21}$-4 fitting for $\gamma_{1}$ all agree with one another for most values of $k_{\mathrm{max}}$, mimicking the shape of the value of $\cssq$ calculated from $P_{\mathrm{n}1}$, while the $B_{\mathrm{n}21}$-3 calculation of $\gamma_{1}$ differs from this curve but is of roughly the same value as $\cssq$ at $k_{\mathrm{max}}<0.07\ihMpc$.  In the right hand panel we present the fittings for $\gamma$ made with the auto bispectrum.  We found that while the joint $\gamma$ tended to give roughly the same result for all fittings, the independent $\gamma_{1}$ and $\gamma_{2}$ sometimes gave results less in keeping with the value we would have expected by comparison with the speed of sound and were omitted from the figure.  The Fisher matrix for the routine $B_{\mathrm{nnn}}$-3 shows $\gamma_{1}$ and $\gamma_{2}$ have a cross-correlation of $-0.98$, making them almost completely degenerate; it is possible that at higher loop order the degeneracy would be broken.  As will be discussed in the next section, while the degeneracy between the two $\gamma$ terms resulted in differing results for the individual counterparameters when they were allowed to vary independently, the overall counterterms still gave good fits to the residuals of the measured bispectra.

\begin{figure}[t]
    \centering  
    \includegraphics[width=0.5\textwidth]{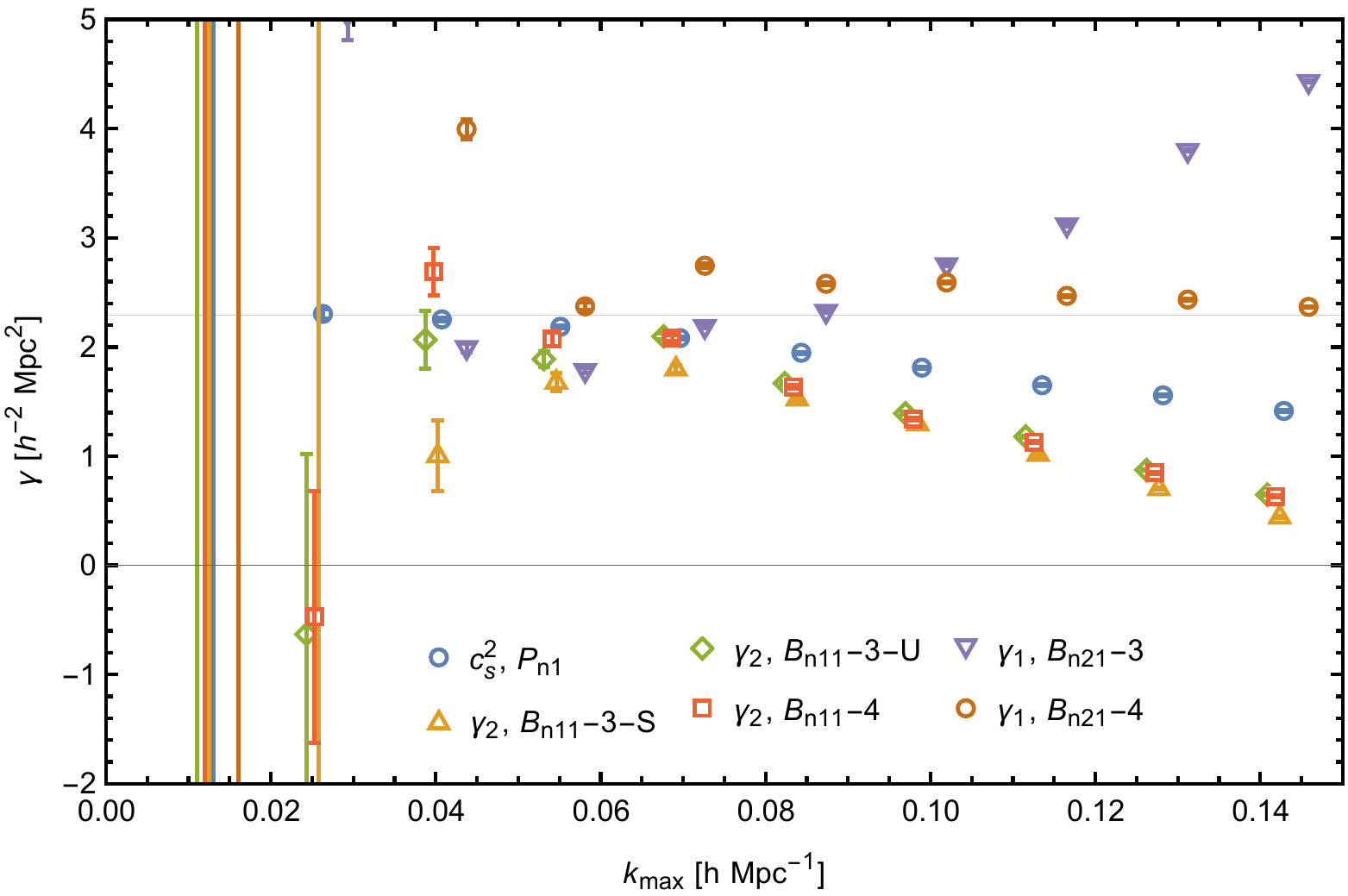}
    \includegraphics[width=0.49\textwidth]{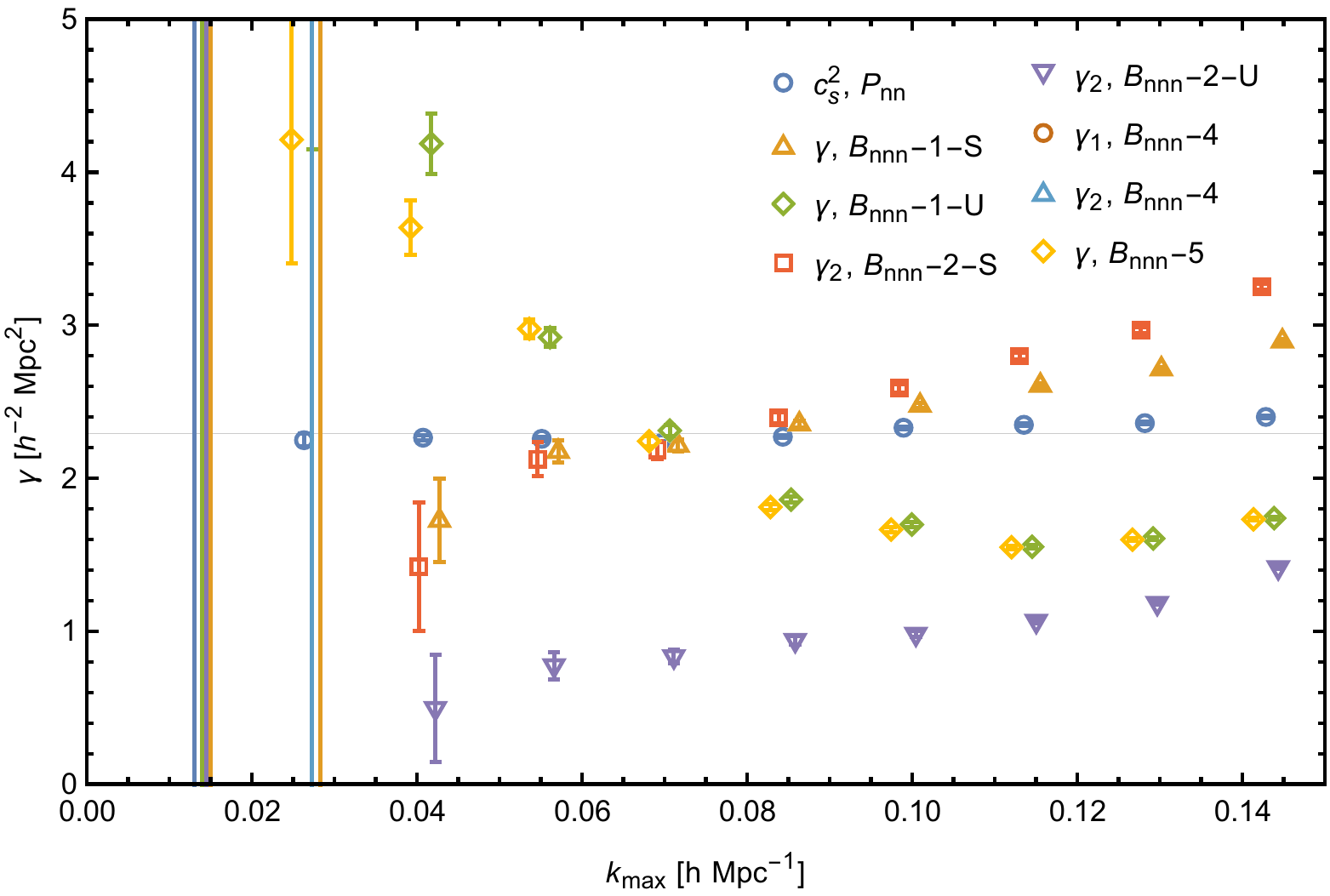}
    \caption{\emph{Left panel:} Propagator measurements of $\gamma_{1}$ and $\gamma_{2}$ alongside the power spectrum based calculation of the speed of sound from $P_{\mathrm{n}1}$.  The results for $\gamma_{2}$ and the $B_{\mathrm{n21}}$-4 fitting for $\gamma_{1}$ closely mimic those of $\cssq$ from the power spectrum propagator while those of $\gamma_{1}$ from $B_{\mathrm{n}21}$-3 follow a different curve, holding similar values to the speed of sound at $k<0.1\ihMpc$.  \emph{Right panel:} Constraints  on $\gamma_{1}$, $\gamma_{2}$, and the joint $\gamma$ from the full bispectrum alongside the speed of sound constrained from $P_{\mathrm{nn}}$.  The calculated values of $\gamma_{1}$ and $\gamma_{2}$ from the $B_{\mathrm{nnn}}$-3 procedures are omitted as the degeneracy of the two parameters made their results differ from the other fits.}
    \label{gammas}
\end{figure}

As expected, the $k_{\mathrm{max}}$ dependence of the results from the auto bispectrum more closely match the $k_{\mathrm{max}}$ dependence of $\cssq$ calculated from $P_{\mathrm{nn}}$, while those from the propagator are more similar to those of $P_{\mathrm{n}1}$.  On the very large scales where the $\cssq$ measurements from the two power spectra agree with one another due to the small contribution from higher loop terms, we also find that our calculations from the auto and propagator bispectra begin to agree, though in both cases we begin to see results from a number of our fitting procedures that differ from those of other procedures.

In Fig.~\ref{e} we plot the inferred values of the counterterm amplitudes $\epsilon_{2}$ and $\epsilon_{3}$ as measured with the symmetry inspired fittings for both the auto bispectrum and the propagator.  The values constrained with the symmetry-inspired ansatz have the opposite sign to those used in the UV inspired fitting.

\begin{figure}[t]
    \centering  
    \includegraphics[width=0.49\textwidth]{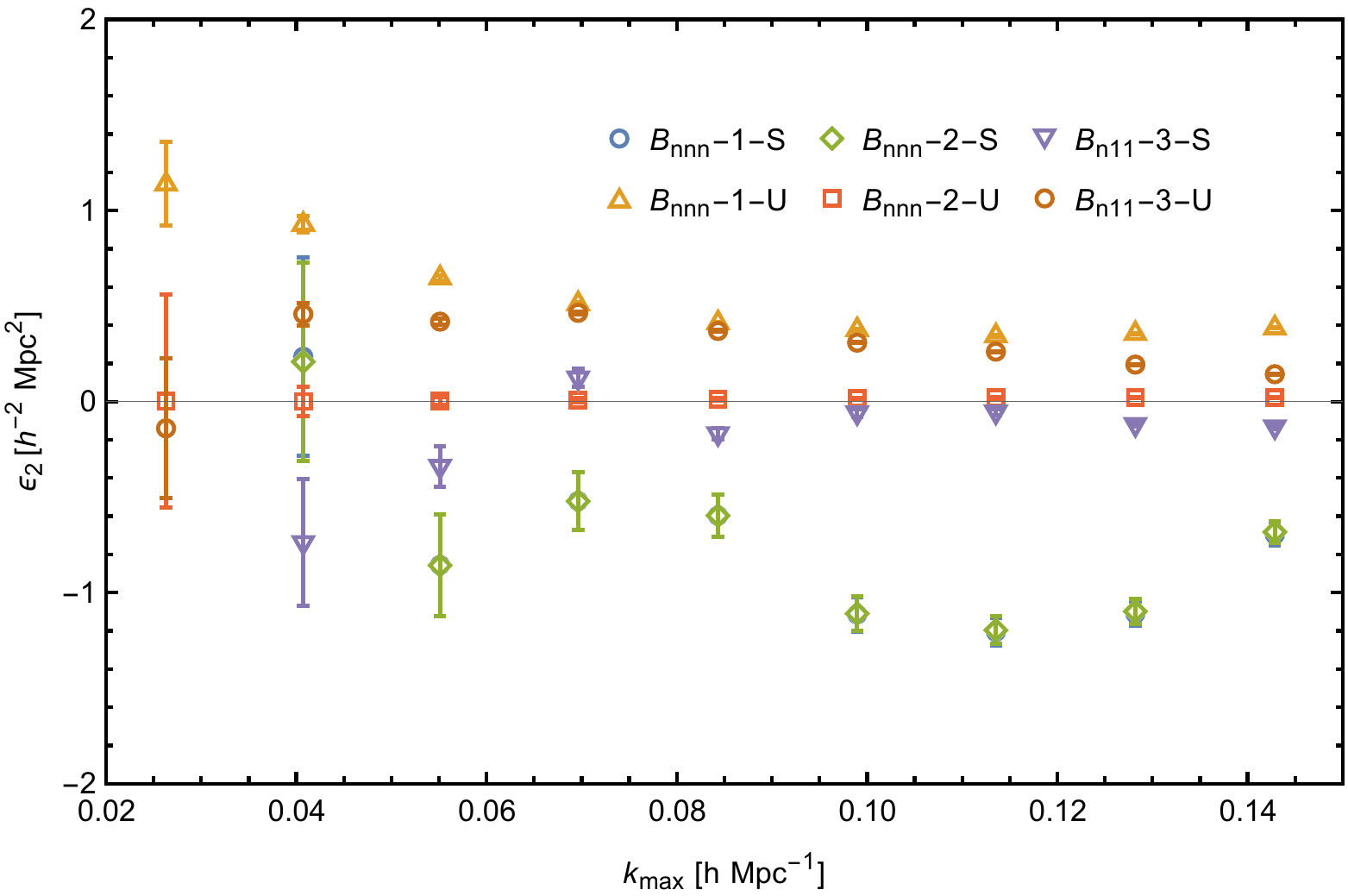}
    \includegraphics[width=0.49\textwidth]{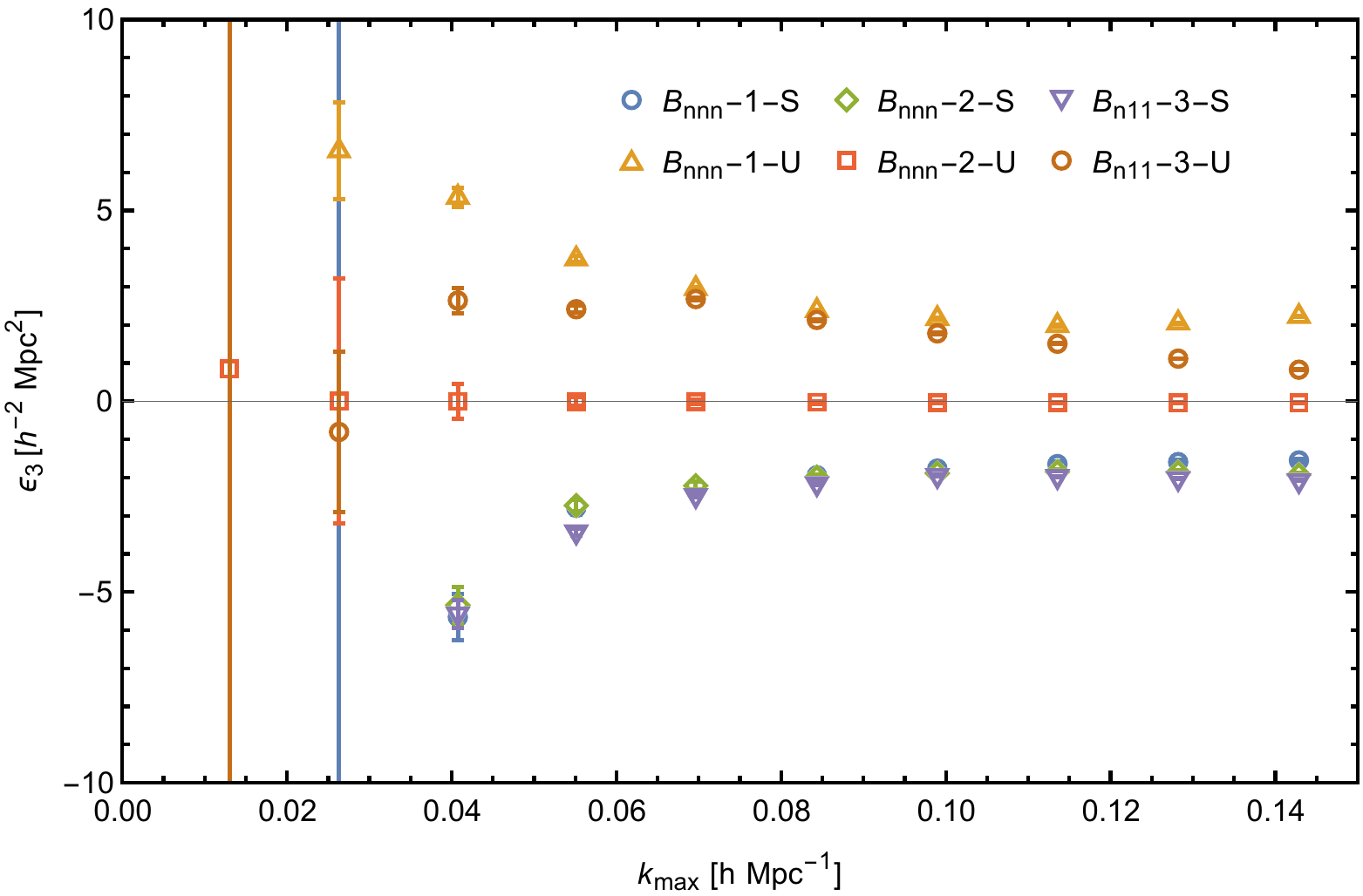}
    \caption{\emph{Left panel:}  The parameter $\epsilon_{2}$ as measured from both the propagator and the auto bispectrum with the symmetry inspired parametrisation.  In the case of UV inspired fits, it is simply a linear function of $\gamma_{2}$.  \emph{Right panel:} The counterparameter $\epsilon_{3}$ as measured from both the propagator and auto bispectrum with symmetry inspired fittings.  In the case of UV inspired fits, it is a linear function of $\gamma_{2}$.  As in Fig.~\ref{gammas}, the results from the $B_{\mathrm{nnn}}$-3 fitting procedures are omitted as the degeneracy of the $\gamma$ terms led to results that differed strongly from those of the other fits.}
    \label{e}
\end{figure}

The complete set of reduced $\chi^{2}$ calculations made with $\Lambda$CDM growth factors is plotted in Fig.~\ref{chi2}.  We can see that many of the fittings give good results, with a $\chi^{2}$ that crosses one roughly where we would expect given that the higher $k_{\mathrm{max}}$ gives a more precise result averaging over more configurations, while the values that approach the non-linear regime begin to suffer from small scale effects that are not accounted for in our model.   However, some of the fittings, most noticeably both fits made with $B_{\mathrm{n}21}$ and both UV inspired fits to $B_{\mathrm{n}11}$, give significantly larger reduced $\chi^{2}$ values at all $k_{\mathrm{max}}$ points of interest.  This indicates that there is not enough freedom in the single parameter, isolated $\gamma$ terms to accurately fit the data.  Indeed, the $\chi^2$ of the UV-based ansatz is significantly larger than the $\chi^2$ of the symmetry-based ansatz in all cases for both the propagator and the auto bispectrum excepting $B_{\mathrm{nnn}}$-3, which we take as evidence that the one parameter UV inspired approximation is unable to accurately capture the bispectrum residuals. 

\begin{figure}[t]
    \centering  
    \includegraphics[width=0.49\textwidth]{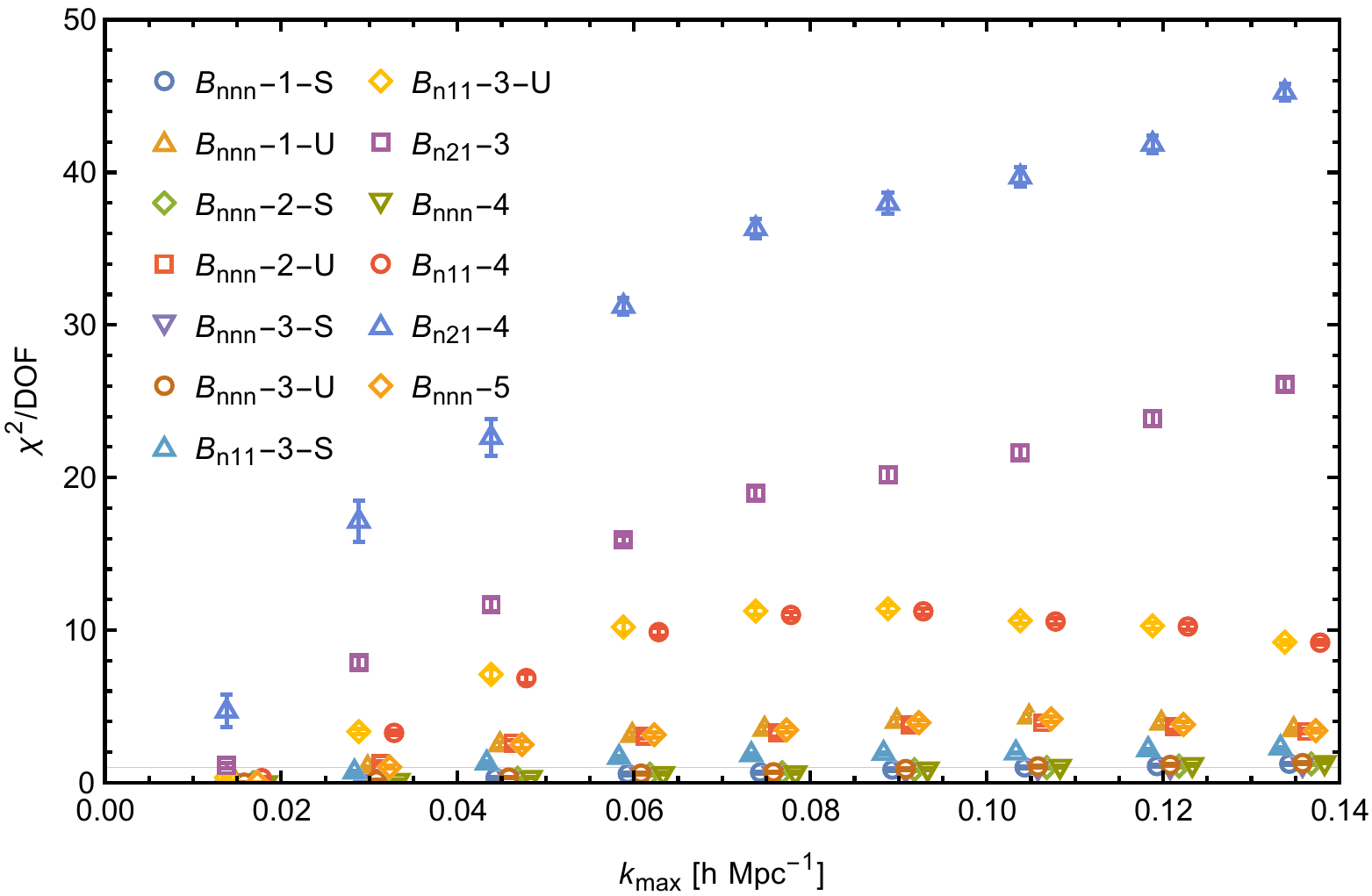}
    \includegraphics[width=0.49\textwidth]{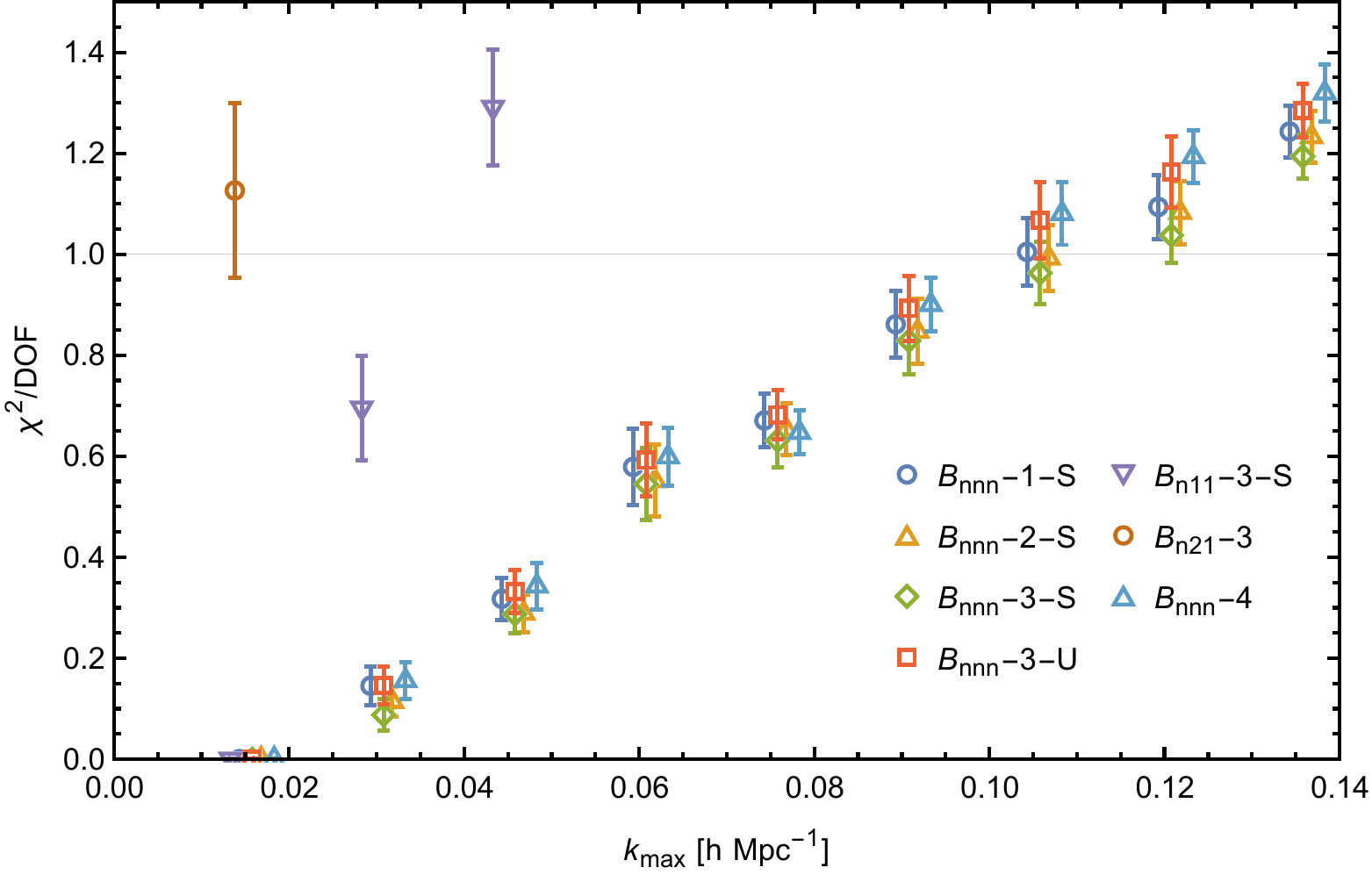}
    \caption{\emph{Left panel:} The reduced $\chi^{2}$ for all fitting procedures of the counterterms with $\Lambda$CDM growth factors for $\delta_{2}$ and $\delta_{3}$.  Noticeably, the one-parameter fittings of isolated $\gamma_{1}$ and $\gamma_{2}$ all produce reduced $\chi^{2}$ values an order of magnitude larger than those of the multiple parameter fittings.  \emph{Right panel:}  The same with a focus on the values close to 1.  The crossing of $\chi^2/\text{DOF}\approx 1$ takes place at about $k_{\mathrm{max}}=0.11\ihMpc$, which is roughly where we would expect the perturbative description of large scale structure to begin breaking down due to higher loop contributions.}
    \label{chi2}
\end{figure}

The performance of the fitted models can also be assessed by comparing the value of $B_{\mathrm{SPT}}+B_{\mathrm{counterterms}}$ to the measured residuals; this is shown for the equilateral configuration in Fig.~\ref{Bres}.  In the top panels of Fig.~\ref{Bres} we plot $B^{\mathrm{s}}_{\mathrm{nnn}}-B^{\mathrm{s}}_{221}-B^{\mathrm{s}}_{311}-B^{\mathrm{s}}_{111}$ against the PT calculations of $B^{\mathrm{s}}_{211}+B^{\mathrm{s}}_{411}+B^{\mathrm{s}}_{321}+B^{\mathrm{s}}_{222}+B^{\mathrm{s}}_{\tilde{1}21}+B^{\mathrm{s}}_{\tilde{2}11}$ with counterkernels calculated according to  a number of fitting procedures.  The left hand panel shows the total values of these calculations while the right hand panels are normalised by the tree-level bispectra, to highlight the effects of the counterterms.  Note that the parameters were constrained using the full bispectrum whereas only the equilateral configuration is shown in this figure.  The central panels show the equivalent calculations for the unsymmetrised $B_{\mathrm{n}11}$ and the bottom panels for $B_{\mathrm{n}21}$.  From this figure, we can see that the results for the UV fitting procedures, 1-U, 2-U, 3-U, and 5 produce results that are at odds with the symmetry inspired procedures; method 4 is the only UV inspired fitting procedure to produce results which are in keeping with the symmetry inspired fits and closely match the residual at low $k$, due to the independent fitting of $\gamma_{1}$ on a measurement of $B_{\tilde{1}21,\mathrm{UV}}$ with modes corresponding exactly to those of the residual.  Together with the results discussed above with reference to Fig.~\ref{e}, this could be taken to mean that the approximation $B_{411,\mathrm{UV}}\approx B_{\tilde{2}11}$ is not sufficient and that the counterkernel $\tilde{F}_{2}$ must be given the full parameter freedom indicated by the symmetries of the EFT in order to accurately regularise the one-loop bispectrum.  Noticeably, in the top and bottom panels of Fig.~\ref{Bres}, which incorporate the regularisation of $\tilde{F}_{1}$, the limit at which regularisation becomes impossible at one-loop order is at $k\sim0.1 \ihMpc$, while in the central panels we can see that for an isolated $\tilde{F}_{2}$ the limit is much lower, at $k\sim0.05\ihMpc$, indicating that this estimator is much more sensitive to IR-resummation effects.

\begin{figure}[p]
    \centering  
    \includegraphics[width=0.49\textwidth]{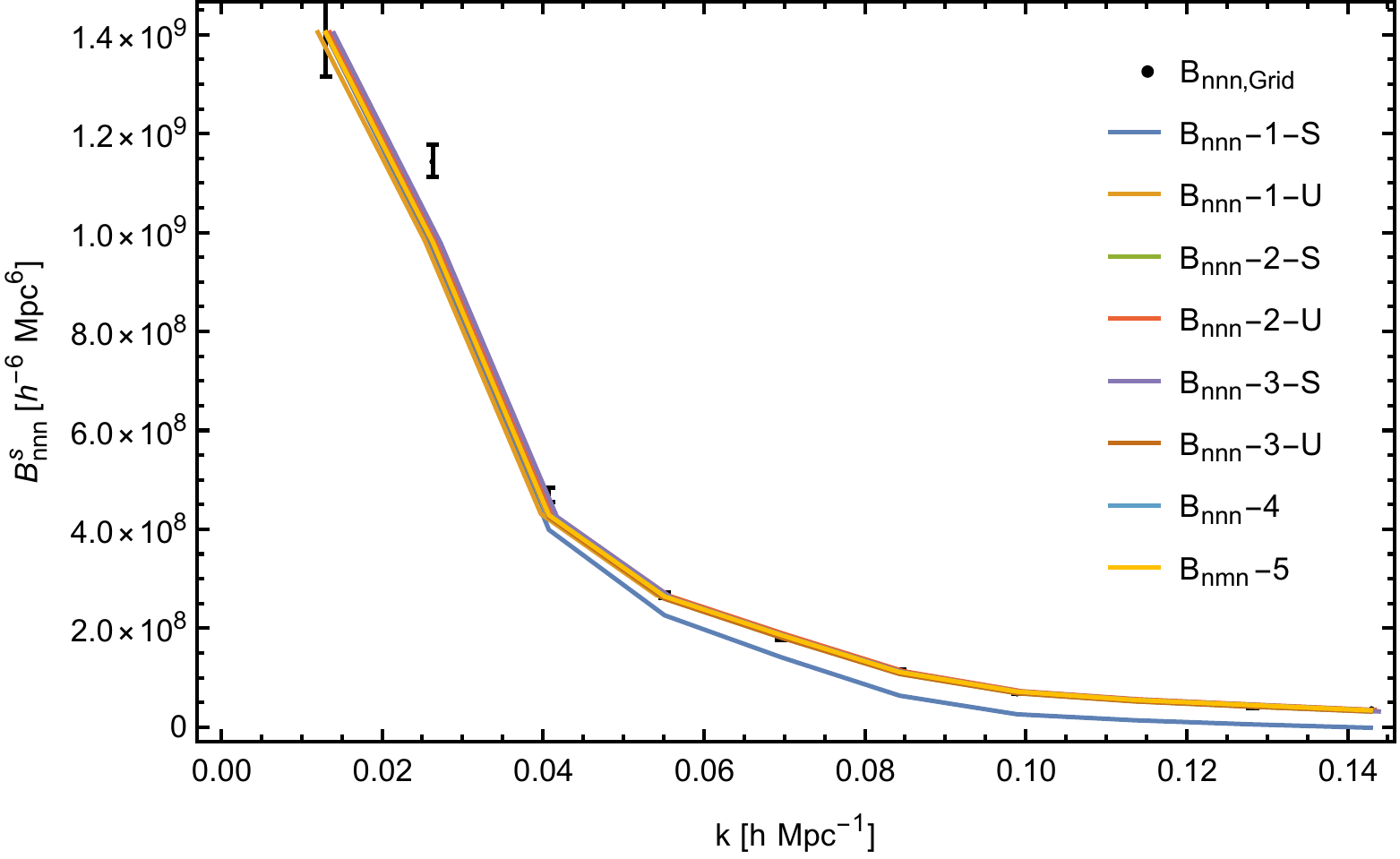}
    \includegraphics[width=0.49\textwidth]{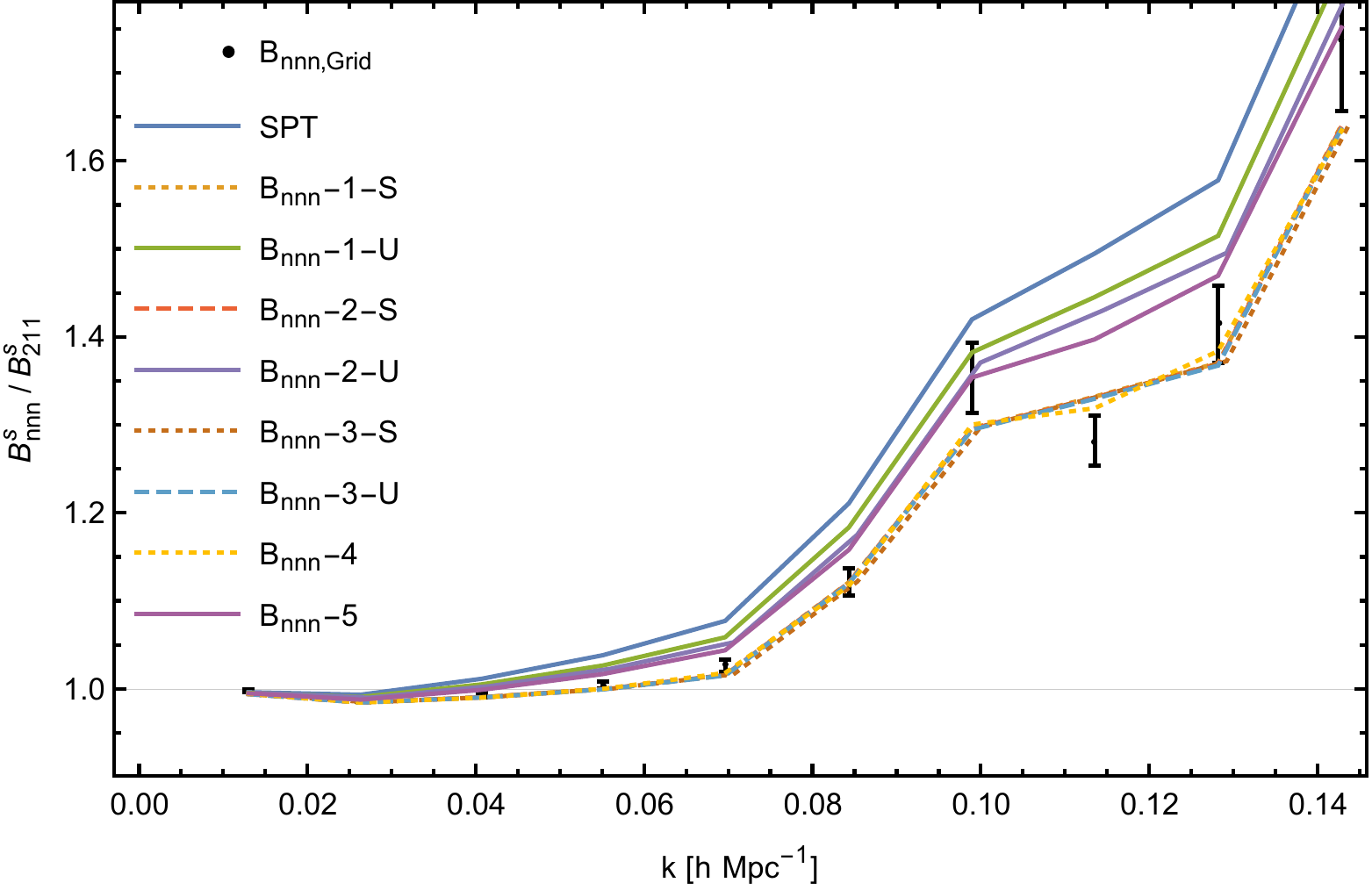}
    \includegraphics[width=0.49\linewidth]{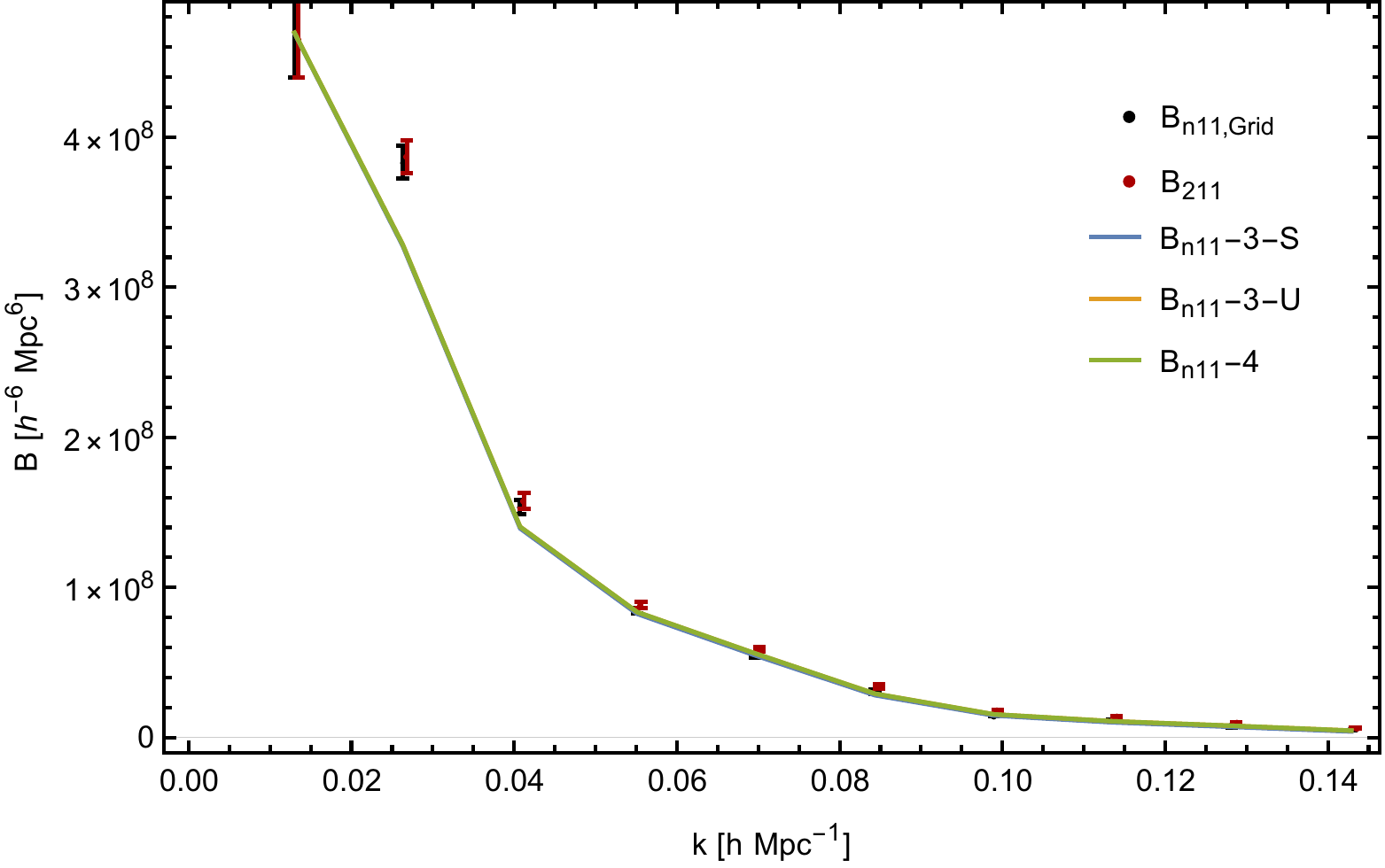}
    \includegraphics[width=0.49\textwidth]{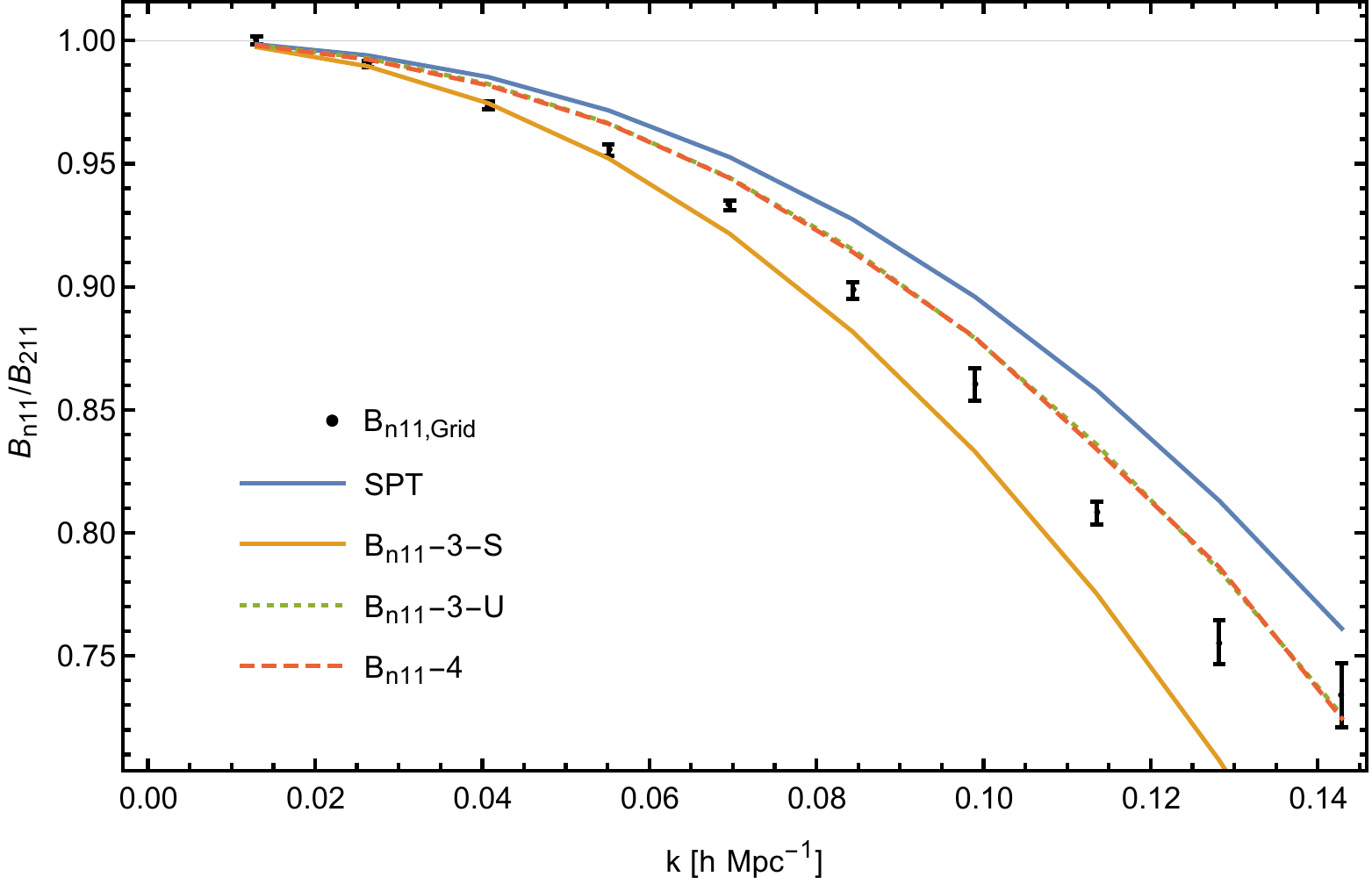}
    \centering  
    \includegraphics[width=0.49\textwidth]{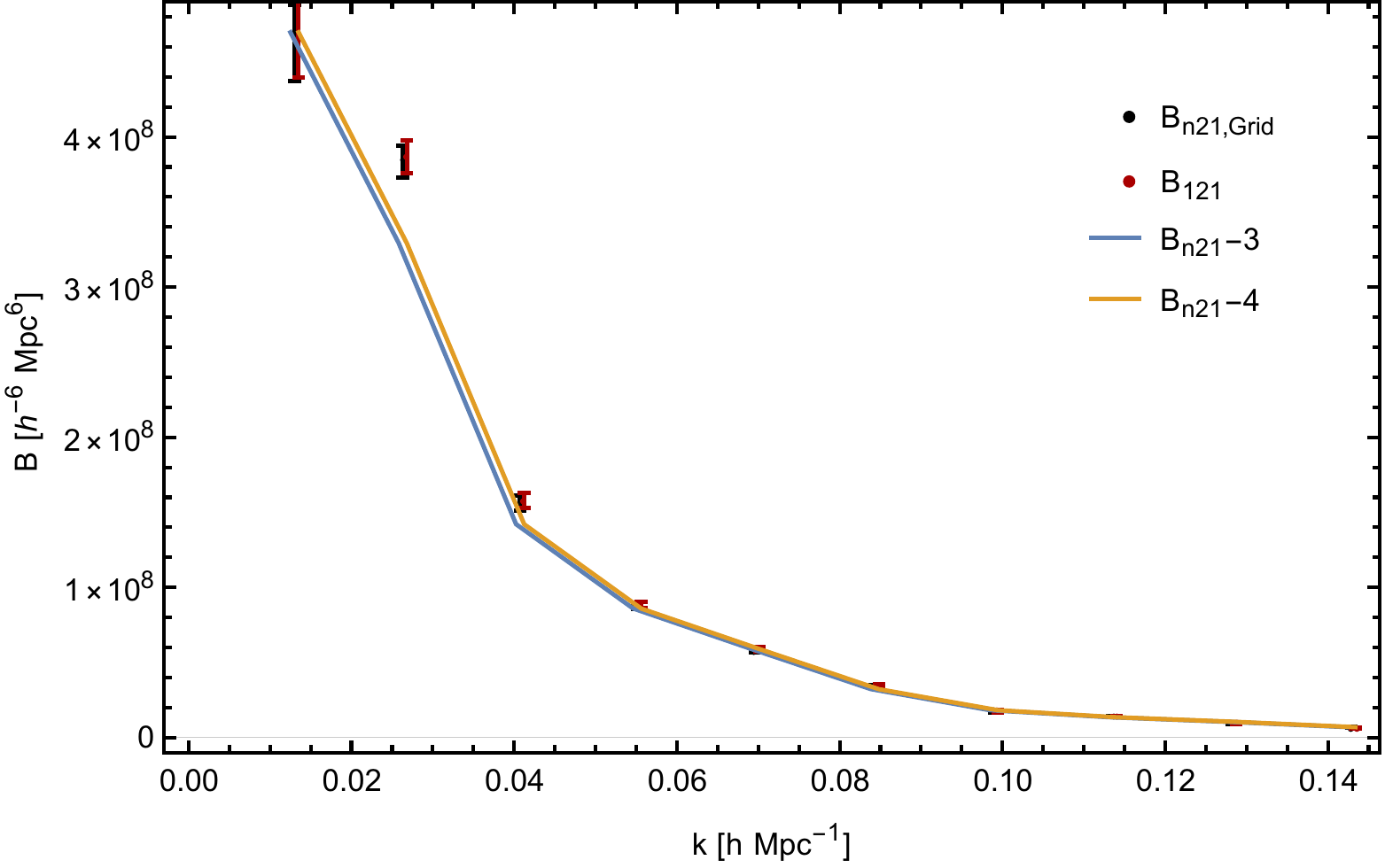}
    \includegraphics[width=0.49\textwidth]{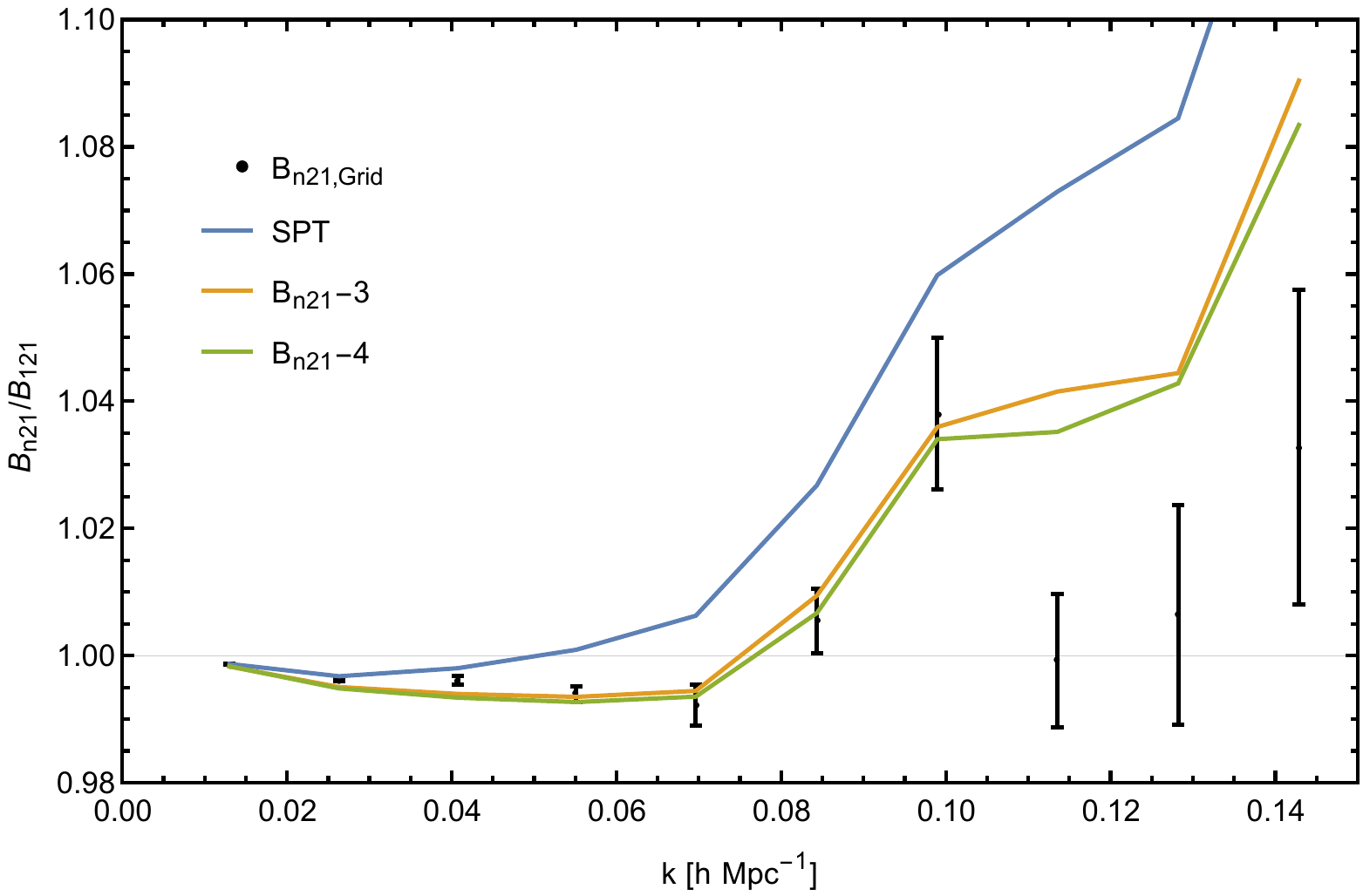}
    \caption{\emph{Top left panel:} The measured equilateral $B_{\mathrm{nnn}}^{\mathrm{s}}$ minus the noise terms against the auto bispectrum up to one-loop calculated from perturbation theory using the values of the counterkernels taken from a variety of fitting procedures.  \emph{Top right panel:} The ratio $B_{\mathrm{nnn}}^{\mathrm{s}}/B_{211}^{\mathrm{s}}$ for both the measured equilateral residual and the calculated equilateral bispectra with a number of fitting procedures as well as the SPT up to one-loop without counterterms.  Many of the fits are good at lower momenta, where the tree-level and one-loop terms dominate, and begin to diverge from the residual at around $k\sim 0.1 \ihMpc$, roughly the limit where two-loop terms would be expected to come to dominate the bispectrum.  \emph{Centre left panel:} The measured equilateral $B_{\mathrm{n}11}$ minus the noise terms against the one-loop bispectrum propagator calculated from perturbation theory using the values of the counterkernels calculated with a number of fitting procedures, together with $B_{211}$.  \emph{Centre right panel:} The ratio $B_{\mathrm{n}11}/B_{211}$ for both the measured and calculated equilateral bispectra with $k_{\mathrm{max}}=0.084\ihMpc$ together with the SPT up to one-loop without counterterms.  The fit works up until roughly this $k$, at which point there becomes a noticeable deviation between the residual and the counterterms due to the increasing involvement of higher loop terms.  \emph{Bottom left panel:} The measured equilateral $B_{\mathrm{n}21}$ minus the noise terms against its one-loop perturbation theory estimator using the counterkernels calculated using methods $B_{\mathrm{n21}}$-3 and $B_{\mathrm{n21}}$-4, together with the SPT up to one loop without counterterms.   \emph{Bottom right panel:} The ratio $B_{\mathrm{n}21}/B_{211}$ for both the measured and calculated equilateral bispectra together with the SPT up to one loop without counterterms.  The fit works until $k=0.1\ihMpc$, indicating that this is the limit at which the exclusion of two-loop terms makes accurate regularisation impossible.}
    \label{Bres}
\end{figure}
\begin{figure}[p]
    \centering  
    \includegraphics[width=0.49\textwidth]{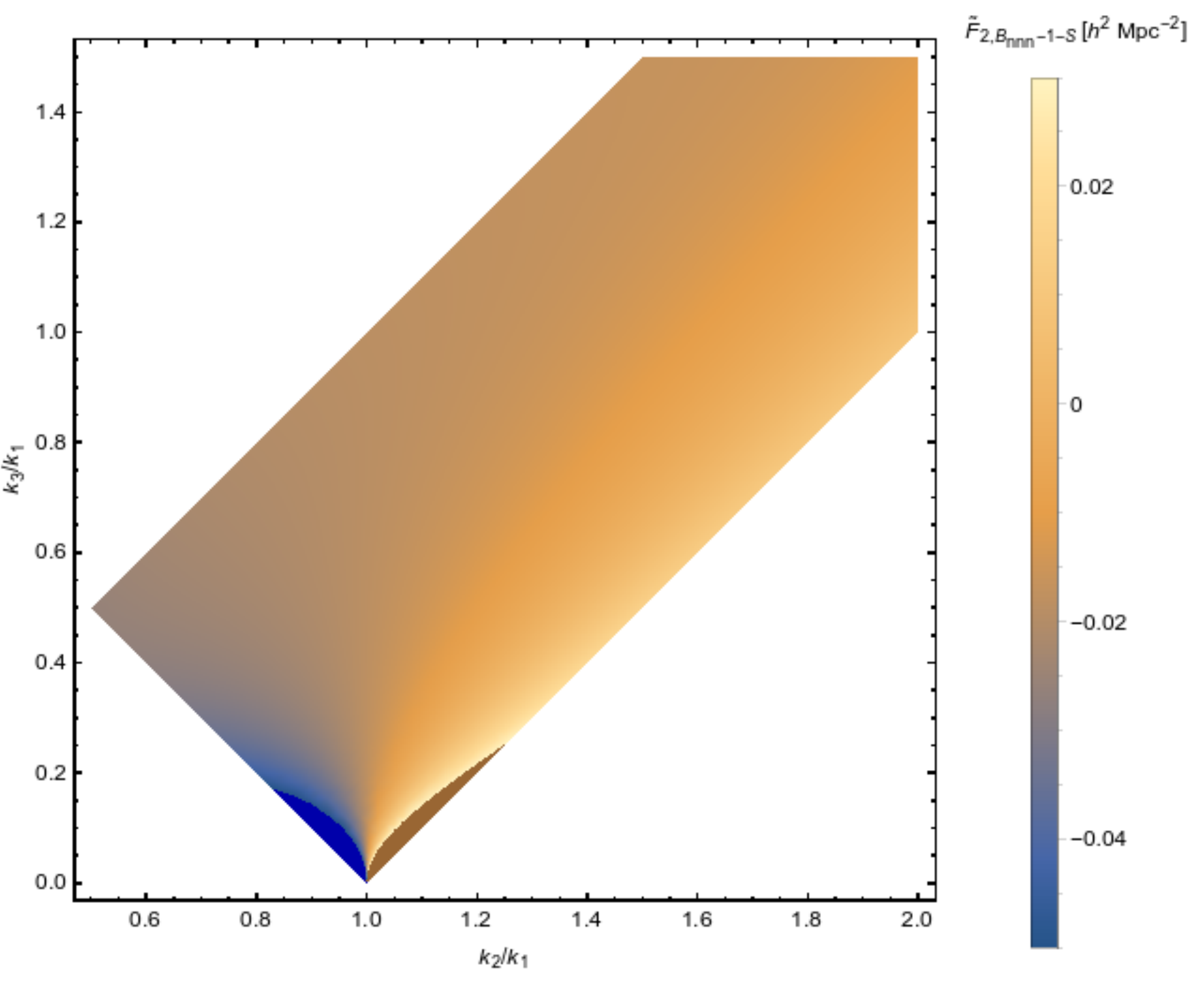}
    \includegraphics[width=0.49\textwidth]{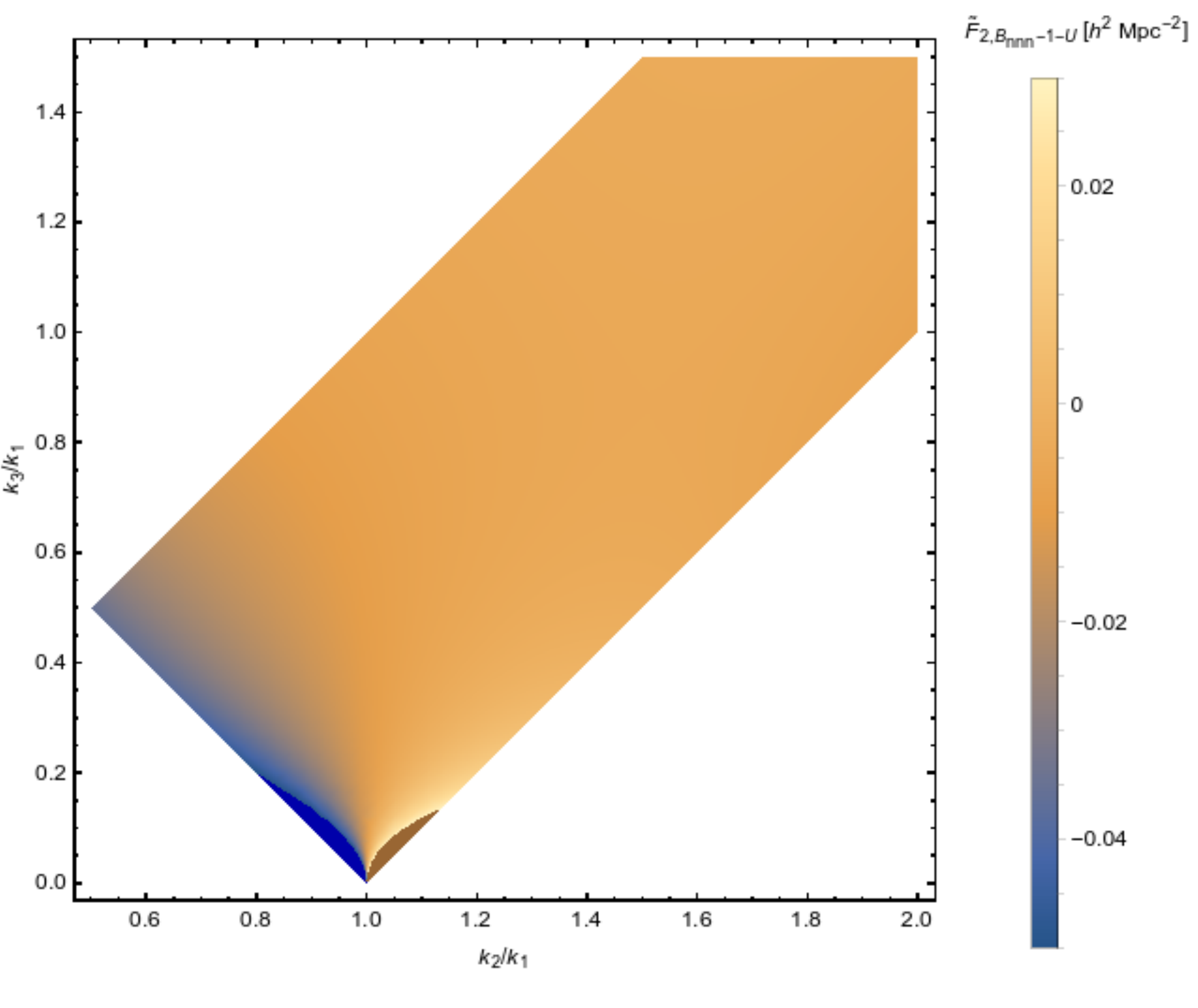}
    \includegraphics[width=0.49\textwidth]{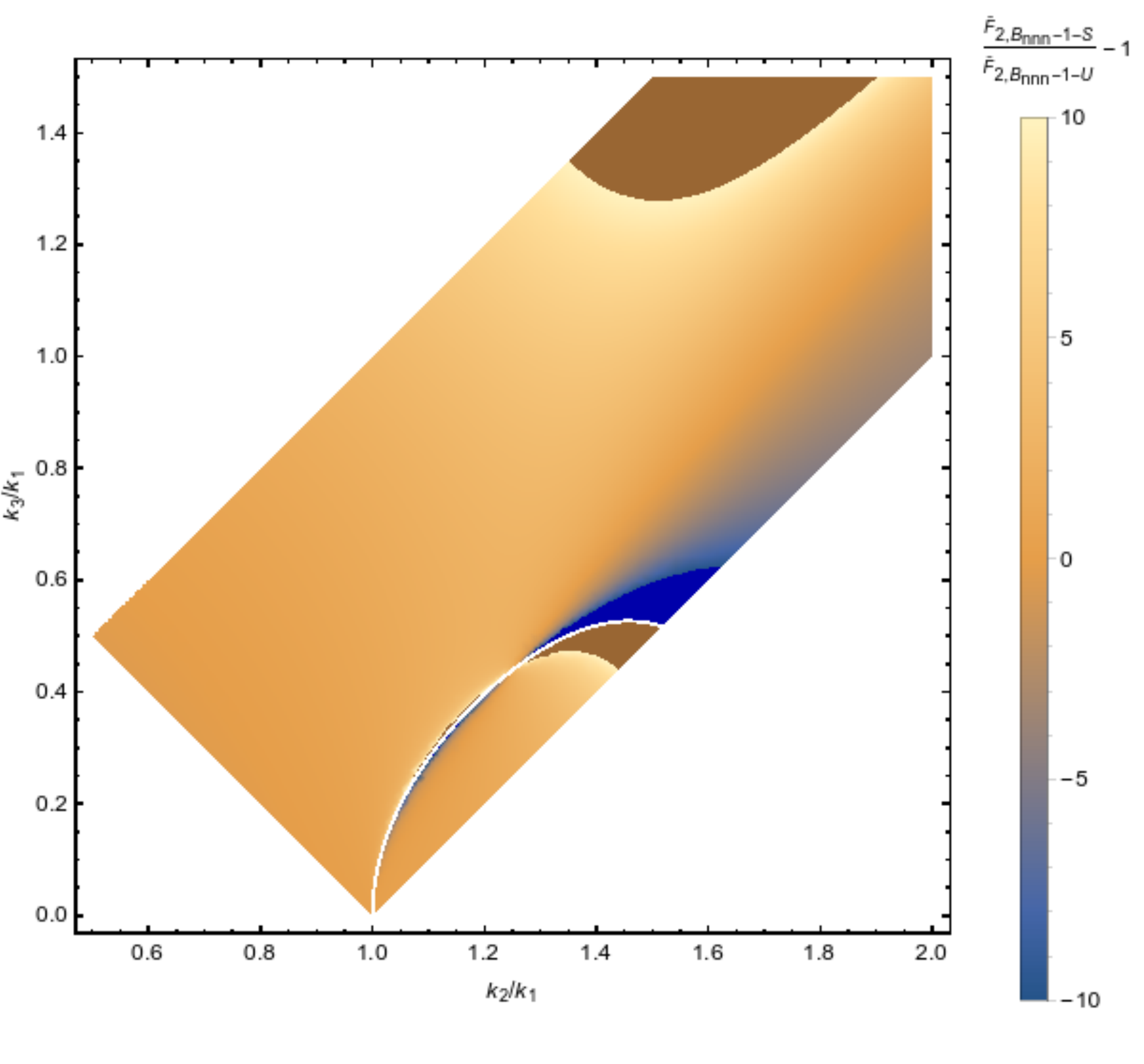}
    \caption{\emph{Top left panel:}  The value of $\tilde{F}_{2}$ as a function of the ratios $k_{2}/k_{1}$ and $k_{3}/k_{2}$ with the $B_{\mathrm{nnn}}$-1-S parametrisation. There is a strong configuration dependence to the calculated counterterms, the sign crossing occurring in the isosceles case and becoming more prominent in the squeezed limit, that being when $k_{3}\rightarrow 0$, when $k_{1}>k_{2}$ and increasingly positive values in the squeezed limit as $k_{1}<k_{2}$.  The dark blue regions show clipping for negative values and the brown regions show clipping for positive values.  \emph{Top right panel:} The value of $\tilde{F}_{2}$ as a function of the ratios $k_{2}/k_{1}$ and $k_{3}/k_{2}$ with the $B_{\mathrm{nnn}}$-1-U parametrisation.  As before, there is a noticeable configuration dependence with the value of $\tilde{F}_{2}$ tending towards increasingly negative values as it descends into the squeezed limit with $k_{1}>k_{2}$ and increasingly positive values as it descends into the same limit with $k_{1}<k_{2}$.  However, the configuration dependence seems to be less strong than that of the symmetry inspired fit, with a much shallower gradient visible throughout most of the plot.  We can clearly see that the UV inspired fitting shows less configuration dependence than the symmetry inspired fit; this is the result of the fixing of the shape functions to be linear functions of $\gamma$, where it is important to remember that the shape functions encode the configuration dependence of the counterterm.  The dark blue regions show clipping for negative values and the brown regions show clipping for positive values.  \emph{Bottom panel:} The fractional deviation of the above two calculations, $\tilde{F}_{2,\mathrm{B_{\mathrm{nnn}}-1-S}}/\tilde{F}_{2,\mathrm{B_{\mathrm{nnn}}-1-U}}-1$.  The dark blue regions show clipping for negative values and the brown regions show clipping for positive values. }
    \label{ratio}
\end{figure}

Interestingly, the results from the $B_{\mathrm{nnn}}$-3 fitting procedure give a good fit to the simulations according to Fig.~\ref{chi2} and regularise the one-loop auto bispectrum  approximately as well as those from the $B_{\mathrm{nnn}}$-1-S fitting, in spite of the differing results these procedures gave for their individual counterparameters.  This indicates that the degeneracy did not affect the ability of the minimisation to give a good fit for the counterterm, merely that allowing the degenerate parameters to vary independently allowed a variety of combinations of counterparameter values to be sampled which were all still able to accurately regularise the auto bispectrum up to $k_{\mathrm{max}}\sim 0.1\ihMpc$.  It is notable that even the $B_{\mathrm{nnn}}$-3-U produced accurate results, in contrast to those of $B_{\mathrm{nnn}}$-1-U, indicating that allowing the $\gamma$ parameters to vary separately compensated for the inaccuracy introduced by fixing the $\epsilon_{i}$ terms.

While Fig.~\ref{Bres} gave us a clear test of the accuracy of the various fittings, we also wish to compare the configuration dependence of the UV and symmetry inspired parametrisations.  In Fig.~\ref{ratio} we plot the calculated values of $\tilde{F}_{2}$ as a function of the ratios of the momentum magnitudes $k_{3}/k_{1}$ and $k_{2}/k_{1}$ for both fitting procedures $B_{\mathrm{nnn}}$-1-S and $B_{\mathrm{nnn}}$-1-U, together with the relative deviation between the shape dependence of the two operators.  As can be seen, the UV approximation shows less configuration dependence, only noticeably decreasing in the squeezed limit as $k_{3}$ becomes much smaller than the other two momentum magnitudes.  As well as producing a less accurate fit to the simulation residuals, the ultraviolet approximation appears to produce less shape dependence in its results, which is understandable as the approximation consists of setting the shape function parameters to be linear functions of $\gamma_{2}$, preventing the counterterm from independently fitting to any given configuration.

\section{Summary \& Discussion}\label{sec:conc}
In this paper, we performed a precise calibration and test of the one-loop bispectrum and its counterterms in the framework of the EFTofLSS, comparing different parametrisation schemes for the counterterms. For the first time, we have considered the bispectrum propagator terms $B_{\mathrm{n}11}$ and $B_{\mathrm{n}21}$ in this context. These cross-bispectra isolate specific counterterms and allow for cross-validation of the fitting procedures and range of validity.  $B_{\mathrm{n}21}$ contains the counterterm to $B_{321}$ in isolation, which is directly related to the power spectrum counterterm for $P_{31}$.  $B_{\mathrm{n}11}$ in turn contains the new bispectrum counterterm regularizing $B_{411}$ in isolation.  The full matter bispectrum contains the symmetrised version of both terms.

To uncover the counterterms on large scales, we evaluate our perturbative predictions for the very modes used to seed the simulations.  This grid based realisation perturbation theory approach allows for the removal of odd correlators from measured clustering statistics and the even correlators share fluctuations with the measurements. The combination of both effects leads to a sufficient reduction on the error bars to allow for the detection of the sub-leading corrections we are after.  We have confirmed that this cosmic variance cancellation significantly reduces the magnitudes of the resultant error bars for the parameter constraints. This cosmic variance cancellation enables the estimation of EFT parameters from smaller simulation volumes. The realisation perturbation theory approach would thus be uniquely suited to constrain the EFT parametrisation of baryonic physics \cite{Lewandowski:2014rca} from numerically demanding, small-volume hydrodynamic simulations.

We have also shown that the use of the commonly used EdS approximation for the growth factors for the second order density field in the tree-level bispectrum, exceeds the one-loop corrections on large scales. We thus implemented the exact \lcdm\ growth factors at quadratic and cubic order.

We found evidence for non-zero correction terms even at order $k^0$ in the EFT. These corrections are artifacts of the numerical integration of the $N$-body system, most likely round off and time stepping issues. Allowing for these bias-like terms leads to more consistent results for the actual EFT counterterms at order $k^2$ on large scales. While the $k^0$ terms are a nuisance parameter for our purposes, their detection on large scales can be used as a diagnostic for  simulation accuracy.

The reduced $\chi^{2}$ clearly shows that the symmetry inspired models tend to provide better fits to the measurements over a wider $k$-range than the UV inspired parametrisations. Yet, we find that the $\chi^2$ degrades past $k_\text{max}=0.09 \ihMpc$, which is in line with what was found in \cite{Baldauf:2015aha} for the one-loop power spectrum and also with what the theoretical errors in Sec.~\ref{sec:theoerr} suggest.  
We find that the measurements of $\gamma_2$ are mostly compatible with the power spectrum value of $c_{\text{s},\infty}^2=1.2 \hMpcsq$.  Future work should investigate the impact of the full \lcdm~time dependence and full covariance matrix.  We have considered the latter but found that our simulation suite is too small to get reliable estimates of the covariance matrix after cosmic variance cancellation.  The leading non-Gaussian covariance of the bispectra without CVC could be estimated from perturbation theory on the grid \cite{Taruya:2020qoy}, but after subtracting the perturbative orders, the remainder is dominated by high orders in perturbation theory and the stochastic contributions. The poor $\chi^2$ obtained in our fits to the propagator terms suggests that IR-resummed operators, such as those motivated by Lagrangian perturbation theory \cite{Baldauf:2015tla,Schmittfull:2018yuk} might be beneficial to improve the agreement between theory and simulations.

In \cite{Schmittfull:2014tca,Abidi:2018eyd} quadratic and cubic bias parameters were obtained by cross-correlating quadratic and cubic operators with the halo field.  While the bias parameters start at $k^0$ in comparison to the EFT counterterms that start at $k^2$, a similar approach might work to constrain EFT parameters.  While this method makes estimation less complex, it buries some of the shape dependence inside the estimator.  We thus preferred to extract and match the full bispectrum in the present study and leave the simplified estimator for future work.
\begin{acknowledgements}
We would like to thank T. Nishimichi, E. Pajer, F. Schmidt, M. Simonovic, A. Taruya and M. Zaldarriaga for fruitful discussions and K. Kornet for excellent computing support. This research made use of the COSMOS supercomputer at DAMTP, Cambridge. TS acknowledges support through the Science and Technology Facilities Council Doctoral Training Centre in Data Intensive Science. TB is supported by the Stephen Hawking Advanced Fellowship at the Center for Theoretical Cosmology.
\end{acknowledgements}

\appendix
\section{Parameter Inference}
\label{fm}
The EFT models used in this study are linear in the parameters. In this Appendix we will briefly review the derivation of the Fisher matrix for a generic linear model before showing how it can be used to calculate the cross-correlation of any two parameters from that model and the error bars of any given parameter.
We will consider a measurement $\vec B_\text{meas}$ with a diagonal covariance matrix and variance $\vec {\Delta B}^{2}_{\mathrm{meas}}$ and a generic theory model that is linear in the parameters such that $\vec B_{\mathrm{model}}=\sum_{i}\alpha_i \vec{B}_{\text{model},i}$. Assuming Gaussian errors, we thus have 
\begin{equation}
\chi^{2}=\sum_{k_s=k_{\mathrm{min}}}^{k_{\mathrm{max}}}\frac{\left[B_{\mathrm{meas},s}-\sum_{i}\alpha_{i}B_{\text{model},i,s}\right]^{2}}{\Delta B^{2}_{\mathrm{meas},s}}~.
\label{chi2fm}
\end{equation}
Taking the first derivative of Eq.~\eqref{chi2fm} with respect to any given parameter $\alpha_{i}$ gives us
\begin{equation}
    \frac{\derd\chi^{2}}{\derd\alpha_{i}}=2 \sum_{k_s=k_{\mathrm{min}}}^{k_{\mathrm{max}}} B_{\text{model},i,s}\frac{B_{\mathrm{meas},s}-\sum_{j}\alpha_{j}B_{\text{model},j,s}}{\Delta B^{2}_{\mathrm{meas},s}}\, .
\end{equation}
Setting this first derivative to zero gives us the linear system
\begin{equation}
    \sum_{k_s=k_{\mathrm{min}}}^{k_{\mathrm{max}}} \frac{B_{\text{model},i,s} B_{\mathrm{meas},s}}{\Delta B^{2}_{\mathrm{meas},s}}
    =\sum_{j}\alpha_{j}
    \sum_{k_s=k_{\mathrm{min}}}^{k_{\mathrm{max}}} \frac{B_{\text{model},i,s}B_{\text{model},j,s}}{\Delta B^{2}_{\mathrm{meas},s}}\, .
\end{equation}
Taking the second derivative of which for any given parameters $\alpha_{i}$ and $\alpha_{j}$ gives us
\begin{equation}
    \frac{\derd^{2}\chi^{2}}{\derd\alpha_{i}\derd\alpha_{j}}=2 \sum_{k_s=k_{\mathrm{min}}}^{k_{\mathrm{max}}} \frac{B_{\text{model},i,s}B_{\text{model},j,s}}{\Delta B^{2}_{\mathrm{meas},s}}\equiv\mathcal{F}_{ij}~,
\end{equation}
where $\mathcal{F}$ is the Fisher matrix of the model in question.  

We can now calculate the cross-correlation of any two parameters by looking at the ratio between the product of their isolated elements in the inverse Fisher matrix and their combined element, as given by 
\begin{equation}
    C_{ij}=\frac{\mathcal{F}_{ij}^{-1}}{\sqrt{\mathcal{F}_{ii}^{-1}\mathcal{F}_{jj}^{-1}}}~.
\end{equation} 
A strong cross-correlation, that being $C_{ij}\approx 1$ or $C_{ij}\approx -1$, means that the parameters $\alpha_{i}$ and $\alpha_{j}$ are degenerate; they correlate so strongly with one another that only one of the two is needed to determine the value of them both and allowing them both to vary freely will result in results for the individual parameters that differ markedly from those of non-degenerate parametrisations of the same functions as the model is effectively being allowed to vary one parameter in two different ways simultaneously.  However, it is important to note that while degenerate parameters may give differing results for their parameters, they can still give accurate fits for the overall model.

We can also calculate the error bars for any given parameter as 
\begin{equation}
\sigma_{i}=\sqrt{\mathcal{F}_{ii}^{-1}}
\end{equation}
and it is this definition that we use for the error bars in the figures for $\gamma_{1,2}$ and $\epsilon_{1,2,3}$ in Sec.~\ref{results}.

\section{The Growth Factor}
\label{app:exgrowth}
Here we review the $\Lambda$CDM growth factors, reproducing the work shown in \cite{Exact}.  

The $n$-th order growth factor $D_{n}$ describes the growth of the density perturbation $\delta_{n}$ as the universe expands.  It is common practice to assume EdS cosmology, such that $D_{n}=a^{n}=1$ for all $D$ at the present time, to simplify calculations.  However, this approximation becomes increasingly inaccurate as our universe is no longer dominated by matter but by dark energy.

The linear growth factor $D_{1}$ is given by the solutions to 
\begin{equation}
\frac{\derd^{2}}{\derd \mathrm{ln}(a^{2})}\frac{D_{1}}{a}+\left(4+\frac{\derd\ln H}{\derd\ln a}\right)\frac{\derd}{\derd\ln a}\frac{D_{1}}{a}+\left[3+\frac{\derd\ln H}{\derd\ln a}-\frac{3}{2}\Omega_\text{m}(a)\right]\frac{D_{1}}{a}=0~.
\end{equation}
with initial conditions chosen to reflect at early times the universe was approximately EdS such that at $a\rightarrow 0$, $D_{1}(a)/a\rightarrow 1$.  The second order solutions are found by solving
\begin{equation}
\begin{split}
&\frac{\derd^{2}}{\derd \ln a^{2}}\frac{D_{2}}{a^{2}}+\left(6+\frac{\derd\ln H}{\derd\ln a}\right)\frac{\derd}{\derd \ln a}\frac{D_{2}}{a^{2}}+\left[8+2\frac{\derd\ln H}{\derd\ln a}-\frac{3}{2}\Omega_\text{m}(a)\right]\frac{D_{2}}{a^{2}}\\
&=\begin{cases}\frac{7}{5}\left[\left(\frac{\derd D_{1}}{\derd a}\right)^{2}+\frac{3}{2}\Omega_\text{m}(a)\left(\frac{D_{1}}{a}\right)^{2}\right]&\mathrm{for} ~D_{2A}~,\\
\frac{7}{2}\left(\frac{\derd D_{1}}{\derd a}\right)^{2}&\mathrm{for} ~D_{2B}~,
\end{cases}
\end{split}
\end{equation}
with initial conditions
\begin{equation}
a(t_{0})=0, ~\frac{D_{2A,B}}{a^{2}}=1, ~\mathrm{and} ~\frac{\derd}{\derd a}\frac{D_{2A,B}}{a^{2}}=0~.
\end{equation}
This gives two solutions, $D_{2A}$ and $D_{2B}$, such that 
\begin{equation}
\delta_{2}(\mathbf{k},a)=D_{2A}(a)A(\mathbf{k})+D_{2B}(a)B(\mathbf{k})~.
\end{equation}  
By inserting the linear density field into Equations~(\ref{eom1} and \ref{eom2}), one can derive the explicit terms for the higher order fields.  The solution for the second order field is given by
\begin{align}
A(\mathbf{k})&=\frac{5}{7}\int \derd^{3}\mathbf{q}~\alpha(\mathbf{q},\mathbf{k}-\mathbf{q})\delta_{1}(\mathbf{q})\delta_{1}(\mathbf{k}-\mathbf{q}),\\
B(\mathbf{k})&=\frac{2}{7}\int \derd^{3}\mathbf{q}~\beta(\bq,\bk-\bq)\delta_{1}(\bq)\delta_{1}(\bk-\bq)~,
\end{align}
The third order growth factors can be found by solving
\begin{equation}
\begin{split}
&\frac{\derd^{2}}{\derd \ln a^{2}}\frac{D_{3}}{a^{3}}+\left(8+\frac{\derd\ln H}{\derd\ln a}\right)\frac{\derd}{\derd\ln a}\frac{D_{3}}{a^{3}}+\left[15+3\frac{\derd\ln H}{\derd \ln a}-\frac{3}{2}\Omega_\text{m}(a)\right]\frac{D_{3}}{a^{3}}\\
&=
\begin{cases}
\frac{18}{7}\left[2\frac{\derd D_{1}}{\derd a}+\frac{3}{2}\Omega_\text{m}(a)\frac{D_{1}}{a}\right]\frac{D_{2A,B}}{a^{2}}+\frac{18}{7}a\frac{\derd D_{1}}{\derd a}\frac{\derd}{\derd a}\frac{D_{2A,B}}{a^{2}}&\mathrm{for} ~D_{3AA,3AB}~,\\
15\frac{\derd D_{1}}{\derd a}\left[a\frac{\derd}{\derd a}\frac{D_{2A}}{a^{2}}+2\frac{D_{2A}}{a^{2}}-\frac{7}{5}\frac{D_{1}}{a}\frac{\derd D_{1}}{\derd a}\right]&\mathrm{for} ~D_{3BA}~,\\
\frac{9}{2}\frac{\derd D_{1}}{\derd a}\left[a\frac{\derd}{\derd a}\frac{\derd D_{2B}}{a^{2}}+2\frac{D_{2B}}{a^{2}}\right]&\mathrm{for} ~D_{3BB}~,
\end{cases}
\end{split}
\end{equation}
in combination with the consistency conditions
\begin{align}
D_{3AA,2}=&3D_1^3-\frac23 D_{3AB,1}-\frac47D_{3BA}-\frac{16}{21}D_{3BB}\\
D_{3AB,2}=&\frac94D_1^3-\frac54 D_{3AA,1}
\end{align}
with initial conditions
\begin{equation}
\frac{D_{3}}{a^{3}}=1, ~\frac{\derd}{\derd a}\frac{D_{3}}{a^{3}}=0~.
\end{equation}
This gives the solution
\begin{equation}
\begin{split}
\delta_{3}(\bk,a)=&D_{3AA,1}(a)C_{AA,1}(\bk)+D_{3AA,2}(a)C_{AA,2}(\bk)+D_{3AB,1}(a)C_{AB,1}(\bk)\\&+D_{3AB,2}(a)C_{AB,2}(\bk)+D_{3BA}(a)C_{BA}(\bk)+D_{3BB}(a)C_{BB}(\bk)~,
\end{split}
\end{equation}
where
\begin{align}
C_{AA,1}(\mathbf{k})&=\frac{7}{18}\int d^{3}\mathbf{q}~\alpha(\mathbf{q},\mathbf{k}-\mathbf{q})\delta_{1}(\mathbf{q})A(\mathbf{k}-\mathbf{q}),\\
C_{AA,2}(\mathbf{k})&=\frac{7}{30}\int d^{3}\mathbf{q}~\alpha(\mathbf{q},\mathbf{k}-\mathbf{q})\delta_{1}(\bk-\bq)A(\bq),\\
C_{AB,1}(\mathbf{k})&=\frac{7}{18}\int d^{3}\mathbf{q}~\alpha(\mathbf{q},\mathbf{k}-\mathbf{q})\delta_{1}(\mathbf{q})B(\mathbf{k}-\mathbf{q}),\\
C_{AB,2}(\mathbf{k})&=\frac{7}{9}\int d^{3}\mathbf{q}~\alpha(\mathbf{q},\mathbf{k}-\mathbf{q})\delta_{1}(\bk-\bq)B(\bq),\\
C_{BA}(\mathbf{k})&=\frac{2}{15}\int d^{3}\mathbf{q}~\beta (\mathbf{q},\mathbf{k}-\mathbf{q})\delta_{1}(\mathbf{q})A(\mathbf{k}-\mathbf{q}),\\
C_{BB}(\mathbf{k})&=\frac{4}{9}\int d^{3}\mathbf{q}~\beta (\mathbf{q},\mathbf{k}-\mathbf{q})\delta_{1}(\bk-\bq)B(\bq),
\end{align}
The differences between the EdS and $\Lambda$CDM growth factors is small, but may be large enough to cause percent scale errors in calculations.  In Sec.~\ref{propanalysis} and \ref{autoanalysis} we compare results for the second and third order density fields with both growth factors and show that the differences can be significant for quantifying the sub-leading corrections induced by one-loop perturbation theory.

\section{Linear Growth Corrections}
\label{app:lingrowcorr}
In this Appendix we briefly discuss the bias-like $k^0$ corrections to the linear and quadratic matter density fields that we introduced to account for inaccurate growth factors in the the $N$-body simulations. 

The time evolution in the simulations has finite precision and can lead to sub-percent level inaccuracies in the density fields measured at $z=0$ ($a=1$). To account for these numerical errors, we introduce corrective terms into the definition of the first and second order density fields and fit for them as independent parameters of the model, giving us the following non-linear density field up to fourth order:
\begin{equation}
    \delta_\text{n}(\vec x,t)=(1+\Delta D_1)\delta_1(\vec x,t)+(1+\Delta D_2)\delta_2(\vec x,t)+\delta_3(\vec x,t)+\delta_4(\vec x,t)~.
\end{equation}
Beyond leading order, this parametrisation might not be the most generic ansatz, but we will leave a more detailed study of the growth systematics in simulations to future inquiry. 

Fig.~\ref{dd} shows the calculated value of the growth factor corrections from both the bispectra and the power spectra. Both the auto power spectrum and propagator measurements asymptote to $\Delta D_1\approx -2.5\times 10^{-4}$ on large scales. The range over which the $\Delta D_1$ measurements agree conforms with the range over which the $\cssq$ measurements agree. The clear detection of this linear growth correction indicates that the linear growth in the simulations has a fractional systematic error of the same magnitude. While an offset this small might not be of any relevance for survey analysis, our ability to detect this offset proves the power of our realisation based perturbation theory approach. It can in fact be used to diagnose the accuracy of the $N$-body code on large scales.

\begin{figure}[h!]
    \centering  
    \includegraphics[width=0.49\textwidth]{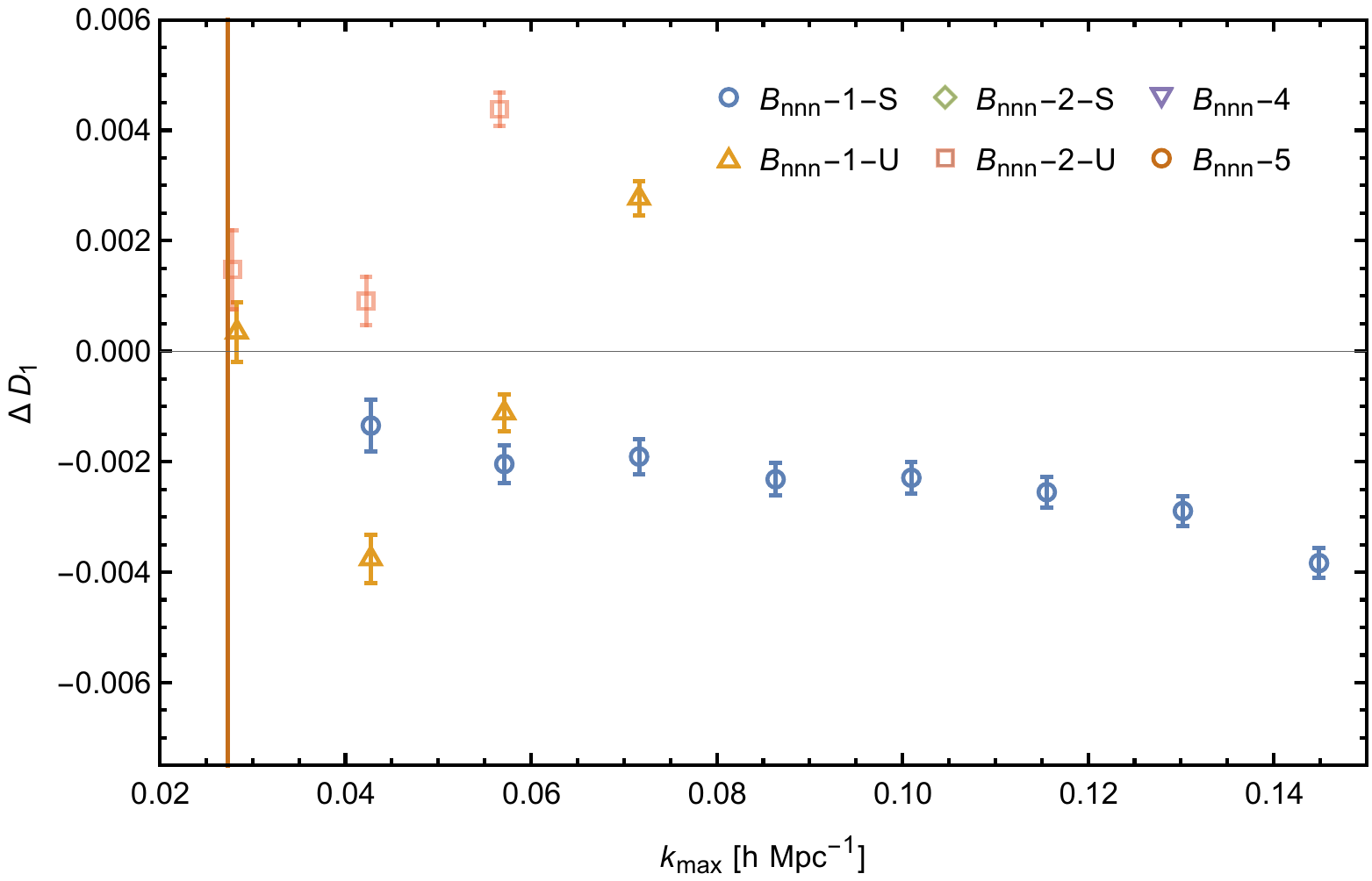}
    \includegraphics[width=0.49\textwidth]{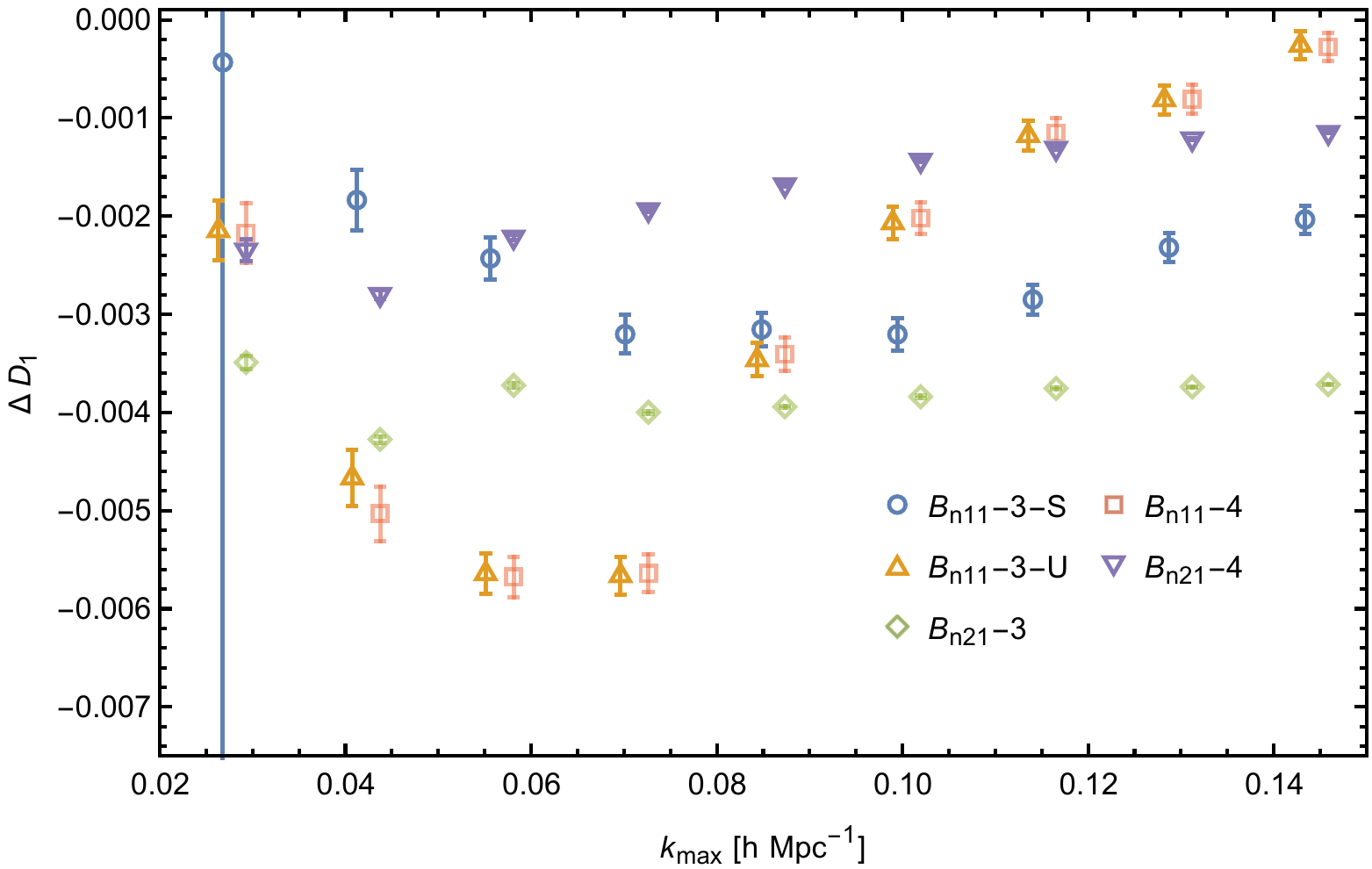}
    \includegraphics[width=0.49\textwidth]{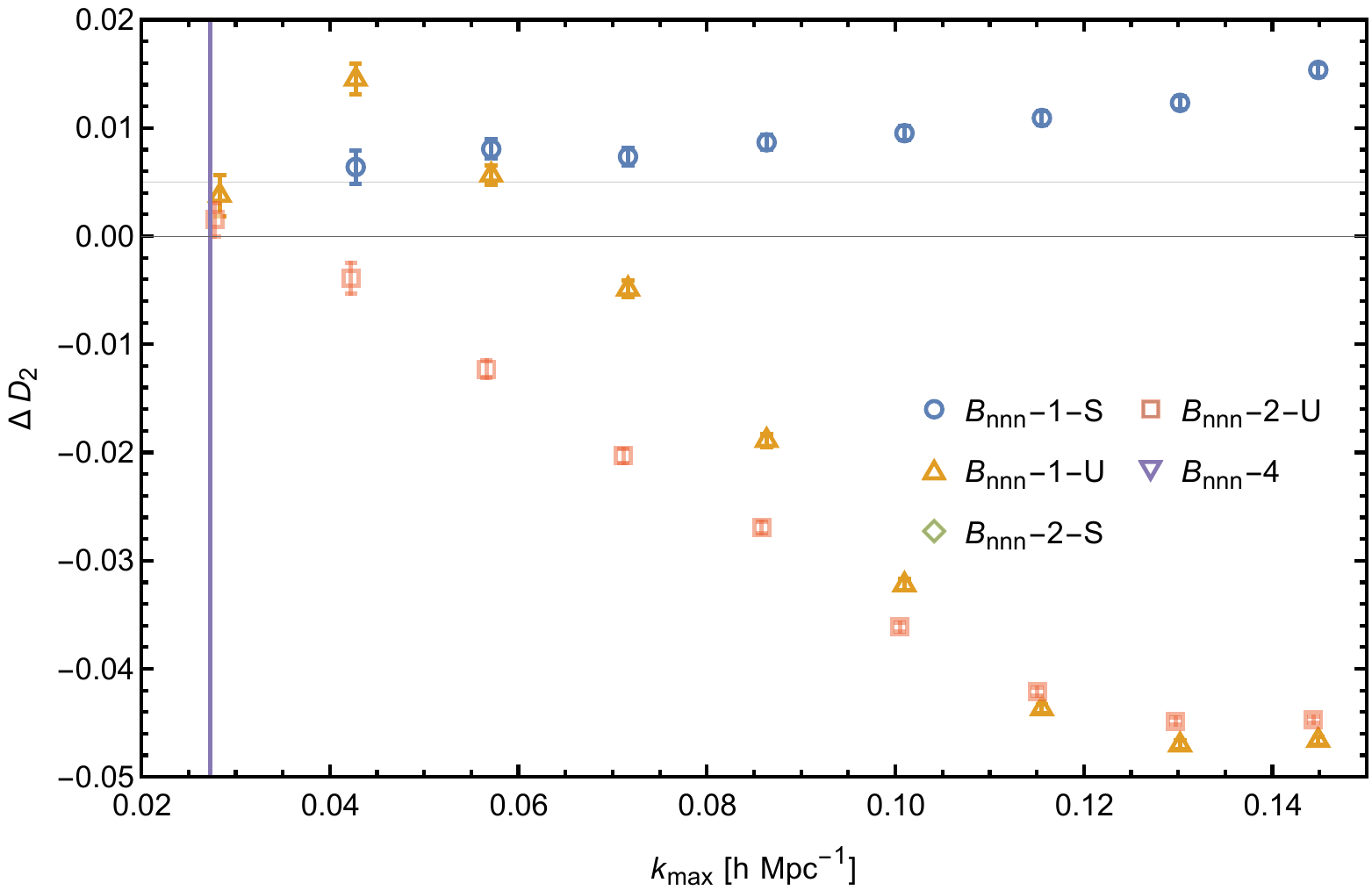}
    \includegraphics[width=0.49\textwidth]{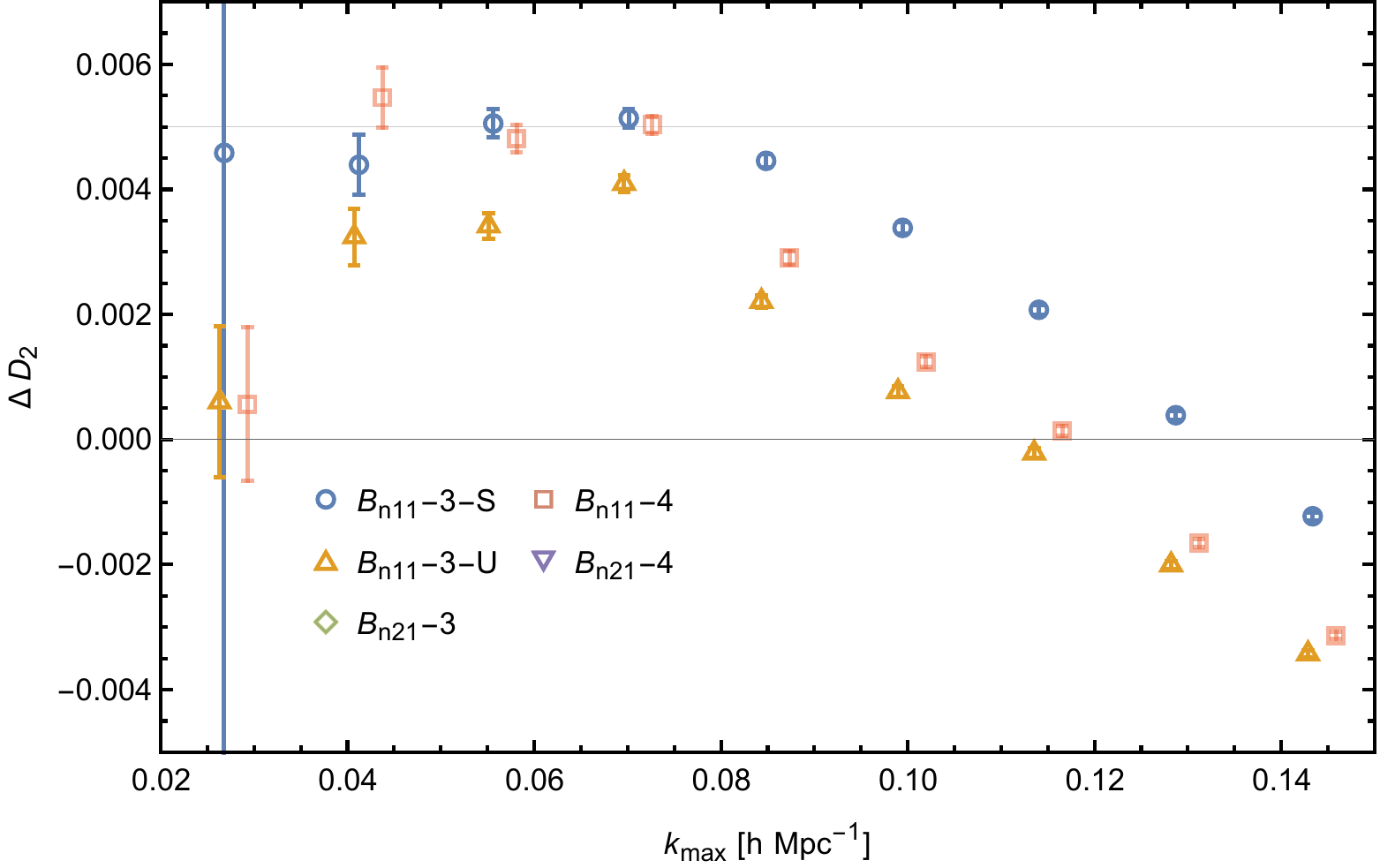}
    \includegraphics[width=0.49\textwidth]{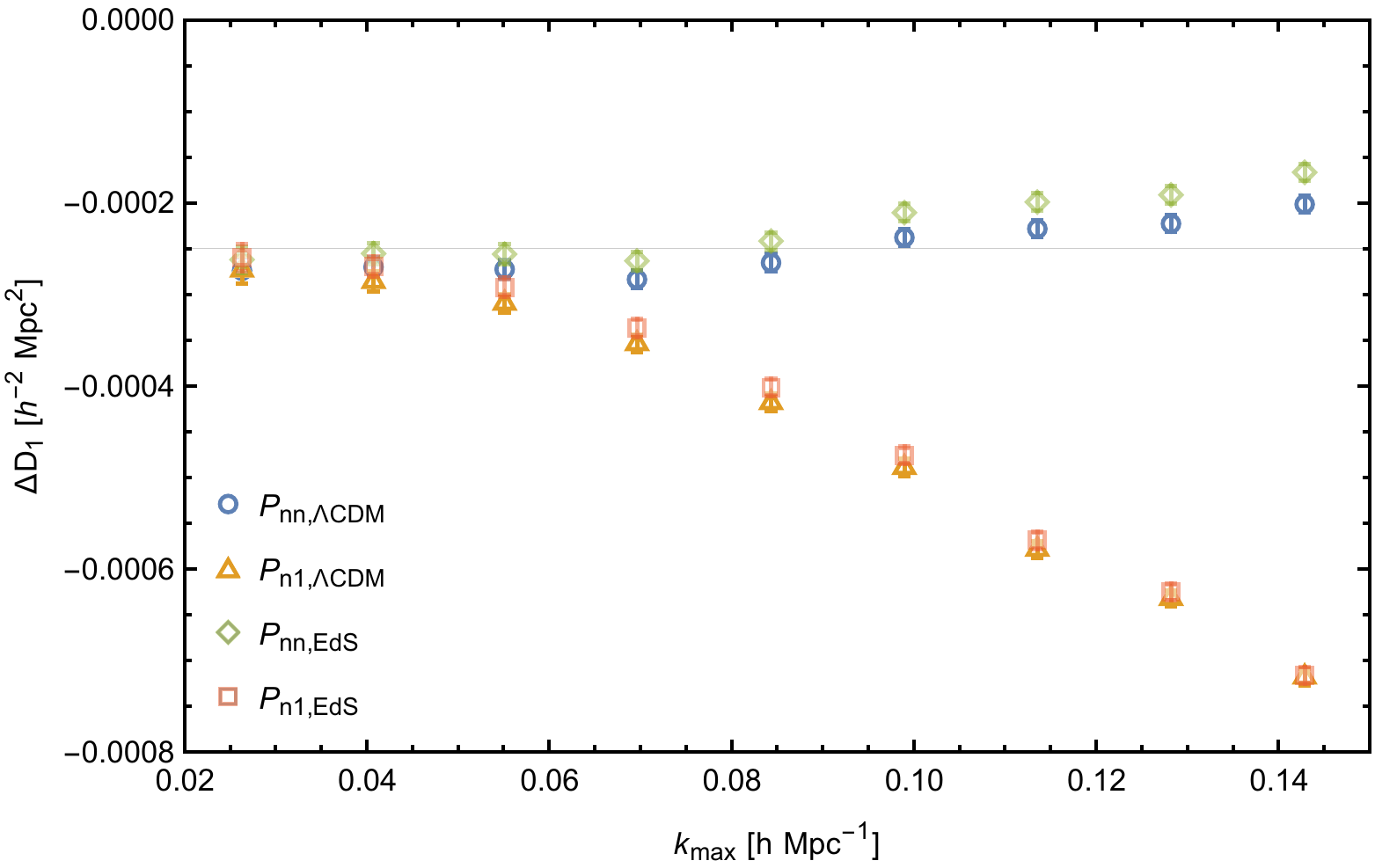}
    \caption{\emph{Top left panel:}  The growth factor correction $\Delta D_{1}$ from the auto bispectrum.  \emph{Top right panel:} The growth factor correction $\Delta D_{1}$ from the propagators.  \emph{Centre left panel:}  The growth factor correction $\Delta D_{2}$ from the auto bispectrum.  The results for fitting procedure $B_{\mathrm{nnn}}$-5 are omitted as they were significantly larger than the others.  \emph{Centre right panel:} The growth factor correction $\Delta D_{2}$ from the propagators.  \emph{Bottom panel:} The growth factor correction $\Delta D_{1}$ as calculated from the power spectra with both $\Lambda$CDM and EdS growth factors.}
    \label{dd}
\end{figure}

The constraints on $\Delta D_1$ from the bispectrum are much less coherent, but it has to be stressed that they are dominated by an odd correlator, the noise term $B_{111}$. The constraints on $\Delta D_2$ are much tighter and point towards $\Delta D_2\approx 0.005$ on large scales, with the constraints from the $B_{n11}$ propagator being the most decisive.

\section{Fitting procedure validation}
\label{d4test}
To validate our fitting and modelling procedure, we use the $B_{411}$ measurement for two different cutoffs as a reference propagator measurement for which we know the exact values of the counterterm amplitude. By applying our fitting procedure to this artificial data set, we can check that our templates are correctly normalised and assess the expected error bars.

We generate our benchmark density field by calculating the difference between the fourth order density fields generated from linear density fields with two different wavenumber cutoffs $\Lambda_1$ and $\Lambda_2$
\begin{equation}
\delta_{4,\mathrm{S}}(k_{1},k_{2},k_{3};\Lambda_{1},\Lambda_{2})=\delta_{4}(k_{1},k_{2},k_{3};\Lambda_{1})-\delta_{4}(k_{1},k_{2},k_{3};\Lambda_{2})\, .
\end{equation}
Since we are calculating a 411 correlator, this is equivalent to calculating the difference between the respective bispectra
\begin{equation}
B_{411,\mathrm{S}}(k_{1},k_{2},k_{3};\Lambda_{1},\Lambda_{2})=B_{411}(k_{1},k_{2},k_{3};\Lambda_{1})-B_{411}(k_{1},k_{2},k_{3};\Lambda_{2})\, .
\end{equation} 
Replacing the residual in the numerator of Eq.~\eqref{chin11} with $B_{411,S}$ gives us 
\begin{equation}
    \chi_{\mathrm{n}11\mathrm{,test}}^{2}(k_{\mathrm{max}})=\sum_{k_{1,2,3}=k_{\mathrm{min}}}^{k_{\mathrm{max}}}\frac{\left[B_{411,\mathrm{S}}(k_{1},k_{2},k_{3};\Lambda_{1},\Lambda_{2})-B_{\tilde{2}11}(k_{1},k_{2},k_{3};\gamma_{2},\epsilon_i)\right]^{2}}{\Delta B^{2}_{\mathrm{n}11}(k_{1},k_{2},k_{3})}.\label{chin11test}
\end{equation}
There is no need for the $\Delta D_1$ and $\Delta D_2$ corrections in this case.
From Eqs.~\eqref{B411UV} and \eqref{B2tUV} we can see that the results of the minimisation of Eq.~\eqref{chin11test} will give values of 
\begin{equation}
\gamma_{2}\approx \frac{61}{210} \left[\sigmadsq(\Lambda_{1})-\sigmadsq(\Lambda_{2})\right]~,
\label{g2s}
\end{equation}
with the $\epsilon_i$ parameters being defined as in Eq.~\eqref{F2tUV}.

The comparison of this analytic calculation with the numerical minimisation can be used as a test of the counterterm implementation and of the minimisation infrastructure. A calculation with $\Lambda_{1}=0.3\ihMpc$ and $\Lambda_{2}=0.2\ihMpc$ shows that they do indeed produce the correct results of $\gamma_{2}$ at the $k_{\mathrm{max}}$ values of interest.

\bibliography{bibliography}

\end{document}